\documentclass{article}%
\usepackage{amsfonts}
\usepackage{amssymb}
\usepackage{amsmath}
\usepackage{graphicx}
\usepackage[onehalfspacing]{setspace}%
\providecommand{\U}[1]{\protect\rule{.1in}{.1in}}

\newtheorem{theorem}{Theorem}

\newtheorem{corollary}{Corollary}

\newtheorem{lemma}[theorem]{Lemma}

\newtheorem{proposition}{Proposition}

\begin{document}

\title{Introduction to Logical Entropy \\and Its Relationship to Shannon Entropy}
\author{David Ellerman\\University of Ljubljana, Slovenia\\Email: david@ellerman.org}
\maketitle

\begin{abstract}
\noindent We live in the information age. Claude Shannon, as the father of the
information age, gave us a theory of communications that quantified an "amount
of information," but, as he pointed out, "no concept of information itself was
defined." Logical entropy provides that definition. Logical entropy\noindent
\ is the natural measure of the notion of information based on distinctions,
differences, distinguishability, and diversity. It is the (normalized)
quantitative measure of the distinctions of a partition on a set--just as the
Boole-Laplace logical probability is the normalized quantitative measure of
the elements of a subset of a set. And partitions and subsets are
mathematically dual concepts--so the logic of partitions is dual in that sense
to the usual Boolean logic of subsets, and hence the name "logical entropy."
The logical entropy of a partition has a simple interpretation as the
probability that a distinction or dit (elements in different blocks) is
obtained in two independent draws from the underlying set. The Shannon entropy
is shown to \textit{also} be based on this notion of
information-as-distinctions; it is the average minimum number of binary
partitions (bits) that need to be joined to make all the \textit{same}
distinctions of the given partition. Hence all the concepts of simple, joint,
conditional, and mutual logical entropy can be transformed into the
corresponding concepts of Shannon entropy by a uniform non-linear dit-bit
transform. And finally logical entropy linearizes naturally to the
corresponding quantum concept. The quantum logical entropy of an observable
applied to a state is the probability that two different eigenvalues are
obtained in two independent projective measurements of that observable on that state.


Classification: MSC 81P45; 94A17

\end{abstract}
\tableofcontents

\section{Introduction}

This paper is an introduction to the concept of logical entropy as the direct
measure of the definition of information in terms of distinctions,
differences, distinguishability, and diversity. The formula for logical
entropy goes back to the early twentieth century, but the current development
comes out of seeing the formula as the quantification of information in a
partition as the normalized number of distinctions or dits (ordered pairs of
elements in different blocks) of the partition. Just as the Laplace-Boole
notion of probability, as the normalized number of elements in a subset,
quantifies the logic of subsets, so logical entropy, as the normalized number
of distinctions in a partition, quantifies the logic of partitions--and hence
the adjective "logical." The logical entropy of a partition is, in fact, a
probability measure--the probability of obtaining a distinction of the
partition in two independent draws from the universe set, just as the logical
Laplace-Boole probability of a subset (or event) is the one-draw probability
of obtaining an element of the subset.

Far from displacing the usual notion of Shannon entropy; the point is to show
that the Shannon entropy of a partition is a different quantification of the
\textit{same} notion of information-as-distinctions, i.e., the average minimum
number of binary partitions (bits) that need to be joined together to make the
\textit{same} distinctions of a partition. In fact, there is a non-linear
dit-to-bit transformation that transforms all the concepts of simple, joint,
conditional and mutual logical entropy into the corresponding formulas for
Shannon entropy, where the latter are especially suited for the theory of
coding and communications.

Edwin Jaynes' MaxEntropy method is intended to generalize the Laplace
indifference principle by determining the `best' probability distribution
consistent with given constraints (e.g., that rule out the uniform
distribution of the indifference principle) by maximizing Shannon entropy
subject to those constraints. We show that maximizing logical entropy subject
to the same constraints gives a different probability distribution. The
logical entropy solution is the closest to the uniform distribution in terms
of the usual notion of (Euclidean) distance while the Jaynes solution is the
closest in terms of the Kullback-Leibler divergence from the uniform
distribution. The notion of information-as-differences also connects to
ordinary statistical theory since the metrical version of logical entropy is
just twice the usual notion of variance (or equals the variance if one counts
unordered pairs), and similarly for the notion of covariance.

There is a quasi-algorithmic method, linearization, that transforms set-based
concepts into vector-space concepts. Applied to the set-based concepts of
`classical' logical entropy, the linearization to Hilbert spaces generates the
quantum versions of logical entropy. The quantum logical entropy of an
observable applied to a quantum state is the probability of getting different
eigenvalues in two independent (projective) measurements of the observable on
that state.

\section{Logical Entropy}

\subsection{Partitions on a set}

A \textit{partition} $\pi=\left\{  B_{1},...,B_{m}\right\}  $ on a finite set
$U=\left\{  u_{1},...,u_{n}\right\}  $ is a set of non-empty subsets
$B_{i}\subseteq U$ called \textit{blocks} that are disjoint and whose union is
all of $U$. A \textit{distinction} or \textit{dit} of $\pi$ is an ordered pair
$\left(  u_{j},u_{k}\right)  \in U\times U$ where $u_{j}$ and $u_{k}$ are in
different blocks of $\pi$. The set of all distinctions of $\pi$ is the
\textit{ditset} $\operatorname{dit}\left(  \pi\right)  \subseteq U\times U$.
An ordered pair $\left(  u_{j},u_{k}\right)  \in U\times U$ is an
\textit{indistinction} or \textit{indit} of $\pi$ if $u_{j}$ and $u_{k}$ are
in the same block of $\pi$, and the set of all indits of $\pi$ is the
\textit{inditset} $\operatorname{indit}\left(  \pi\right)  =\cup_{j=1}%
^{m}\left(  B_{j}\times B_{j}\right)  $. A binary relation $E\subseteq U\times
U$ is an \textit{equivalence relation} on $U$ if it is reflexive (i.e., for
all $u\in U$, $\left(  u,u\right)  \in E$), \textit{symmetric} (i.e., for all
$\left(  u,u^{\prime}\right)  \in E$, $\left(  u^{\prime},u\right)  \in E$),
and transitive (i.e., if $\left(  u,u^{\prime}\right)  \in E$ and $\left(
u^{\prime},u^{\prime\prime}\right)  \in E$, then $\left(  u,u^{\prime\prime
}\right)  \in E$). The inditset $\operatorname{indit}\left(  \pi\right)  $ of
a partition on $U$ is an equivalence relation on $U$. Given an equivalence
relation $E$ on $U$, two elements are said to be \textit{equivalent}, $u\sim
u^{\prime}$, if $\left(  u,u^{\prime}\right)  \in E$. Let $\left[  u\right]
_{E}\subseteq U$ be the set of elements of $U$ equivalent to $u\in U$, i.e.,
an \textit{equivalence class} of $E$. The set of equivalence classes of $E$ is
a partition on $U$ and the inditset of that partition is $E$. Hence the notion
of an equivalence relation and an inditset of a partition are equivalent notions.

Since each ordered pair $\left(  u_{j},u_{k}\right)  \in U\times U$ is either
an dit of $\pi$ or an indit of $\pi$ but not both, the ditset
$\operatorname{dit}\left(  \pi\right)  =U\times U-\operatorname{indit}\left(
\pi\right)  $ is the complement of the inditset in $U\times U=U^{2}$. As a
binary relation $\operatorname{dit}\left(  \pi\right)  \subseteq U\times U$,
the ditsets of a partition are called a \textit{partition relation} or an
\textit{apartness relation}. Partition relations $P\subseteq U\times U$ can be
characterized as being irreflexive (i.e., for any $u\in U$, $\left(
u,u\right)  \notin P$), symmetric, and anti-transitive (i.e., for any $\left(
u_{j},u_{k}\right)  \in P$ and for any sequence $u_{j}=u_{j_{0}},u_{j_{1}%
},...,u_{j_{k}},u_{j_{k+1}}=u_{k}$ of elements of $U$, there is a pair
$\left(  u_{j_{i}},u_{j_{i+1}}\right)  \in P$). Every ditset of a partition is
a partition relation and vice-versa.

Given another partition $\sigma=\left\{  C_{1},...,C_{k}\right\}  $ on $U$,
the partition $\pi$ \textit{refines} $\sigma$, written $\sigma\precsim\pi$, if
for every block $B\in\pi$, there is a block $C\in\sigma$ such that $B\subseteq
C$. Intuitively, $\pi$ is obtained from $\sigma$ by splitting up some of the
blocks of $\sigma$ which creates more distinctions. Indeed, $\sigma\precsim
\pi$ if and only if (iff) $\operatorname{dit}\left(  \sigma\right)
\subseteq\operatorname{dit}\left(  \pi\right)  $. The refinement relation on
the partitions on $U$ is a \textit{partial order }in the sense that it is
reflexive, anti-symmetric (i.e., if $\sigma\precsim\pi$ and $\pi\precsim
\sigma$ then $\sigma=\pi$), and transitive. The partial order has a maximal or
top partition and a minimal or bottom partition. The top is the
\textit{discrete partition} $\mathbf{1}_{U}=\left\{  \left\{  u\right\}
\right\}  _{u\in U}$ where all the blocks are singletons, and the bottom is
the \textit{indiscrete partition} or "\textit{blob}" $\mathbf{0}_{U}=\left\{
U\right\}  $ with only one block $U$. Both the join (least upper bound) and
meet (greatest lower bound) of two partitions always exist so the refinement
partial order is a lattice $\Pi\left(  U\right)  $.\footnote{In some of the
older literature, the partial order is written in the opposite way as
`unrefinement,' so that interchanges the top and bottom and the join and meet
(\cite{Birk:lt}; \cite{grat:glt}).} Only the join operation is used here, but
all the Boolean operations on subsets can be extended to partitions to form
the logic of partitions (\cite{ell:lop}; \cite{ell:intropartitions}) that is
the dual counterpart to the Boolean logic of subsets (which is usually
presented in the special case of propositional logic). Given $\pi$ and
$\sigma$, the\textit{\ join} $\pi\vee\sigma$ is the partition on $U$ whose
blocks are all the non-empty intersections $B\cap C$ for $B\in\pi$ and
$C\in\sigma$.

One of the easiest ways to see the dual pairing of the concepts of a subset
and a partition is to consider a function $f:X\rightarrow Y$ from a set $X$ to
a set $Y$. The \textit{image} is the subset $f\left(  X\right)  =\left\{  y\in
Y:\exists x\in X,f\left(  x\right)  =y\right\}  $ of the codomain $Y$, and the
\textit{inverse-image} or \textit{coimage} is the partition $\left\{
f^{-1}\left(  y\right)  \right\}  _{y\in f\left(  X\right)  }$ on the domain
$X$.\footnote{In category theory, the notion of a subset generalizes to the
notion of a subobject or `part' and the "dual notion (obtained by reversing
the arrows) of `part' is the notion of partition." \cite[p, 85]{law:sfm}}%

\begin{center}
\includegraphics[
height=2.1459in,
width=4.5484in
]%
{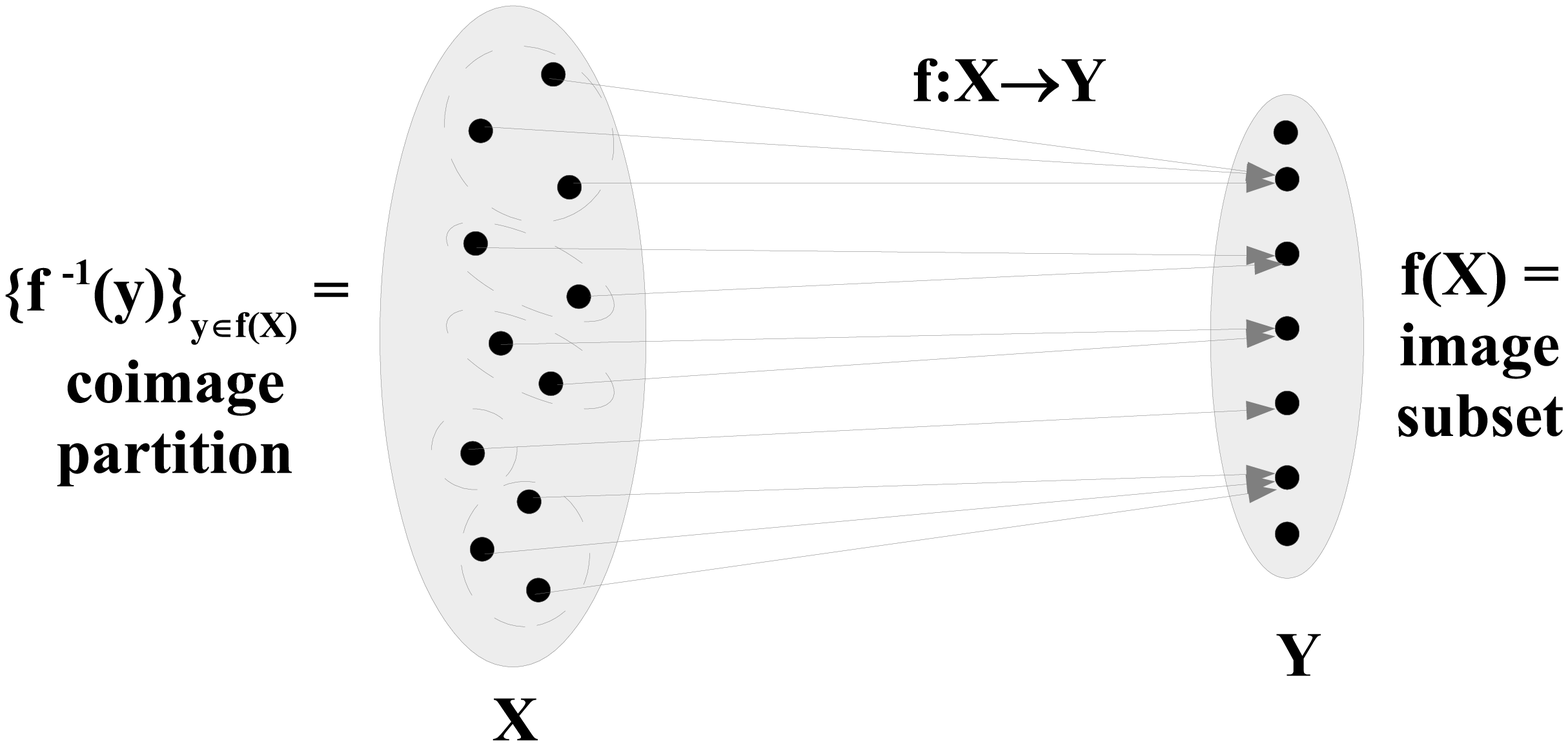}%
\end{center}

\begin{center}
Figure 1: Image subset and inverse-image partition of a function
$f:X\rightarrow Y$.
\end{center}

\subsection{\noindent Logical entropy: the quantification of distinctions}

The set of all subsets of a set $U$, the powerset $\wp\left(  U\right)  $,
also forms a lattice under the inclusion partial order with the top $U$, the
bottom $\emptyset$, and the join and meet being set union and intersection
respectively. Given the duality between subsets and partitions, it is natural
to see what concepts carry over from subsets to partitions. In particular, the
quantitative measure of a subset $S\subseteq U$ is its cardinality $\left\vert
S\right\vert $, and the normalized cardinality of a subset $S$ is the logical
notion of probability $\Pr\left(  S\right)  =\frac{\left\vert S\right\vert
}{\left\vert U\right\vert }$ developed by Boole and Laplace (where each point
$u\in U$ is considered equiprobable). Gian-Carlo Rota in his Fubini Lectures
\cite{rota:fubini} and in his lectures at MIT on probability theory
\cite{rota:prob-guidi} argued that information or entropy should be to
partitions what probability was to subsets, i.e.,%
\begin{equation}
\frac{\text{probability}}{\text{subsets }}\approx\frac{\text{information}%
}{\text{partitions}}. \tag{2.1}%
\end{equation}

\noindent The quantitative notion attached to a subset is its number of
elements $\left\vert S\right\vert $, so the question is; what is the
quantitative notion associated with a partition? The duality between subsets
and partitions can be analyzed back to its conceptual building blocks which
are the dual notions of \textit{elements} (\textit{its}) of a subset and the
\textit{distinctions} (\textit{dits}) of a partition \cite{ell:canonicity}.
Hence, the natural notion of information in a partition would, by this
reasoning, be the normalized number of distinctions, and that is our
definition of the \textit{logical entropy of a partition} $\pi$;%
\begin{align}
h\left(  \pi\right)   &  =\frac{\left\vert \operatorname{dit}\left(
\pi\right)  \right\vert }{\left\vert U\times U\right\vert }=\frac{\left\vert
U\times U\right\vert -\left\vert \operatorname{indit}\left(  \pi\right)
\right\vert }{\left\vert U\times U\right\vert }\tag{2.2}\\
&  =1-\frac{\left\vert \cup_{i}\left(  B_{i}\times B_{i}\right)  \right\vert
}{\left\vert U\times U\right\vert }=1-\sum_{i=1}^{m}\left(  \frac{\left\vert
B_{i}\right\vert }{\left\vert U\right\vert }\right)  ^{2}=1-\sum_{i}\Pr\left(
B_{i}\right)  ^{2}\nonumber
\end{align}

\noindent where $\Pr\left(  B_{i}\right)  =\frac{\left\vert B_{i}\right\vert
}{\left\vert U\right\vert }$ is the probability of a random draw from $U$ will
give an element of $B_{i}$ (with equiprobable points).

When there are point probabilities $p=\left(  p_{1},...,p_{n}\right)  $ for
$p_{j}$ as the probability of the outcome $u_{j}\in U$ with $\sum_{j=1}%
^{n}p_{j}=1$, then $\Pr\left(  B_{i}\right)  =\sum\left\{  p_{j}:u_{j}\in
B_{i}\right\}  $ in the formula for logical entropy. This also gives the
definition of logical entropy for any probability distribution $p=\left(
p_{1},...,p_{n}\right)  $,%
\begin{equation}
h\left(  p\right)  =1-\sum_{j=1}^{n}p_{j}^{2}. \tag{2.3}%
\end{equation}

Logical entropy always has an ultra-simple and logical interpretation. Logical
information theory is built on the idea that information is about
distinctions, differences, distinguishability, and diversity. The notion of
difference requires \textit{two} things in order to have a difference. Hence,
given a partition $\pi=\left\{  B_{1},...,B_{m}\right\}  $ or a probability
distribution $p=\left(  p_{1},...,p_{n}\right)  $, the obvious measure for
idea of information as distinctions or difference is the probability that in
\textit{two} independent samples or draws from $U$ or from the distribution
$p$, one will obtain elements in different blocks of $\pi$, i.e., a
distinction of $\pi$, or different outcomes $p_{j},p_{k}$ for $j\neq k$. And
that is precisely the interpretation of logical entropy, the "probability of
difference." The probability of obtaining elements from the same block of
$\pi$ is $\sum_{i}\Pr\left(  B_{i}\right)  ^{2}$ so the probability of getting
elements from different blocks is $h\left(  \pi\right)  =1-\sum_{i}\Pr\left(
B_{i}\right)  ^{2}$. And similarly for the logical entropy of a probability
distribution $h\left(  p\right)  =1-\sum_{j}p_{j}^{2}$. Another way to express
this result is the formula:%
\begin{equation}
1=1^{2}=\left(  p_{1}+...+p_{n}\right)  \left(  p_{1}+...+p_{n}\right)
=\sum_{j=1}^{n}p_{i}^{2}+\sum_{j\neq k}p_{j}p_{k} \tag{2.4}%
\end{equation}

\noindent so that:%
\begin{equation}
h\left(  p\right)  =1-\sum_{j=1}^{n}p_{i}^{2}=\sum_{j=1}^{n}p_{j}\left(
1-p_{j}\right)  =\sum_{j\neq k}p_{j}p_{k}=2\sum_{j<k}p_{j}p_{k} \tag{2.5}%
\end{equation}

\noindent for $j,k=1,...,n$. Thus, to be more specific, logical entropy is the
probability of getting an ordered pair of distinct indices $p_{j}$ and $p_{k}$
for $j\neq k$--which is twice the probability of getting an unordered pair of
different indices such as $p_{j}$ and $p_{k}$ for $j<k$.

There is a simple way to picture the logical entropy. Given partition
$\pi=\left\{  \left\{  u_{1},u_{2}\right\}  ,\left\{  u_{3}\right\}  ,\left\{
u_{4}\right\}  \right\}  $ with the corresponding point probabilities
$p=\left(  p_{1},p_{2},p_{3},p_{4}\right)  $. Since the sum of the
probabilities is $1$, the logical entropy $h\left(  \pi\right)  $ can be
pictured in a $1\times1$ box, Figure 2, as the shaded area outside the boxed diagonal.%

\begin{center}
\includegraphics[
height=2.3441in,
width=3.1133in
]%
{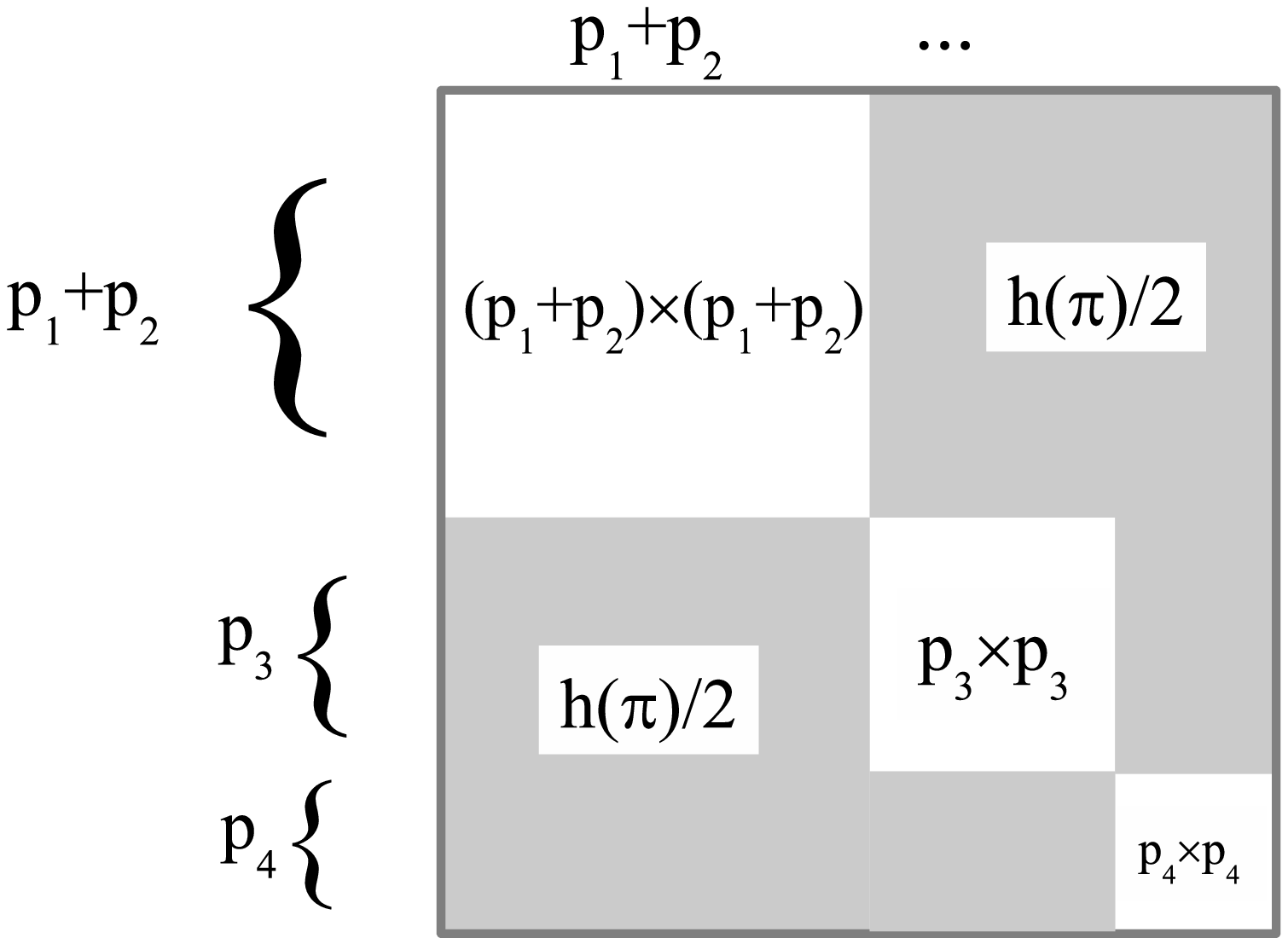}%
\end{center}

\begin{center}
Figure 2: Logical entropy box diagram.
\end{center}

Logical entropy is also a measure, indeed, a probability measure, in the usual
sense of measure theory \cite[p. 30]{halmos:measure} (although terminology
differs) which includes being non-negative. A finitely additive set function
(the values on disjoint sets add together) that can take negative values is
usually called a "signed measure" \cite[p. 118]{halmos:measure} (or a "charge"
\cite{rao-kps:charges}), and, as we will see, Shannon mutual information can
be negative.

Partitions often arise as the inverse-images of random variables
$X:U\rightarrow%
\mathbb{R}
$. To use an example that we will have use of later, consider the throw of one
fair die followed by the throw of a second fair die. All that is recorded is
whether the face up on each die was even or odd, i.e., its parity (or
$\operatorname{mod}(2)$ value). With even represented by $0$ and odd by $1$,
then the space of possible outcomes for the throws of the dice is $U=\left\{
\left(  0,0\right)  ,\left(  0,1\right)  ,\left(  1,0\right)  ,\left(
1,1\right)  \right\}  $. Let $X:U\rightarrow2=\left\{  0,1\right\}  $
represent the outcome of the first die, the $X$-die, and $Y:U\rightarrow2$ the
outcome of the second die, the $Y$-die. For instance, the point $\left(
1,0\right)  \in U$ represents that the first die came up with odd parity $1$
and the second die with even parity $0$.

A measure on a finite set is determined by just an assignment of a
non-negative number to each point in the set. The set on which logical entropy
is a (probability) measure is $U\times U$ so it can again be represented in a
box diagram with the equiprobable outcome pairs in $U$ along each edge. Each
square in Figure 3, representing a pair $\left(  \left(  x,y\right)  ,\left(
x^{\prime},y^{\prime}\right)  \right)  $ of pairs, has the probability weight
of $\frac{1}{4}\times\frac{1}{4}=\frac{1}{16}$ assigned to it. The
inverse-image partition of the random variable $X$ is%
\begin{equation}
X^{-1}=\left\{  X^{-1}\left(  0\right)  ,X^{-1}\left(  1\right)  \right\}
=\left\{  \left\{  \left(  0,0\right)  ,\left(  0,1\right)  \right\}
,\left\{  \left(  1,0\right)  ,\left(  1,1\right)  \right\}  \right\}  .
\tag{2.6}%
\end{equation}

\noindent The ditset $\operatorname{dit}\left(  X^{-1}\right)  $ is the set of
pairs of pairs, i.e., points in $U\times U$, that differ in the first
coordinate:%
\begin{equation}
\operatorname{dit}\left(  X^{-1}\right)  =\left\{  \left(  \left(  0,0\right)
,\left(  1,0\right)  \right)  ,\left(  \left(  0,0\right)  ,\left(
1,1\right)  \right)  ,\left(  \left(  0,1\right)  ,\left(  1,0\right)
\right)  ,\left(  \left(  0,1\right)  ,\left(  1,1\right)  \right)
,\ldots\right\}  \tag{2.7}%
\end{equation}

\noindent where the ellipsis $\ldots$ represents the pairs of pairs with the
reversed order. The shaded squares in Figure 3 box diagram are the ones
included in the logical entropy $h(X)$ since they are the ones which differ in
the first coordinate of the ordered pairs of outcomes, i.e.,, the pairs where
the first die's outcomes had different parities. Each outcome $\left(
x,y\right)  $ has probability $p\left(  x,y\right)  =\frac{1}{4}$ and the only
squares that count for the logical entropy of $X$ are the ones for $\left(
\left(  x,y\right)  ,\left(  x^{\prime},y^{\prime}\right)  \right)  $ where
$x\neq x^{\prime}$.%

\begin{center}
\includegraphics[
height=2.5966in,
width=2.5576in
]%
{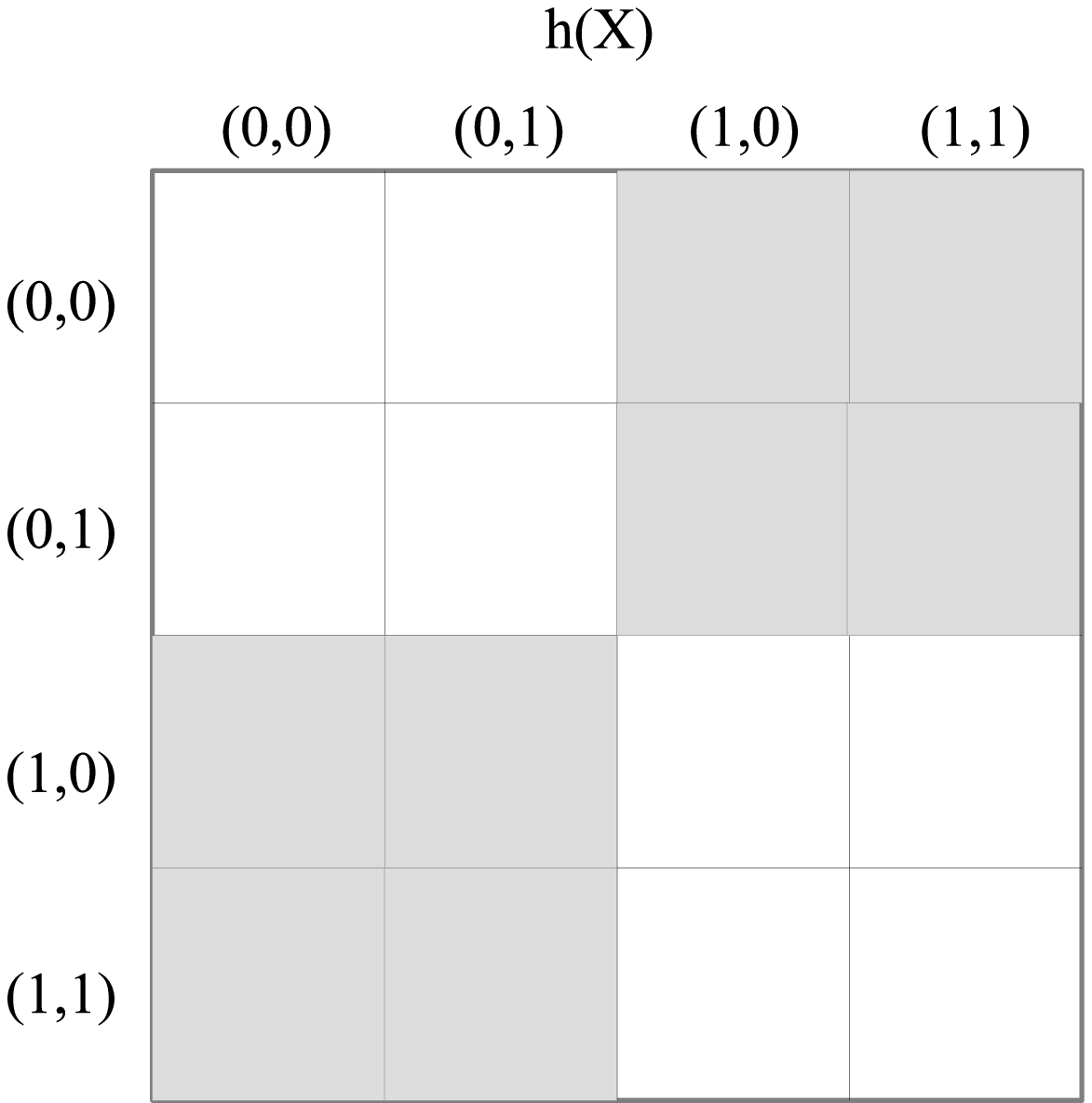}%
\end{center}

\begin{center}
Figure 3: Box diagram for $h\left(  X\right)  =\sum\left\{  p\left(
x,y\right)  p\left(  x^{\prime},y^{\prime}\right)  :x\neq x^{\prime}\right\}
=\frac{8}{16}=\frac{1}{2}$

which can also be seen as a Venn diagram.
\end{center}

The logical entropy of the random variable $Y:U\rightarrow2$ is computed and
represented in Figure 4 in the same manner except that the relevant pairs of
pairs are those that differ in the second coordinate representing the parity
of the second die.%

\begin{center}
\includegraphics[
height=2.633in,
width=2.5932in
]%
{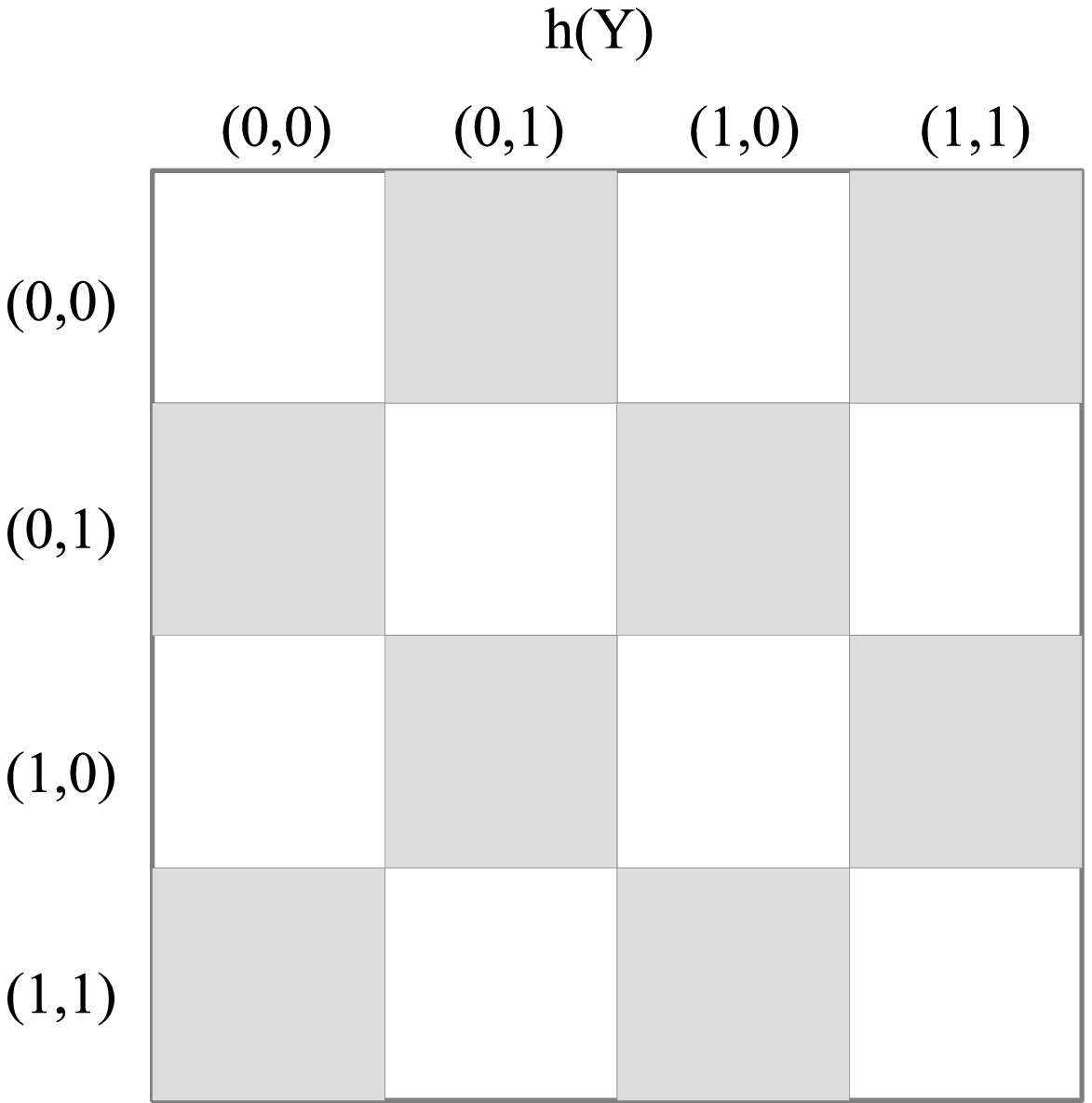}%
\end{center}

\begin{center}
Figure 4: Box/Venn diagram for $h\left(  Y\right)  =\sum\left\{  p\left(
x,y\right)  p\left(  x^{\prime},y^{\prime}\right)  :y\neq y^{\prime}\right\}
=\frac{1}{2}$.
\end{center}

\noindent The logical entropy $h\left(  X\right)  $ for $X$ (the parity of the
outcome for the first die) and $h(Y)$ for $Y$ (the parity of outcome for the
second die) is the probability that on two independent throws of the relevant
die, one will obtain outcomes of different parity.

\subsection{History of the logical entropy formula}

The concept of information as a measure of differences goes back to 1641, the
year before Isaac Newton was born, when the polymath John Wilkins (1614-1672)
anonymously published one of the earliest books on cryptography,
\textit{Mercury or the Secret and Swift Messenger}. This book not only pointed
out the fundamental role of differences but noted that any (finite) set of
different things could be encoded by words in a binary code.

\begin{quotation}
\noindent\noindent For in the general we must note, That whatever is capable
of a competent Difference, perceptible to any Sense, may be a sufficient Means
whereby to express the Cogitations. It is more convenient, indeed, that these
Differences should be of as great Variety as the Letters of the Alphabet; but
it is sufficient if they be but twofold, because Two alone may, with somewhat
more Labour and Time, be well enough contrived to express all the rest.
\cite[Chap. XVII, p. 69]{wilkins:merc}
\end{quotation}

\noindent Wilkins explains that a five letter binary code would be sufficient
to code the letters of the alphabet since $2^{5}=32$.

\begin{quotation}
\noindent\noindent Thus any two Letters or Numbers, suppose $A.B$. being
transposed through five Places, will yield Thirty Two Differences, and so
consequently will superabundantly serve for the Four and twenty Letters...
.\cite[Chap. XVII, p. 69]{wilkins:merc}
\end{quotation}

\noindent In James Gleick's 2011 book, \textit{The Information: A History, A
Theory, A Flood}, he noted that:

\begin{quotation}
\noindent\noindent Any difference meant a binary choice. Any binary choice
began the expressing of cogitations. Here, in this arcane and anonymous
treatise of $1641$, the essential idea of information theory poked to the
surface of human thought, saw its shadow, and disappeared again for [three]
hundred years. \cite[p. 161]{gleick:info}\footnote{Gleick is referring to the
old Pennsylvania Dutch superstition that on February 2 each year, if a
groundhog emerges from its den and sees its shadow, then it will go back in
for six more weeks.} \noindent
\end{quotation}

\noindent The idea that information is about differences was also expressed by
the polymath, Gregory Bateson, who noted that (the transmission of)
"[i]nformation consists of differences that make a difference." \cite[p.
99]{bateson:mind}

The formula that is a measure of differences, $h\left(  p\right)  =1-\sum
_{j}p_{j}^{2}$ (or its complementary form $1-h\left(  p\right)  =\sum_{j}%
p_{j}^{2}$), goes back at least to Corrado Gini (1884-1965) who published it
as an \textit{index of mutability} \cite{gini:vem} in 1912 (not to be confused
with Gini's better-known index of inequality). Some of the immediate following
history of the formula was connected to cryptology as foreshadowed by Wilkins.
William F. Friedman, an American cryptologist, devoted a 1922 book
(\cite{fried:ioc}) to the \textquotedblleft index of
coincidence\textquotedblright\ (i.e., $\sum p_{i}^{2}$). Solomon Kullback
worked as an assistant to Friedman and wrote a book on cryptology which used
the index. \cite{kull:crypt}

During World War II, Alan M. Turing worked for a time in the Government Code
and Cypher School at the Bletchley Park facility in England. Probably unaware
of the earlier work, Turing used $\rho=\sum p_{i}^{2}$ in his cryptoanalysis
work and called it the \textit{repeat rate} since it is the probability of a
repeat in a pair of independent draws from a population with those
probabilities. Polish cryptographers had independently used the repeat rate in
their work on the Enigma \cite{rej:polish}. After WWII, Edward H. Simpson, a
British statistician, proposed $\sum_{B\in\pi}p_{B}^{2}$ as a measure of
species concentration (the opposite of diversity) where $\pi$ is the partition
of animals or plants according to species and where each animal or plant is
considered as equiprobable. And Simpson gave the interpretation of this
homogeneity measure as \textquotedblleft the probability that two individuals
chosen at random and independently from the population will be found to belong
to the same group.\textquotedblright\cite[p. 688]{simp:md} Hence $1-\sum
_{B\in\pi}p_{B}^{2}$ is the probability that a random ordered pair will belong
to different species, i.e., will be distinguished by the species partition.
The biodiversity literature \cite{ric:unify} refers to the formula as
\textquotedblleft Simpson's index of diversity\textquotedblright\ or
sometimes, the \textquotedblleft Gini-Simpson diversity
index.\textquotedblright\ {In the bioinformatics literature, Masatoshi Nei
\cite{nei:div} introduced the logical entropy formula as a measure of gene
diversity.}

But the Simpson story has a twist. Simpson along with I. J. Good worked at
Bletchley Park during WWII, and, according to Good, \textquotedblleft E. H.
Simpson and I both obtained the notion [the repeat rate] from
Turing.\textquotedblright\ \cite[p. 395]{good:turing} When Simpson published
the index in 1948, he (again, according to Good) did not acknowledge Turing
\textquotedblleft fearing that to acknowledge him would be regarded as a
breach of security.\textquotedblright\ \cite[p. 562]{good:div} Perhaps logical
entropy should be called "Turing entropy" to compete with the `big names'
attached to Shannon entropy and von Neumann entropy. But given its frequent
discovery and rediscovery, Good also negated that idea.

\begin{quotation}
\noindent If $p_{1},p_{2},...,p_{n}$ are the probabilities of mutually
exclusive and exhaustive events, any statistician of this century who wanted a
measure of homogeneity would have taken about two seconds to suggest $\sum
p_{i}^{2}$, which I shall call $\rho$. ... Thus it is unjust to associate
$\rho$ with any one person. It would be better to use such names as "repeat
rate" or "quadratic index of homogeneity" for $\rho$ and perhaps "quadratic
index of heterogeneity or diversity" for $1-\rho$. \cite[pp. 561-2]{good:div}
\end{quotation}

\noindent Thus the name "logical entropy" seems appropriate, particularly in
view of Stigler's Law of Eponymy, i.e., \textquotedblleft No scientific
discovery is named after its original discoverer\textquotedblright%
\ \cite{stigler:table}, and since it is the quantitative measure associated
with partitions in the logic of partitions just as finite probability is the
quantitative measure associated with subsets in the usual Boolean logic of subsets.

In economics, Albert O. Hirschman \cite[p. 159]{hirsch:np} suggested in 1945
using $\sqrt{\sum p_{i}^{2}}$ as an index of trade concentration (where
$p_{i}$ is the relative share of trade in a certain commodity or with a
certain partner). A few years afterwards, Orris Herfindahl \cite{her:conc}
independently suggested using $\sum p_{i}^{2}$ as an index of industrial
concentration (where $p_{i}$ is the relative share of the $i^{th}$ firm in an
industry). In the literature on industrial economics, the index $H=\sum
p_{i}^{2}$ is variously called the Hirschman-Herfindahl index, the HH index,
or just the H index of concentration.

Another way to look at logical entropy is that two elements from $U=\left\{
u_{1},...,u_{n}\right\}  $ are either identical or distinct. Gini
\cite{gini:vem} introduced $d_{ij}=1-\delta_{ij}$ (the complement of the
Kronecker delta function) as the "distance"\ between the $i^{th}$ and $j^{th}$
elements where $d_{ij}=1$ for $i\not =j$ and $d_{ii}=0$. Then Gini's index of
mutability, $h\left(  p\right)  =\sum_{i,j}d_{ij}p_{i}p_{j}$, is the average
(logical) distance between a pair of independently drawn elements. But one
might generalize by allowing other non-negative distances $d_{ij}=d_{ji}$ for
$i\not =j$ (but always $d_{ii}=0$) so that $Q=\sum_{i,j}d_{ij}p_{i}p_{j}$
would be the average distance between a pair of independently drawn elements
from $U$. In 1982, C. R. (Calyampudi Radhakrishna) Rao introduced precisely
this concept as \textit{quadratic entropy} \cite{rao:div}. {The logical
entropy is also the quadratic special case of the Tsallis--Havrda--Charvat
entropy (\cite{havrda:kibernetica}, \cite{tsallis:generalization}). }

\v{C}aslav Brukner and Anton Zeilinger have also developed the logical entropy
formula $1-\sum_{i=1}^{n}p_{i}^{2}$ in their treatment of quantum information
(\cite{brukner-z:invar}, \cite{brukner-z:fundamentals}) and have also used the
normalized form of the (Euclidean) distance squared of a probability
distribution from the uniform distribution, which is closely related to the
logical entropy since: $\sum_{i=1}^{n}\left(  p_{i}-\frac{1}{n}\right)
^{2}=\left(  1-\frac{1}{n}\right)  -h\left(  p\right)  $.

\subsection{Compound notions of logical entropy}

We now consider a joint probability distribution $\left\{  p\left(
x,y\right)  \right\}  $ on the finite sample space $X\times Y$ (where to avoid
trivialities, assume $\left\vert X\right\vert ,\left\vert Y\right\vert \geq
2$), with the marginal distributions $\left\{  p\left(  x\right)  \right\}  $
and $\left\{  p\left(  y\right)  \right\}  $ where $p\left(  x\right)
=\sum_{y\in Y}p\left(  x,y\right)  $ and $p\left(  y\right)  =\sum_{x\in
X}p\left(  x,y\right)  $. The setting is a pair of random variables $X$ and
$Y$ where we also consider $X$ as the set of possible values $x$ of the r.v.
$X$ and similarly for the r.v. $Y$. Then the joint probability distribution is
$p\left(  x,y\right)  =\Pr\left(  X=x,Y=y\right)  $, and the marginals are
$p\left(  x\right)  =\Pr\left(  X=x\right)  $, and $p\left(  y\right)
=\Pr\left(  Y=y\right)  $. For notational simplicity, the entropies can be
considered as functions of the random variables or of their probability
distributions, e.g., $h\left(  \left\{  p\left(  x\right)  \right\}  \right)
=h\left(  X\right)  $ and $h\left(  \left\{  p\left(  y\right)  \right\}
\right)  =h\left(  Y\right)  $. Logical entropy is characterized in terms of
the probability that in two independent draws $\left(  x,y\right)  $ and
$\left(  x^{\prime},y^{\prime}\right)  $ from the sample space, one will get
different outcomes. Hence in this case,%
\begin{align}
h\left(  X\right)   &  =\sum_{x,y}\left\{  p\left(  x,y\right)  p\left(
x^{\prime},y^{\prime}\right)  :x\neq x^{\prime}\right\} \tag{2.8}\\
h\left(  Y\right)   &  =\sum_{x,y}\left\{  p\left(  x,y\right)  p\left(
x^{\prime},y^{\prime}\right)  :y\neq y^{\prime}\right\}  . \tag{2.9}%
\end{align}

\noindent Then the joint entropy $h\left(  X,Y\right)  $ is just the logical
entropy $h\left(  \left\{  p\left(  x,y\right)  \right\}  _{\left(
x,y\right)  \in X\times Y}\right)  $ of the joint probability distribution
which can also be characterized as:%
\begin{equation}
h\left(  X,Y\right)  =\sum_{x,y}\left\{  p\left(  x,y\right)  p\left(
x^{\prime},y^{\prime}\right)  :x\neq x^{\prime}\text{ or }y\neq y^{\prime
}\right\}  . \tag{2.10}%
\end{equation}

In the previous even-odd dice example of throwing an $X$-die and a $Y$-die,
each die had an outcome set of $\left\{  0,1\right\}  $ so $X\times Y=\left\{
\left(  0,0\right)  ,\left(  0,1\right)  ,\left(  1,0\right)  ,\left(
1,1\right)  \right\}  =U$. The space on which the probabilities are assigned
is $U\times U=\left(  X\times Y\right)  \times\left(  X\times Y\right)  $ so
the probability assigned to each point $\left(  \left(  x,y\right)  ,\left(
x^{\prime},y^{\prime}\right)  \right)  $ is $p\left(  x,y\right)  p\left(
x^{\prime},y^{\prime}\right)  $. The points in the space $\left(  X\times
Y\right)  ^{2}$ whose probabilities add up to give $h\left(  X,Y\right)  $ are
just the union of the points for $h\left(  X\right)  $, i.e., where $x\neq
x^{\prime}$, and for $h\left(  Y\right)  $, i.e., where $y\neq y^{\prime}$.
Since each point in $\left(  X\times Y\right)  ^{2}$ is represented by a
square with probability $\frac{1}{16}$, the shaded squares for $h\left(
X,Y\right)  $ are just the union of the squares for $h\left(  X\right)  $ and
$h\left(  Y\right)  $ as shown in Figure 5.%

\begin{center}
\includegraphics[
height=1.297in,
width=4.3087in
]%
{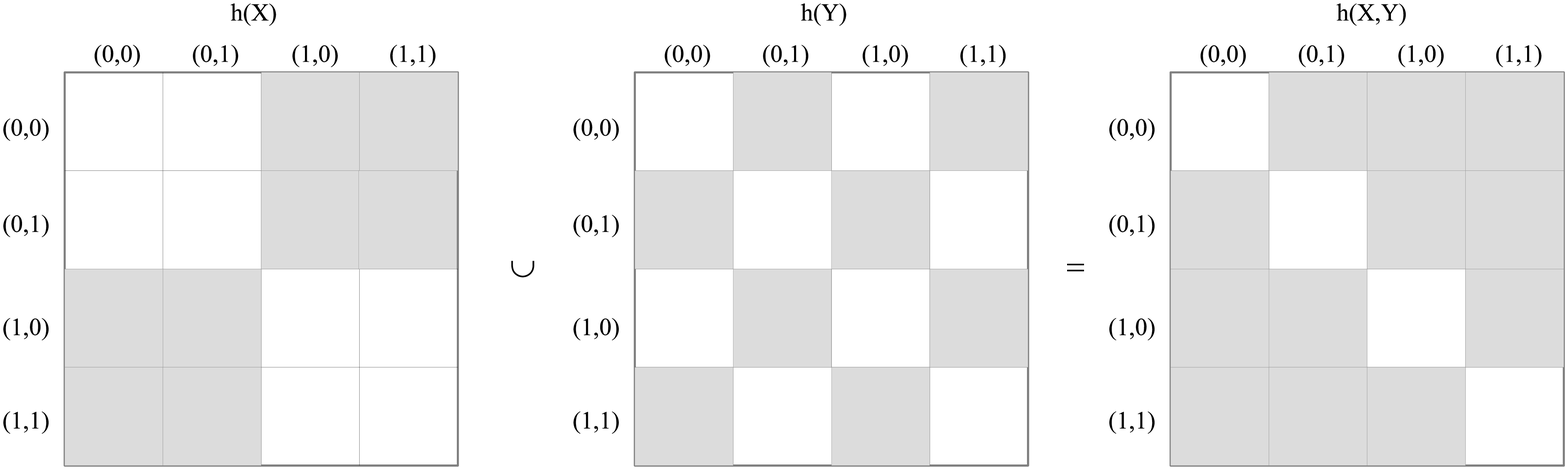}%
\end{center}

\begin{center}
Figure 5: Union of Box/Venn diagrams for $h\left(  X\right)  $ and $h\left(
Y\right)  $ gives

the box diagram for joint entropy $h\left(  X,Y\right)  =\frac{12}{16}%
=\frac{3}{4}$.
\end{center}

\noindent The usual interpretation carries over to the compound notions such
as the joint entropy; in two independent throws of the pair of dice, the
probability that one will get a different parity in the $X$-die \textit{or} in
the $Y$-die (or both) is $h\left(  X,Y\right)  =\frac{3}{4}$.

In a Venn diagram that is merely illustrative, the logical entropies would be
represented as circles and the union of the circles would represent the joint
entropy as in Figure 6.%

\begin{center}
\includegraphics[
height=1.4605in,
width=2.0171in
]%
{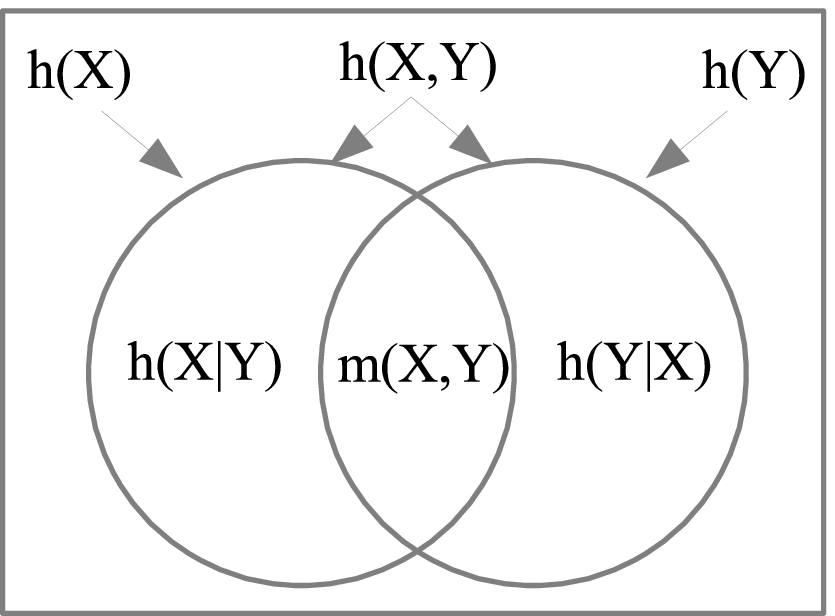}%
\end{center}

\begin{center}
Figure 6: Illustrative Venn diagram for the compound logical entropies.
\end{center}

Figure 6 also illustrates the `formulas' for the other compound logical
entropies. The \textit{conditional logical entropy}%
\begin{equation}
h\left(  X|Y\right)  =\sum_{x,y}\left\{  p\left(  x,y\right)  p\left(
x^{\prime},y^{\prime}\right)  :x\neq x^{\prime}\text{ and }y=y^{\prime
}\right\}  \tag{2.11}%
\end{equation}

\noindent represents the distinctions made by $X$ (i.e., the cases where the
two throws of $X$-die had different parities) after the distinctions made by
$Y$ are taken away (so $y=y^{\prime}$), and vice-versa for $h\left(
Y|X\right)  $. And the \textit{mutual logical information}%
\begin{equation}
m\left(  X,Y\right)  =\sum_{x,y}\left\{  p(x,y)p(x^{\prime},y^{\prime}):x\neq
x^{\prime}\text{ and }y\neq y^{\prime}\right\}  \tag{2.12}%
\end{equation}

\noindent is the probability that in the two throws of the pair of dice, the
pair of pairs of outcomes will have different parity in the $X$-die
\textit{and} in the $Y$-die--as one can easily see from the shaded squares for
$m\left(  X,Y\right)  $ in Figure 7.%

\begin{center}
\includegraphics[
height=1.3334in,
width=4.3223in
]%
{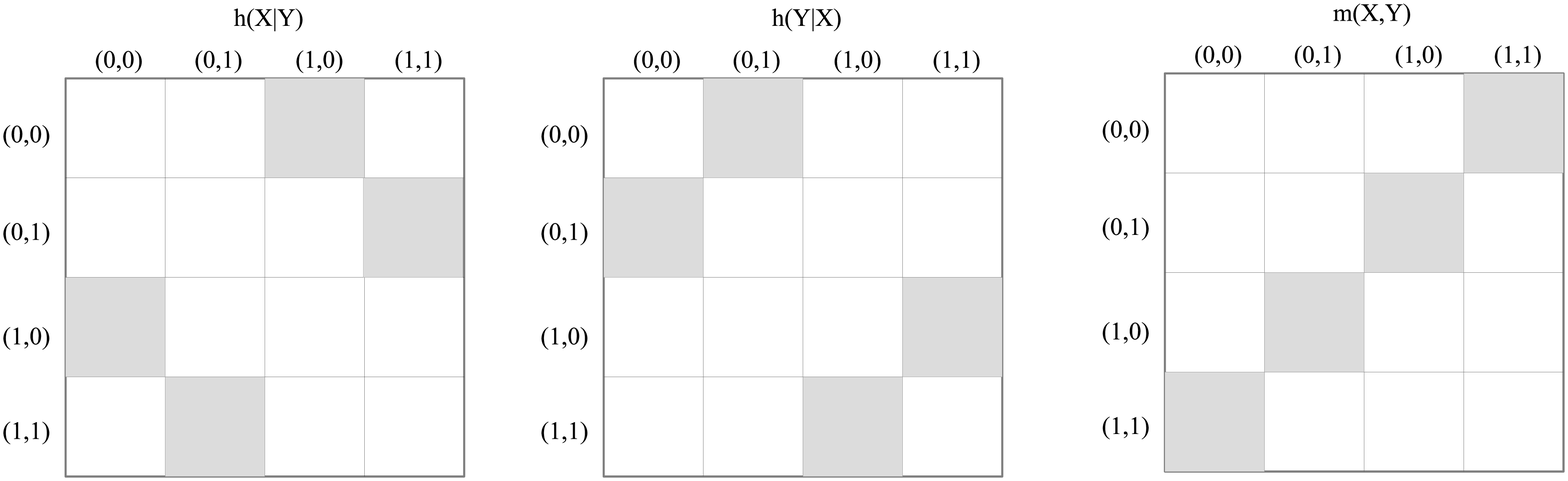}%
\end{center}

\begin{center}
Figure 7: Box diagrams representing the two conditional logical entropies and
the mutual logical information all with the value $\frac{1}{4}$.
\end{center}

These specific box/Venn diagrams illustrate general relationships such as the
two conditional entropies and mutual information all being disjoint and adding
to the joint entropy. In general (not just for this example), the compound
logical entropies stand in the relationships shown by the areas in the
illustrative Figure 6:%
\begin{align}
h\left(  X,Y\right)   &  =h\left(  X|Y\right)  +h\left(  Y|X\right)  +m\left(
X,Y\right) \tag{2.13}\\
h\left(  X\right)   &  =h\left(  X|Y\right)  +m\left(  X,Y\right) \tag{2.14}\\
h\left(  Y\right)   &  =h\left(  Y|X\right)  +m\left(  X,Y\right) \tag{2.15}\\
h\left(  X,Y\right)   &  =h\left(  X\right)  +h\left(  Y\right)  -m\left(
X,Y\right)  . \tag{2.16}%
\end{align}

\section{Shannon entropy}

\subsection{The basic definitions}

Both the logical and the Shannon entropies are defined for probability
distributions regardless of whether the distribution is derived from the
blocks of a partition $\Pr\left(  B_{i}\right)  $ or the values of a random
variable $\Pr\left(  X=x\right)  $. Hence we can start the treatment of
\textit{Shannon entropy} (\cite{shannon:comm}; \cite{shannonweaver:comm})
defined on a probability distribution $p=\left(  p_{1},...,p_{n}\right)  $:%
\begin{equation}
H\left(  p\right)  =-\sum_{i=1}^{n}p_{i}\log_{2}\left(  p_{i}\right)
=\sum_{i=1}^{n}p_{i}\log_{2}\left(  \frac{1}{p_{i}}\right)  \tag{3.1}%
\end{equation}

\noindent where for $p_{i}=0$, $p_{i}\log_{2}\left(  \frac{1}{p_{i}}\right)  $
is defined to be $0$. Henceforth, the logs are to base $2$ unless otherwise
specified. For a random variable $X$ with $p\left(  x\right)  =\Pr\left(
X=x\right)  $, then:%
\begin{equation}
H\left(  X\right)  =\sum_{x\in X}p\left(  x\right)  \log\left(  \frac
{1}{p\left(  x\right)  }\right)  . \tag{3.2}%
\end{equation}

\noindent Given a joint probability distribution $p\left(  x,y\right)  $ on
$X\times Y$, the\textit{\ joint Shannon entropy} is:%
\begin{equation}
H\left(  X,Y\right)  =\sum_{\left(  x,y\right)  \in X\times Y}p\left(
x,y\right)  \log\left(  \frac{1}{p\left(  x,y\right)  }\right)  . \tag{3.3}%
\end{equation}

The conditional Shannon entropy $H(X|Y)$ is defined as the average of the
Shannon entropies for conditional probability distributions. Given the joint
distribution $\left\{  p\left(  x,y\right)  \right\}  $ on $X\times Y$, then
for a specific $y_{0}\in Y$, then the conditional probability distribution is
$p\left(  x|y_{0}\right)  =\frac{p\left(  x,y_{0}\right)  }{p\left(
y_{0}\right)  }$ which has the Shannon entropy: $H\left(  X|y_{0}\right)
=\sum_{x\in X}p\left(  x|y_{0}\right)  \log\left(  \frac{1}{p\left(
x|y_{0}\right)  }\right)  $. Then the \textit{Shannon conditional entropy} is
defined as the \textit{average} of these entropies:%
\begin{equation}
H\left(  X|Y\right)  =\sum_{y\in Y}p\left(  y\right)  \sum_{x}\frac{p\left(
x,y\right)  }{p\left(  y\right)  }\log\left(  \frac{p\left(  y\right)
}{p\left(  x,y\right)  }\right)  =\sum_{x,y}p\left(  x,y\right)  \log\left(
\frac{p\left(  y\right)  }{p\left(  x,y\right)  }\right)  . \tag{3.4}%
\end{equation}

Since the Venn diagram for any measure like logical entropy satisfies a
relationship like $h\left(  X\right)  +h\left(  Y\right)  -h\left(
X,Y\right)  =m\left(  X,Y\right)  $, Shannon defined the \textit{mutual
Shannon information} as:%
\begin{equation}
I\left(  X;Y\right)  =\sum_{x\in X,y\in Y}p\left(  x,y\right)  \left[
\log\left(  \frac{1}{p\left(  x\right)  }\right)  +\log\left(  \frac
{1}{p\left(  y\right)  }\right)  -\log\left(  \frac{1}{p\left(  x,y\right)
}\right)  \right]  . \tag{3.5}%
\end{equation}

\noindent Then it is perhaps no surprise that these compound Shannon entropies
satisfy the Venn diagram relationship \textit{as if} the Shannon entropy was
defined as a measure on a set. Hence one finds in the textbooks on Shannon's
theory of communications, a Venn diagram like Figure 8 to serve at least as a
mnemonic about the interrelationships.%

\begin{center}
\includegraphics[
height=1.4173in,
width=1.9417in
]%
{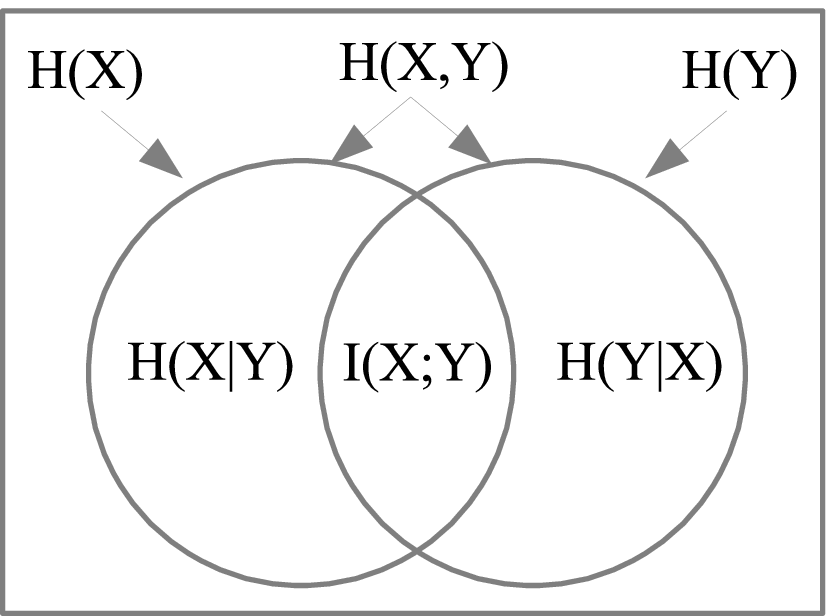}%
\end{center}

\begin{center}
Figure 8: Venn diagram mnemonic for the compound Shannon entropies.
\end{center}

\subsection{Shannon's communications theory and `information theory'}

This paper presents a different version of `information theory' than the
received version. There is no difference in the part of information theory
where Shannon entropy actually does its work, namely the theory of coding and
communication. Shannon himself did not name his original paper or book as
"information theory" but rather as the "mathematical theory of communication"
(\cite{shannon:comm}; \cite{shannonweaver:comm}). Thus the notion that the
theory of communications (including coding theory) was an "information theory"
was a creation of the science press, science popularizers, and textbook
writers. Shannon even reacted against the "bandwagon" that inflated
"information theory" far beyond the actual technical results of communications theory.

\begin{quotation}
\noindent Information theory has, in the last few years, become something of a
scientific bandwagon. Starting as a technical tool for the communication
engineer, it has received an extraordinary amount of publicity in the popular
as well as the scientific press. In part, this has been due to connections
with such fashionable fields as computing machines, cybernetics, and
automation; and in part, to the novelty of its subject matter. As a
consequence, it has perhaps been ballooned to an importance beyond its actual
accomplishments. Our fellow scientists in many different fields, attracted by
the fanfare and by the new avenues opened to scientific analysis, are using
these ideas in their own problems. Applications are being made to biology,
psychology, linguistics, fundamental physics, economics, the theory of
organization, and many others. In short, information theory is currently
partaking of a somewhat heady draught of general popularity. \cite[p.
462]{shannon:bandwagon}
\end{quotation}

\noindent Shannon repeated the points in a 1961 interview with Myron Tribus.

\begin{quotation}
\noindent In 1961 Professor Shannon, in a private conversation, made it quite
clear to me that he considered applications of his work to problems outside of
communication theory to be suspect and he did not attach fundamental
significance to them. \cite[p. 1]{tribus:30years}
\end{quotation}

Moreover, while Shannon noted that while his entropy formula indicates the
"amount of information" (i.e., the average numbers of binary distinctions
needed to distinguish all the "messages"), "no concept of information itself
was defined" \cite[p. 458]{shannon:topics} in communications theory. Perhaps
the most common idea about Shannon entropy is that it a measure of "amount of
uncertainty." But there are many other interpretations.

\begin{quotation}
\noindent Other terms used to convey an intuitive feeling for entropy include
randomness, disorganization, \textquotedblleft mixed-up-ness\textquotedblright%
\ (Gibbs), missing information, in-complete knowledge, complexity, chaos,
ignorance, and uncertainty. \cite[p. 9]{ramshaw;bgs}
\end{quotation}

\noindent There is also the view that entropy and information were in fact
opposites or complements; "Gain in entropy always means loss of information,
and nothing more." \cite[p. 573]{lewis:symmetry} That view was later
popularized by Leon Brillouin who claimed that:

\begin{quotation}
\noindent information must be considered as a negative term in the entropy of
a system; in short, information is negentropy. ... Entropy measures the lack
of information. \cite[p. xii]{brillouin;negent}
\end{quotation}

However, there is no need for this conceptual chaos; the (simple) Shannon
entropy is another way to quantify the notion of information-as-distinctions.
That is, Shannon entropy is the minimum average number of binary partitions
(bits) that need to be joined in order to make the distinctions that
distinguish all the "messages." And simple logical entropy is the direct
measure of distinctions.

\subsection{Is Shannon entropy a `measure'?}

Shannon entropy and a host of other entropy "formulas" (sans interpretation)
are routinely called "measures" of information \cite{aczel-d:measures}. A
prominent information theorist, Lorne Campbell, has noted in 1965 the
analogies between Shannon entropy and measures (in the usual non-negative sense).

\begin{quotation}
\noindent Certain analogies between~ entropy~ ~and measure ~have ~been ~noted
~by ~various ~authors. These analogies provide a convenient mnemonic for the
various relations between entropy, conditional entropy, joint entropy, and
mutual information. It is interesting ~to ~speculate ~whether ~these
~analogies ~have a deeper foundation. It would seem to be quite significant if
entropy did admit an interpretation as the measure of some set. \cite[p.
112]{camp:meas}
\end{quotation}

We only need be concerned with the simplest case of a measure
\cite{halmos:measure} on a finite set where for any finite set $U$, a
\textit{measure} $\mu$ is a function from the powerset of $U$ (the subsets of
$U$) to the reals $\mu:\wp\left(  U\right)  \rightarrow%
\mathbb{R}
$ such that:

\begin{enumerate}
\item $\mu\left(  \emptyset\right)  =0$,

\item for any $E\subseteq U$, $\mu\left(  E\right)  \geq0$, and

\item for any disjoint subsets $E_{1}$ and $E_{2}$, $\mu(E_{1}\cup E_{2}%
)=\mu\left(  E_{1}\right)  +\mu\left(  E_{2}\right)  $.
\end{enumerate}

\noindent The whole measure is determined by the values on singletons and
simply summed over larger finite subsets.

It would be desirable for Shannon entropy to be a measure in this technical
sense so:

\begin{quotation}
\noindent\noindent\noindent\noindent that $H\left(  \alpha\right)  $ and
$H\left(  \beta\right)  $ are measures of sets, that $H\left(  \alpha
,\beta\right)  $ is the measure of their union, that $I\left(  \alpha
,\beta\right)  $ is the measure of their intersection, and that $H\left(
\alpha|\beta\right)  $ is the measure of their difference. The possibility
that $I\left(  \alpha,\beta\right)  $ is the entropy of the \textquotedblleft
intersection\textquotedblright\ of two partitions is particularly interesting.
This \textquotedblleft intersection,\textquotedblright\ if it existed, would
presumably contain the information common to the partitions $\alpha$ and
$\beta$.\cite[p. 113]{camp:meas}
\end{quotation}

\noindent Logical entropy satisfies all those desiderata.

There are some differences in the use of the word "measure." It would seem
that the usual notion of a measure is always non-negative
(\cite{halmos:measure}; \cite{doob:measure}) and then there is an extended
notion of a "signed measure" that can take on negative values. Other authors
define a "measure" to allow negative values and then define a "positive
measure" to have only non-negative values. The most general usage, adopted
here, is that a measure is non-negative and the generalized notion to allow
negative value is a "signed measure." This is important since logical entropy
is defined as a measure, indeed a probability measure, while Shannon entropy
is not defined as a measure on a set. Given any Venn diagram of Shannon
entropies, then, as with any Venn diagram, an \textit{ex post} measue or
signed measure can always be trivially constructed. Both measures and signed
measures can be represented as additive set functions ( \cite[Part 8, Chap. 1,
Prob. 26]{polya-szego:vol-II}; \cite{hu:info}; \cite[Chap. 2]{ryser:comb-math}%
; \cite{takacs:incl-excl}) that satisfy the inclusion-exclusion principle (or
overcount-undercount relationships) that can be associated with Venn diagrams
(if we allow negative areas).

For logical entropy, consider a set $U=\left\{  u_{1},...,u_{n}\right\}  $
with point probabilities $\left\{  p_{i}\right\}  _{i=1}^{n}$ and a random
variable $X:U\rightarrow%
\mathbb{R}
$ which induces a partition $X^{-1}$ on $U$ and similarly for $Y:U\rightarrow%
\mathbb{R}
$. The set on which the logical entropy measure is defined is $U\times U$ and
the value assigned to a point $\left(  u_{j},u_{k}\right)  \in U\times U$ is
$\mu\left(  \left\{  \left(  u_{j},u_{k}\right)  \right\}  \right)
=p_{j}p_{k}$. The logical entropy associated with the random variable is:%
\begin{equation}
h\left(  X\right)  =\mu\left(  \operatorname{dit}\left(  X^{-1}\right)
\right)  =\sum_{u_{j},u_{k}\in U}\left\{  p_{j}p_{k}:\left(  u_{j}%
,u_{k}\right)  \in\operatorname{dit}\left(  X^{-1}\right)  \right\}  ,
\tag{3.6}%
\end{equation}

\noindent namely the sum of all the products $p_{j}p_{k}$ for which $X\left(
u_{j}\right)  \neq X\left(  u_{k}\right)  $. Thus it is interpreted as the
probability that on two independent trials, the random variable $X$ will give
different values. That illustrates how logical entropy measures differences.
If the values of $X$ have no differences, i.e., if it is constant, then
$X^{-1}=\mathbf{0}_{U}$ and $h\left(  \mathbf{0}_{U}\right)  =0$. The more
refined the inverse-image partition $X^{-1}$, the higher the logical entropy.
Then all the usual Venn diagram relationships hold such as%
\begin{align}
h\left(  X\right)   &  =\sum\left\{  p_{j}p_{k}:\left(  u_{j},u_{k}\right)
\in\operatorname{dit}\left(  X^{-1}\right)  \right\} \nonumber\\
&  =\sum_{u_{j},u_{k}\in U}\left\{  p_{j}p_{k}:\left(  u_{j},u_{k}\right)
\in\operatorname{dit}\left(  X^{-1}\right)  \text{ and }\left(  u_{j}%
,u_{k}\right)  \notin\operatorname{dit}\left(  Y^{-1}\right)  \right\}
\nonumber\\
&  +\sum_{u_{j},u_{k}\in U}\left\{  p_{j}p_{k}:\left(  u_{j},u_{k}\right)
\in\operatorname{dit}\left(  X^{-1}\right)  \text{ and }\left(  u_{j}%
,u_{k}\right)  \in\operatorname{dit}\left(  Y^{-1}\right)  \right\}
\nonumber\\
&  =h\left(  X|Y\right)  +m\left(  X,Y\right)  \tag{3.7}%
\end{align}

\noindent and all of Campbell's desiderata are satisfied. For instance, the
logical conditional entropy is the measure on the difference of the sets for
$h\left(  X\right)  $ and $h\left(  Y\right)  $:%
\begin{equation}
h\left(  X|Y\right)  =\sum_{u_{j},u_{k}\in U}\left\{  p_{j}p_{k}:\left(
u_{j},u_{k}\right)  \in\operatorname{dit}\left(  X^{-1}\right)
-\operatorname{dit}\left(  Y^{-1}\right)  \right\}  . \tag{3.8}%
\end{equation}

To see why Shannon entropy is not in general a (non-negative) measure,
consider the previous even-odd dice example of two random variables
$X,Y:U\rightarrow2$ for $U=\left\{  \left(  0,0\right)  ,\left(  0,1\right)
,\left(  1,0\right)  ,\left(  1,1\right)  \right\}  $ where $X$ was the parity
of the first die thrown and $Y$ the parity of a second die thrown. Each point
$\left(  x,y\right)  \in U=X\times Y$ has probability $p\left(  x,y\right)
=\frac{1}{4}$ and marginal distributions have $p\left(  x\right)  =\frac{1}%
{2}=p\left(  y\right)  $. A two-variable joint distribution $p\left(
x,y\right)  $ has the \textit{independence} property if $p\left(  x,y\right)
=p\left(  x\right)  p\left(  y\right)  $ for all $\left(  x,y\right)  \in U$.
Hence the two r.v.s $X$ and $Y$ are independent. One of the original `selling
points' of Shannon entropy was that for independent r.v.s, $H\left(
X,Y\right)  =H\left(  X\right)  +H\left(  Y\right)  $, i.e., independent r.v.s
have `no information in common' so that $I(X,Y)=0$. It might be noted that
having an overlap of $H\left(  X\right)  $ and $H\left(  Y\right)  $ of $0$ is
not the same as the Venn diagrams for $H\left(  X\right)  $ and $H\left(
Y\right)  $ not overlapping.

Consider a third random variable $Z:U\rightarrow2$ whose value is the parity
of the sum $X+Y$ so $Z\left(  \left(  0,0\right)  \right)  =Z\left(  \left(
1,1\right)  \right)  =0$ and $Z\left(  \left(  0,1\right)  \right)  =Z\left(
\left(  1,0\right)  \right)  =1$. Then $\Pr\left(  Z=0\right)  =\Pr\left(
Z=1\right)  =\frac{1}{2}=p\left(  z\right)  $ and $p\left(  x,z\right)
=p\left(  x\right)  p\left(  z\right)  $ for all $x,z\in\left\{  0,1\right\}
=2$ so $X$ and $Z$ are also independent and similarly for $Y$ and $Z$. Thus
the three variables are pair-wise independent but they are not mutually
independent for the simple reason that if you know the values of any two of
them, you know the value of the third variable. Hence in the Venn diagram for
the Shannon entropies of $X$, $Y$, and $Z$, each pair of areas must have zero
overlap but the three areas must intersect in non-zero overlap. The only way
this can happen is for the three-way overlap to be negative and the two-way
overlaps be the sum of that negative triple overlap and the equal positive
remaining two-way overlap so all the two-way overlaps are zero as shown in
Figure 9.%

\begin{center}
\includegraphics[
height=2.2569in,
width=2.6533in
]%
{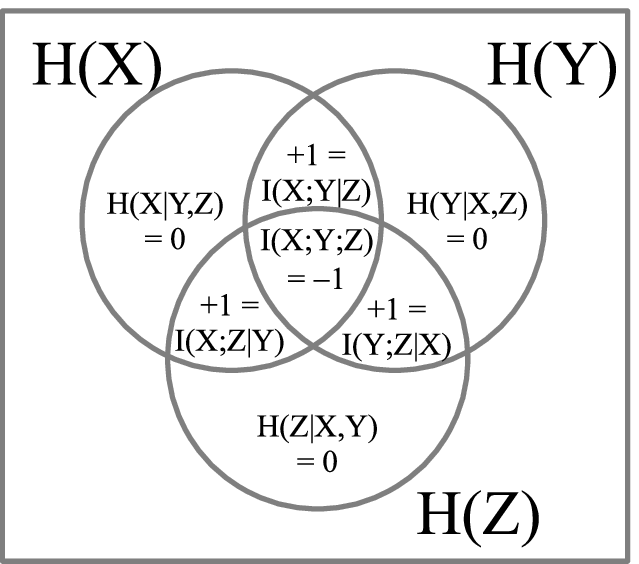}%
\end{center}

\begin{center}
Figure 9: Venn diagram for three-way negative Shannon mutual information
$I\left(  X;Y;Z\right)  $.
\end{center}

Thus the intuitively satisfactory idea of the two-way overlaps for independent
variables being zero (`no information in common') leads to the interpretive
`problem' of three-way mutual information being possibly negative. Shannon
dealt with this problem in the simplest possible way; he never defined mutual
information for more than two variables. Or, as perhaps the most definitive
monograph on information theory casually put it; "There isn't really a notion
of mutual information common to three random variables." \cite[p.
49]{cover:eit} But the three-way definition is automatically given by the
usual inclusion-exclusion formulas that hold for measures and signed measures.
For two variables, $H\left(  X,Y\right)  =H\left(  X\right)  +H\left(
Y\right)  -I\left(  X;Y\right)  $, and for three variables, it is:%
\begin{equation}
H\left(  X,Y,Z\right)  =H\left(  X\right)  +H\left(  Y\right)  +H\left(
Z\right)  -I\left(  X;Y\right)  -I\left(  X;Z\right)  -I\left(  Y;Z\right)
+I\left(  X;Y;Z\right)  \tag{3.9}%
\end{equation}

\noindent where Shannon defined all the terms in the equation except the last
one $I\left(  X;Y;Z\right)  $ which is thus determined. The Shannon entropy
for each variable $X$, $Y$, and $Z$, is:%
\begin{equation}
H\left(  X\right)  =H\left(  Y\right)  =H\left(  Z\right)  =p\left(  0\right)
\log\left(  1/p\left(  0\right)  \right)  +p\left(  1\right)  \log\left(
1/p\left(  1\right)  \right)  =\frac{1}{2}\log\left(  2\right)  +\frac{1}%
{2}\log\left(  2\right)  =1. \tag{3.10}%
\end{equation}

\noindent And all the two-way overlaps have the values:%
\begin{equation}
I\left(  X;Y\right)  =I\left(  X;Z\right)  =I\left(  Y;Z\right)  =0.
\tag{3.11}%
\end{equation}

The three-way joint entropy is the Shannon entropy of the probability
distribution $p\left(  x,y,z\right)  $ which is easily computed in Table 1.
Since the values of any two variables determine the third, the probabilities
are either $\frac{1}{4}$ if the third value agrees with the values of the
other two or $0$ otherwise.

\begin{center}%
\begin{tabular}
[c]{|c|c|c|c|c|}\hline
$X$ & $Y$ & $Z$ & $p\left(  x,y,z\right)  $ & $p\left(  x,y,z\right)
\log\left(  1/p\left(  x,y,z\right)  \right)  $\\\hline\hline
$0$ & $0$ & $0$ & $\frac{1}{4}$ & $\frac{1}{4}\times2=\frac{1}{2}$\\\hline
$0$ & $0$ & $1$ & $0$ & $0$\\\hline
$0$ & $1$ & $0$ & $0$ & $0$\\\hline
$0$ & $1$ & $1$ & $\frac{1}{4}$ & $\frac{1}{2}$\\\hline
$1$ & $0$ & $0$ & $0$ & $0$\\\hline
$1$ & $0$ & $1$ & $\frac{1}{4}$ & $\frac{1}{2}$\\\hline
$1$ & $1$ & $0$ & $0$ & $0$\\\hline
$1$ & $1$ & $1$ & $\frac{1}{4}$ & $\frac{1}{2}$\\\hline
\end{tabular}

Table 1: Probability distribution $p\left(  x,y,z\right)  $ and computation of
$H\left(  X,Y,Z\right)  $.
\end{center}

\noindent The sum of the last column gives the three-way joint Shannon entropy
of $H\left(  X,Y,Z\right)  =2$. Hence the inclusion-exclusion formula gives:%
\begin{align}
I\left(  X;Y;Z\right)   &  =H\left(  X,Y,Z\right)  -H\left(  X\right)
-H\left(  Y\right)  -H\left(  Z\right)  +I\left(  X;Y\right)  +I\left(
X;Z\right)  +I\left(  Y;Z\right) \nonumber\\
&  =2-1-1-1+0+0+0=-1. \tag{3.12}%
\end{align}

\noindent Thinking in term of underlying points, the three-way overlap has
points that are common to $H\left(  X\right)  $, $H\left(  Y\right)  $, and
$H\left(  Z\right)  $, so some of the points must have negative values.
\textit{Thus all three, }$H\left(  X\right)  $\textit{, }$H\left(  Y\right)
$\textit{, and }$H\left(  Z\right)  $\textit{, cannot be the value of a
(non-negative) measure on some set}. Moreover, the intuitive appeal of
$I\left(  X;Y\right)  =0$ as meaning "no information in common" for
independent variables is lessened when it turns out to mean not disjoint or
non-overlapping areas but that the positive information in each of the three
two-way overlap of these independent random variables must be balanced by
"negative information" in the three-way overlap, which, as Csiszar and
Kr\"{o}ner remark, has "no natural intuitive meaning." \cite[p. 53]%
{csis-korn:infotheory}

Finally, we might consider how this example is treated by logical entropy. The
r.v. $Z$ has a logical entropy $h\left(  Z\right)  $ as the sum of the shaded
$\frac{1}{16}$ squares in Figure 10.%

\begin{center}
\includegraphics[
height=2.5618in,
width=2.5373in
]%
{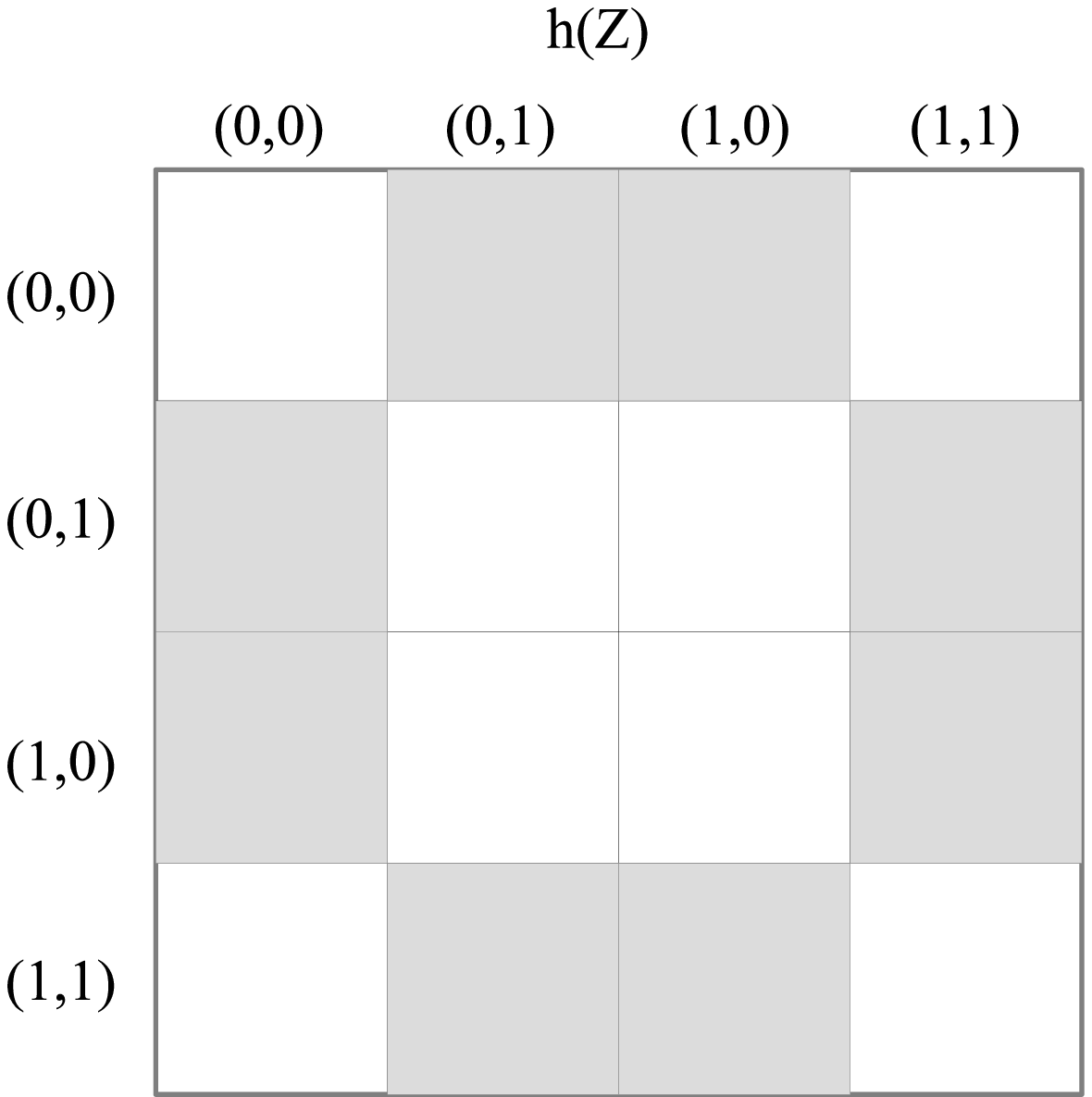}%
\end{center}

\begin{center}
Figure 10: Venn diagram for logical entropy $h\left(  Z\right)  =\frac{8}%
{16}=\frac{1}{2}$.
\end{center}

\noindent The two-way mutual logical information, say for $m\left(
X,Y\right)  $, is given by the shaded squares that are in common, i.e., the
two-way overlap of $h\left(  X\right)  $ and $h\left(  Y\right)  $, and the
three-way mutual logical entropy $m\left(  X,Y,Z\right)  $ is given by the
shaded squares in common to all three. But since logical entropy is a measure
(in the usual non-negative sense), we can compute the three-way mutual
information by using the undercount-overcount formula:%
\begin{equation}
m\left(  X,Y,Z\right)  =h\left(  X,Y,Z\right)  -h\left(  X\right)  -h\left(
Y\right)  -h\left(  Z\right)  +m\left(  X,Y\right)  +m\left(  X,Z\right)
+m\left(  Y,Z\right)  . \tag{3.12}%
\end{equation}

\noindent The three-way joint logical entropy includes all squares except the
diagonal so its value is $\frac{12}{16}=\frac{3}{4}$. The single logical
entropies are all $\frac{1}{2}$ and the two-way mutual informations are all
$\frac{4}{16}=\frac{1}{4}$ so the formula yields:%
\begin{equation}
m\left(  X,Y,Z\right)  =\frac{3}{4}-\frac{1}{2}-\frac{1}{2}-\frac{1}{2}%
+\frac{1}{4}+\frac{1}{4}+\frac{1}{4}=\frac{6}{4}-\frac{3}{2}=0. \tag{3.13}%
\end{equation}

\noindent The two-way logical mutual informations for independent variables
are not zero since logical entropy is a probability distribution--so for
independent variables, it is multiplicative, e.g., $m\left(  X,Y\right)
=h\left(  X\right)  h\left(  Y\right)  $. The calculation of the three-way
mutual logical information can be intuitively checked by considering the three
areas for $h\left(  X\right)  $, $h\left(  Y\right)  $, and $h\left(
Z\right)  $ in Figure 11.%

\begin{center}
\includegraphics[
height=1.347in,
width=4.3799in
]%
{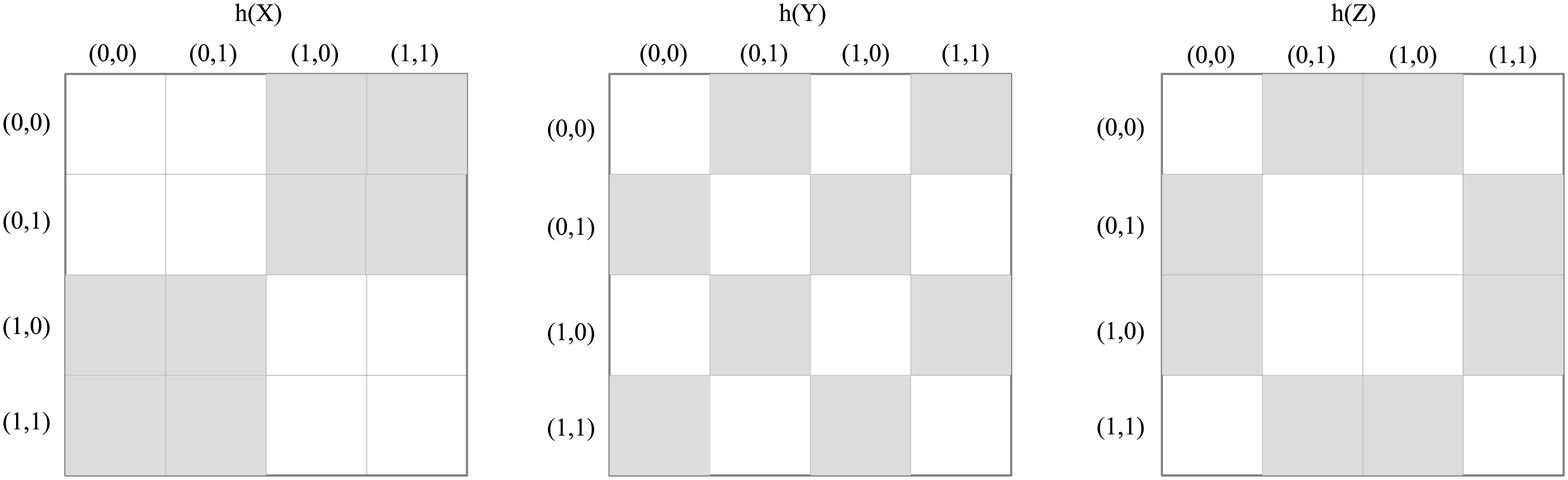}%
\end{center}

\begin{center}
Figure 11: The box diagrams for $h\left(  X\right)  $, $h\left(  Y\right)  $,
and $h\left(  Z\right)  $.
\end{center}

\noindent It can then be checked by inspection that there is no shaded square
common to all three diagrams so the three-way overlap is zero.

It is interesting to note that all \textit{two}-way mutual logical
informations such as $m\left(  X,Y\right)  $, $m\left(  X,Z\right)  $, and
$m\left(  Y,Z\right)  $, are in general never zero when all point
probabilities are positive. This is the result of the Common-Dits Theorem that
any \textit{two} non-empty ditsets have a non-empty
intersection.\footnote{This is a restatement of the graph-theoretic result
that the complement of any disconnected graph is connected \cite[p.
30]{wil:gt}. In terms of inditsets or equivalence relations $E$ and
$E^{\prime}$, if $E\cup E^{\prime}=U\times U$, then $E=U\times U$ or
$E^{\prime}=U\times U$.}

\begin{theorem}
[3.1 Common Dits]Any two non-empty ditsets intersect, i.e., have some dits in common.
\end{theorem}

\noindent\textbf{Proof}: A ditset $\operatorname{dit}\left(  \pi\right)
=\emptyset$ iff $\pi=\mathbf{0}_{U}$, the indiscrete partition or blob.
Consider any two non-empty ditsets $\operatorname{dit}\left(  \pi\right)  $
and $\operatorname{dit}\left(  \sigma\right)  $. Since $\pi$ is not the blob
$\mathbf{0}_{U}$, consider two elements $u$ and $u^{\prime}$ distinguished by
$\pi$ but identified by $\sigma$; otherwise $\left(  u,u^{\prime}\right)
\in\operatorname*{dit}\left(  \pi\right)  \cap\operatorname*{dit}\left(
\sigma\right)  $ and we are finished. Since $\sigma$ is also not the blob,
there must be a third element $u^{\prime\prime}$ not in the same block of
$\sigma$ as $u$ and $u^{\prime}$, as shown in Figure 12.%

\begin{center}
\includegraphics[
height=1.5622in,
width=1.6054in
]%
{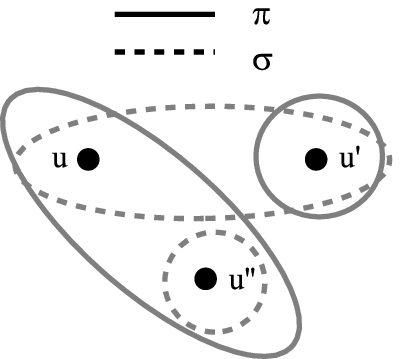}%
\end{center}

\begin{center}
Figure 12: Solid circles = blocks of $\pi$, dashed circles = blocks of
$\sigma$, and $\left(  u^{\prime},u^{\prime\prime}\right)  $ as a common dit
to $\pi$ and $\sigma$.
\end{center}

\noindent But since $u$ and $u^{\prime}$ are in different blocks of $\pi$, the
third element $u^{\prime\prime}$ must be distinguished from one or the other
or both in $\pi$, e.g., distinguished from $u^{\prime}$ in both partitions as
in Figure 12. Hence $\left(  u,u^{\prime\prime}\right)  $ or $\left(
u^{\prime},u^{\prime\prime}\right)  $ must be distinguished by both partitions
and thus must be in $\operatorname*{dit}\left(  \pi\right)  \cap
\operatorname*{dit}\left(  \sigma\right)  $. $\square$

The three non-trivial partitions on a three-element set show that there are no
common dits to all three of them (as in the dice example), only to each pair
of partitions.

\subsection{The connection between the logical and Shannon entropies}

One question lingers. If, as we have seen, Shannon entropy is not defined as a
measure in the usual non-negative sense, then what accounts for the compound
Shannon entropies satisfying the Venn diagram relationships? As one author
surmised: \textquotedblleft Shannon carefully contrived for this `accident' to
occur\textquotedblright\ \cite[p. 153]{rozeboom:abstract}, and Campbell asked
"whether these analogies have a deeper foundation." \cite[p. 112]{camp:meas}
Since Shannon arranged or "contrived" for the compound entropies to satisfy
the Venn diagram relationships for two random variables, they can be extended
to any number of variables using the inclusion-exclusion formulas
(\cite{mcgill:psycho}; \cite{fano:trans}). As we have seen, mutual information
can be negative for three or more variables.

But there is an interesting twist to the story. Information theorists do not
\textit{define} Shannon entropy as a signed measure on a given set. But such a
set can be trivially constructed \textit{ex post} in the manner shown by Hu
\cite{hu:info} and Yeung \cite{yeung:outlook} but the underlying mathematical
fact about additive set functions goes back at least to the 1925 first edition
of Polya-Szego's book \cite{polya-szego:vol-II}. In the Venn diagram showing
all possible overlaps of three "circles" for three random variables, there are
$2^{3}=8$ atomic areas ($2^{n}$ in the general finite case of $n$ random
variables), each of which can trivially be taken as a single element in a set.
Then numbers (positive or negative) can be assigned arbitrarily to those
points and then summed to get the values attached to the circles. Then all the
Venn diagram relationships are automatically satisfied. Since these sets are
constructed in terms of the independently defined Shannon entropies, the set
and value assignments may change when more variables come into play. In the
even-odd dice example, as long as only $X$ and $Y$ are considered, then all
the compound Shannon entropies are non-negative and a set with a
(non-negative) measure on it can be constructed to yield the values of $H(X)$
and $H\left(  Y\right)  $. But when the variable $Z$ is brought into
consideration, then the underlying set must be reconstructed to have a
negative-valued point representing $I\left(  X;Y;Z\right)  $ so that the
\textit{signed} measure on that set will give the values of all the compound
Shannon entropies.\footnote{The construction is easy; take the seven atomic
areas inside the three circles in Figure 9 as containing one point having the
value assigned to that atomic area. The eighth area outside the three circles
can have an arbitrary value--at least until another random variable appears.}
This serves to underline the fact that Shannon entropy is not \textit{defined}
as a measure on a set in the first place.

In contrast, the logical entropy defines the set beforehand, namely $U\times
U$, and the values assigned to the points is determined beforehand, namely
$p_{i}p_{j}$ is assigned to $\left(  u_{i},u_{j}\right)  $, and then the
simple and compound logical entropies are defined by collections of those
points and their values. Nothing changes when new random variables are
considered; it just means considering a different set of points. Thus it is
not a simple matter of saying logical entropy is a (non-negative) measure and
Shannon entropy is a signed measure. Logical entropy is defined as a
probability measure on a set given beforehand, and the Shannon entropies are
only a signed measure on a set \textit{ex post} constructed for the purpose
after all the numerical values are independently given in the Venn diagram
formulas for a given set of random variables.

There is, however, a deeper connection between the two entropies since there
is a transform from all the compound logical entropy formulas to the
corresponding compound Shannon entropy formulas that preserves the Venn
diagram relationships. To understand this transform, consider the canonical
examples of $2^{m}$ equiprobable points in $U$. For any $m$ such as $m=3$, the
binary partitions required to distinguish the $2^{3}=8$ "messages" (or leaves)
can be pictured in the (upside-down) binary tree of Figure 13.%

\begin{center}
\includegraphics[
height=2.1366in,
width=3.5547in
]%
{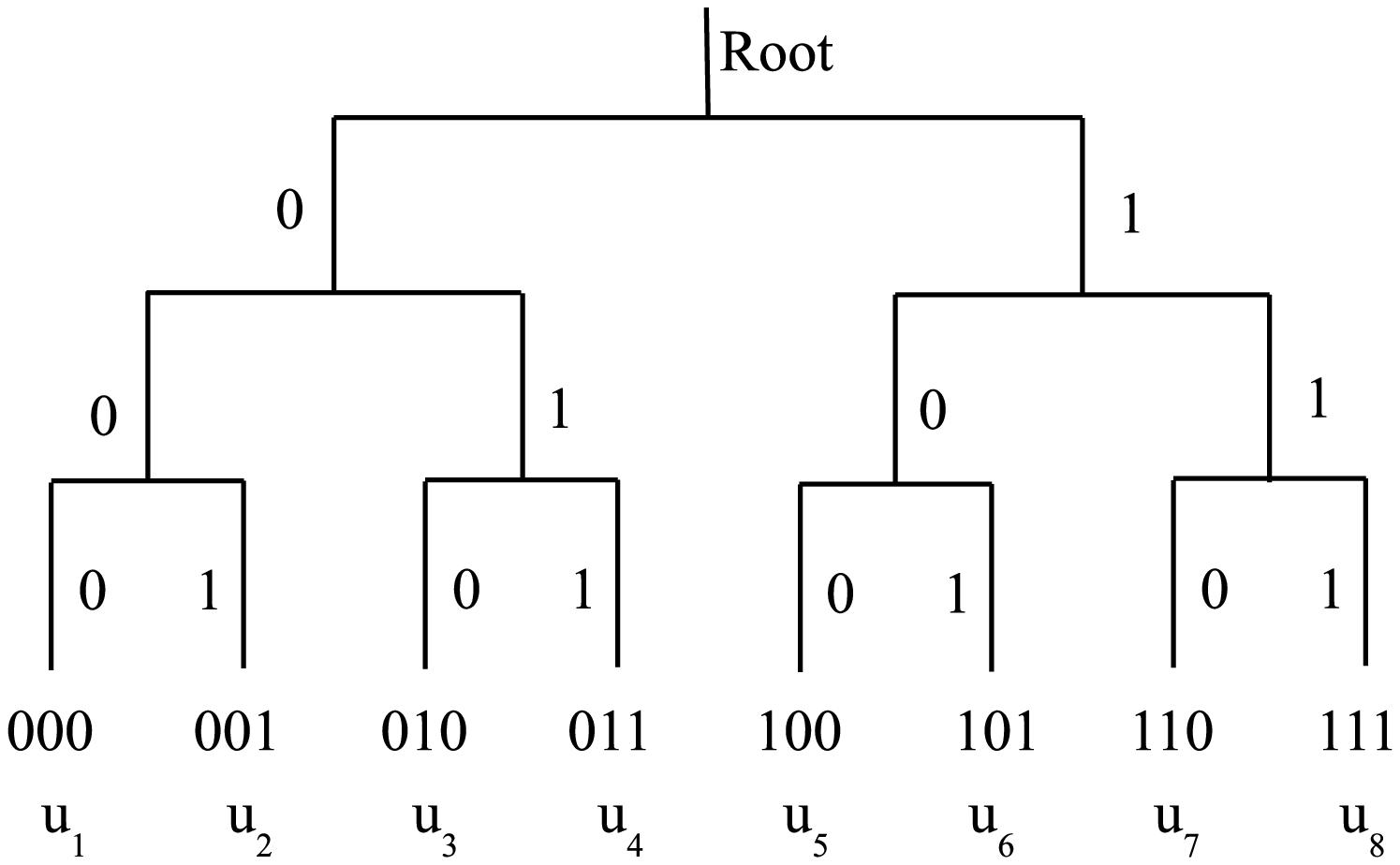}%
\end{center}

\begin{center}
Figure 13: Three equiprobable binary partitions distinguish the $2^{3}=8$
leaves on the tree.
\end{center}

\noindent It was previously asserted that logical entropy and Shannon entropy
are two different ways to quantify the definition of
information-as-distinctions. Logical entropy is the direct (normalized) count
of the distinctions or dits in a partition and Shannon entropy is the minimum
average number of binary partitions that need to be joined together to make
\textit{the same distinctions}.

This connection can be easily demonstrated using the Figure 13 example. Let
$\pi=\left\{  \left\{  u_{1}\right\}  ,...,\left\{  u_{8}\right\}  \right\}  $
be the discrete partition on the equi-probable outcomes (messages or leaves in
the tree) in $U=\left\{  u_{1},...,u_{8}\right\}  $. The number of
distinctions $\left\vert \operatorname{dit}\left(  \pi\right)  \right\vert $
is $\left\vert U\times U-\Delta\right\vert =64-8=56$ (where $\Delta$ is the
diagonal of self-pairs $\left\{  \left(  u_{1},u_{1}\right)  ,...,\left(
u_{8},u_{8}\right)  \right\}  $), and the logical entropy $h\left(
\pi\right)  $ of $\pi$ is $\frac{\left\vert \operatorname{dit}\left(
\pi\right)  \right\vert }{\left\vert U\times U\right\vert }=\frac{56}%
{64}=\frac{7}{8}=1-\frac{1}{8}$, the probability that in two independent draws
from $U$, different elements of $U$ are obtained, i.e., the probability
$1-\frac{1}{8}$ that the second draw isn't the same as the first draw. Since
the outcomes $u_{i}$ are the leaves of the tree in Figure 13, one could image
a marble rolling down from the root (like on a Galton board) and then going
one way or the other with equal probability at each branching. The logical
entropy is the probability that two such marbles will end up in different leaves.

We need to show that the Shannon entropy of $\pi$ is the minimum number of
binary partitions (corresponding to yes-or-no questions in the game of
20-questions) necessary to make all the same distinctions of $\pi$. Recall
that given $\pi$ and $\sigma$ partitions on $U$, the\textit{\ join} $\pi
\vee\sigma$ is the partition on $U$ whose blocks are all the non-empty
intersections $B\cap C$ for $B\in\pi$ and $C\in\sigma$. Since
$\operatorname{dit}\left(  \pi\vee\sigma\right)  =\operatorname{dit}\left(
\pi\right)  \cup\operatorname{dit}\left(  \sigma\right)  $, the join of
partitions accumulates the distinctions made by each of the partitions.
Gian-Carlo Rota formulated the problem as the Devil selecting a particular
$u_{i}$ or message and not revealing it to the questioner but having to answer
any yes-or-no question truthfully. Or less colorfully, "To determine an
object, we need to ensure that the responses to the sequence of questions
uniquely identifies the object from the set of possible objects" \cite[p.
120]{cover:eit}. But the problems of 1) making all the distinctions, and 2)
uniquely determining any given outcome or message, are equivalent. If the
binary partitions or binary questioning does not distinguish $u_{i}$ from
$u_{j}$, then it would not determine the hidden message if it happened to be
$u_{i}$ or $u_{j}$--and if the questioning cannot determine the message if it
was $u_{i}$ or $u_{j}$, then the corresponding binary partitions do not
distinguish $u_{i}$ and $u_{j}$. This means that the usual Shannon
interpretation about the minimum average number of binary questions necessary
to uniquely determine the message is also the minimum average number of binary
partitions necessary to make all the distinctions between messages.

This can be illustrated with the example of Figure 13. The first binary
partition $\pi^{1}$ corresponds to the first branching point in Figure 13 and
the first binary digit in the codes (reading from left to right in the code
words):%
\begin{equation}
\pi^{1}=\left\{  \left\{  u_{1},...,u_{4}\right\}  ,\left\{  u_{5}%
,...,u_{8}\right\}  \right\}  \tag{3,14}%
\end{equation}

\noindent where $u_{1},...,u_{4}$ have $0$ as the first digit and
$u_{5},...,u_{8}$ have $1$ as the first digit. The binary partition $\pi^{1}$
corresponds to the yes-or-no question, "Is the first letter in the code for
$u_{i}$ a $0$?". The partition has $16$ distinctions from $\left\{
u_{1},...,u_{4}\right\}  \times\left\{  u_{5},...,u_{8}\right\}  $ and another
$16$ from the reverse ordering for a total of $32$ distinctions.

The second binary partition $\pi^{2}$ in effect asks about the second digit in
the codes for the $u_{i}$, and it is:%
\begin{equation}
\pi^{2}=\left\{  \left\{  u_{1},u_{2},u_{5},u_{6}\right\}  ,\left\{
u_{3},u_{4},u_{7},u_{8}\right\}  \right\}  \tag{3.15}%
\end{equation}

\noindent so the join is:%
\begin{equation}
\pi^{1}\vee\pi^{2}=\left\{  \left\{  u_{1},u_{2}\right\}  ,\left\{
u_{3},u_{4}\right\}  ,\left\{  u_{5},u_{6}\right\}  ,\left\{  u_{7}%
,u_{8}\right\}  \right\}  . \tag{3.16}%
\end{equation}

\noindent Comparing $\pi^{1}\vee\pi^{2}$ to $\pi^{1}$, we see the splitting
$\left\{  u_{1},...,u_{4}\right\}  $ into $\left\{  u_{1},u_{2}\right\}  $ and
$\left\{  u_{3},u_{4}\right\}  $ so that creates the new distinctions
$\left\vert \left\{  u_{1},u_{2}\right\}  \times\left\{  u_{3},u_{4}\right\}
\right\vert \times2=8$ and similarly for $\left\{  u_{5},u_{6}\right\}  $ and
$\left\{  u_{7},u_{8}\right\}  $, so $\pi^{2}$ makes $8+8=16$ new distinctions
for $32+16=48$ distinctions. Equivalently, one could compute the distinctions
of $\pi^{1}\vee\pi^{2}$ from scratch to get the same total.

The third binary partition $\pi^{3}$ in effect asks about the third digit in
the codes for the $u_{i}$, and it is:%
\begin{equation}
\pi^{3}=\left\{  \left\{  u_{1},u_{3},u_{5},u_{7}\right\}  ,\left\{
u_{2},u_{4},u_{6},u_{8}\right\}  \right\}  \tag{3.17}%
\end{equation}

\noindent and the final join is:%
\begin{equation}
\pi^{1}\vee\pi^{2}\vee\pi^{3}=\left\{  \left\{  u_{1}\right\}  ,...,\left\{
u_{8}\right\}  \right\}  =\pi. \tag{3.18}%
\end{equation}

\noindent Since $\pi^{3}$ distinguishes each of the four pairs in $\pi^{1}%
\vee\pi^{2}$, it introduces $\left\vert \left\{  u_{1}\right\}  \times\left\{
u_{2}\right\}  \right\vert \times4\times2=8$ new distinctions for a total of
$48+8=56$ distinctions. Thus the three partitions together make the same $56$
distinctions, but Shannon entropy counts the number of those binary partitions
(bits) necessary to make the distinctions instead of counting the distinctions
or dits themselves. This illustrates that Shannon entropy $H\left(  p\right)
=\sum_{i=1}^{8}\frac{1}{8}\log_{2}\left(  \frac{1}{1/8}\right)  =\log
_{2}\left(  \frac{1}{1/8}\right)  =3$ is \textit{also} quantifying
distinctions in the sense of counting the minimum number of binary partitions,
namely $3$ in this case, needed to make the same distinctions. Hence Shannon
entropy is a different quantification of the \textit{same} notion of
information, \textit{information-as-distinctions }(of a partition). It is not
just a quantification of the "amount of uncertainty"--whatever that may be.

Moreover, the example shows how to transform the dit-quantification of
information-as-distinctions (logical entropy) into the bit-quantification of
information-as-distinctions (Shannon entropy). In this canonical case (all
$p_{i}=\frac{1}{2^{n}}$), $h\left(  p\right)  =1-p_{i}$ and $H\left(
p\right)  =\log_{2}\left(  \frac{1}{p_{i}}\right)  $ so the dit-count and
bit-count are precisely related: $h(p)=1-\frac{1}{2^{H\left(  p\right)  }}$
and $H\left(  p\right)  =\log_{2}\left(  \frac{1}{1-h\left(  p\right)
}\right)  $. In general, the two entropies are the probability averages
$\sum_{i}p_{i}\left(  \cdots\right)  $ of those canonical values $1-p_{i}$ and
$\log_{2}\left(  \frac{1}{p_{i}}\right)  $. Hence the transform
\begin{equation}
1-p_{i}\rightsquigarrow\log_{2}\left(  \frac{1}{p_{i}}\right)  \tag{3.19}%
\end{equation}

\noindent transforms logical entropy into Shannon entropy in general:%
\begin{equation}
\fbox{$h(p)=\sum_ip_i\left(  1-p_i\right)  \rightsquigarrow H\left(  p\right)
=\sum_ip_i\log_2\left(  \frac{1}{p_i}\right)  $} \tag{3.20}%
\end{equation}

\begin{center}
The Dit-Bit Transform.
\end{center}

Since the dit-bit transform works for the simple entropies, let us consider
the conditional entropies where Shannon constructed $H\left(  X|Y\right)  $ as
the average of the Shannon entropies for the conditional probability
distributions for $y\in Y$,%
\begin{equation}
H\left(  X|Y\right)  =\sum_{y}p\left(  y\right)  \sum_{x}\frac{p\left(
x,y\right)  }{p\left(  y\right)  }\log\left(  \frac{p\left(  y\right)
}{p\left(  x,y\right)  }\right)  =\sum_{x,y}p\left(  x,y\right)  \log\left(
\frac{p\left(  y\right)  }{p\left(  x,y\right)  }\right)  . \tag{3.21}%
\end{equation}

\noindent First, we express the logical conditional entropy as a probability
average:%
\begin{align}
h\left(  X|Y\right)   &  =h\left(  X,Y\right)  -h\left(  Y\right)  =\sum
_{x,y}p\left(  x,y\right)  \left(  1-p\left(  x,y\right)  \right)  -\sum
_{y}p\left(  y\right)  \left(  1-p\left(  y\right)  \right) \nonumber\\
&  =\sum_{x,y}p\left(  x,y\right)  \left[  \left(  1-p\left(  x,y\right)
\right)  -\left(  1-p\left(  y\right)  \right)  \right]  \tag{3.22}%
\end{align}

\begin{center}

\end{center}

\noindent and then we make the substitutions of the dit-bit transform:
$1-p\left(  x,y\right)  \rightsquigarrow\log\left(  1/p\left(  x,y\right)
\right)  $ and $1-p\left(  y\right)  \rightsquigarrow\log\left(  1/p\left(
y\right)  \right)  $ to get:%
\begin{equation}
\sum_{x,y}p\left(  x,y\right)  \left[  \log\left(  1/p\left(  x,y\right)
\right)  -\log\left(  1/p\left(  y\right)  \right)  \right]  =\sum
_{x,y}p\left(  x,y\right)  \log\left(  \frac{p\left(  y\right)  }{p\left(
x,y\right)  }\right)  =H\left(  X|Y\right)  . \tag{3.23}%
\end{equation}

\noindent The other dit-bit transforms go in the same manner at indicated in
Table 2.

\begin{center}%
\begin{tabular}
[c]{c|c|}\cline{2-2}
& The Dit-Bit Transform: $1-p_{i}\rightsquigarrow\log\left(  \frac{1}{p_{i}%
}\right)  $\\\hline\hline
\multicolumn{1}{|r|}{$h\left(  p\right)  =$} & $\sum_{i}p_{i}\left(
1-p_{i}\right)  $\\\hline
\multicolumn{1}{|r|}{$H\left(  p\right)  =$} & $\sum_{i}p_{i}\log\left(
1/p_{i}\right)  $\\\hline\hline
\multicolumn{1}{|c|}{$h\left(  X,Y\right)  =$} & $\sum_{x,y}p\left(
x,y\right)  \left[  1-p\left(  x,y\right)  \right]  $\\\hline
\multicolumn{1}{|c|}{$H\left(  X,Y\right)  =$} & $\sum_{x,y}p\left(
x,y\right)  \log\left(  \frac{1}{p\left(  x,y\right)  }\right)  $%
\\\hline\hline
\multicolumn{1}{|r|}{$m\left(  X,Y\right)  =$} & $\sum_{x,y}p\left(
x,y\right)  \left[  \left[  1-p\left(  x\right)  \right]  +\left[  1-p\left(
y\right)  \right]  -\left[  1-p\left(  x,y\right)  \right]  \right]  $\\\hline
\multicolumn{1}{|r|}{$I(X,Y)=$} & $\sum_{x,y}p\left(  x,y\right)  \left[
\log\left(  \frac{1}{p\left(  x\right)  }\right)  +\log\left(  \frac
{1}{p\left(  y\right)  }\right)  -\log\left(  \frac{1}{p\left(  x,y\right)
}\right)  \right]  $\\\hline\hline
\end{tabular}

Table 2: The dit-bit transform from the compound logical entropies to the
corresponding Shannon entropies.
\end{center}

\noindent As one can see, the preservation of the Venn diagram relationships
is built into the dit-bit transformation. For instance,%
\begin{equation}
m\left(  X,Y\right)  =\sum_{x,y}p\left(  x,y\right)  \left[  \left[
1-p\left(  x\right)  \right]  +\left[  1-p\left(  y\right)  \right]  -\left[
1-p\left(  x,y\right)  \right]  \right]  =h\left(  X\right)  +h\left(
Y\right)  -h\left(  X,Y\right)  \tag{3.24}%
\end{equation}

\noindent transforms to:%
\begin{align}
I\left(  X;Y\right)   &  =\sum_{x,y}p\left(  x,y\right)  \left[  \log\left(
\frac{1}{p\left(  x\right)  }\right)  +\log\left(  \frac{1}{p\left(  y\right)
}\right)  -\log\left(  \frac{1}{p\left(  x,y\right)  }\right)  \right]
\nonumber\\
&  =H\left(  X\right)  +H\left(  Y\right)  -H\left(  X,Y\right)  \tag{3.25}%
\end{align}

\noindent so that Venn diagram relationships are preserved. The dit-bit
transform thus provides the "deeper foundation" \cite[p. 112]{camp:meas}
sought more than a half-century ago by Lorne Campbell for the Shannon
entropies satisfying the Venn diagram relationships in spite of not being
defined as a measure on a set.

A basic inequality in Shannon's communications theory is that for positive
$x$, $1-x\leq\ln\left(  1/x\right)  $. Substituting logs to base $2$ for
natural logs, it is still true that $1-p_{i}\leq\log_{2}\left(  1/p_{i}%
\right)  $ for $0<p_{i}\leq1$ as shown in Figure 14.%

\begin{center}
\includegraphics[
height=2.3356in,
width=3.5115in
]%
{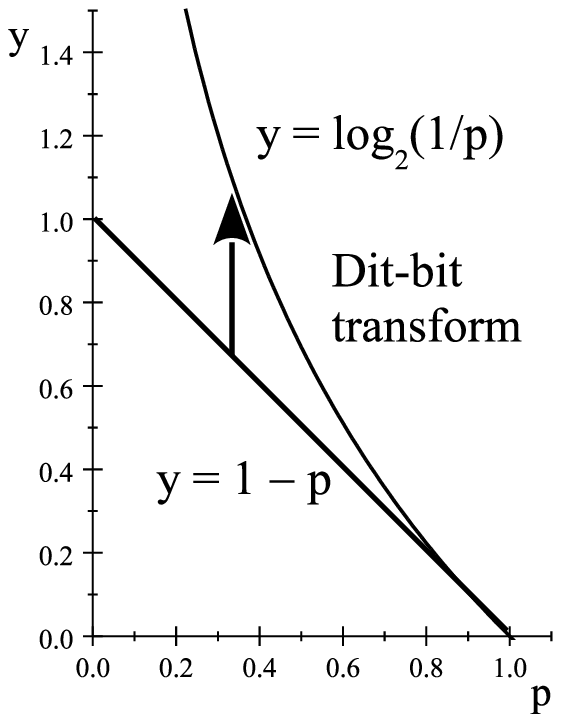}%
\end{center}

\begin{center}
Figure 14: Dit-bit transform and inequality: $1-p\leq\log_{2}\left(
1/p\right)  $ for $0<p\leq1$.
\end{center}

The dit-bit transform is just replacing the left-hand side with the right-hand
side of the inequality so the transform is highly nonlinear--unlike converting
units of measurement like feet and meters. Hence Shannon is correct when he
terms his entropy as the "amount of information" \cite[p. 458]{shannon:topics}
denominated in bits. The logical entropy $h\left(  \pi\right)  $ of a
partition is a direct measure of the distinctions made by a partition and the
Shannon entropy $H\left(  \pi\right)  $ of a partition is statistically the
minimum average number of binary partitions that must be joined to make all
the distinctions of the partition.

\subsection{Boltzmann and Shannon entropies: A conceptual connection?}

When Shannon showed his formula to John von Neumann, then von Neumann
suggested calling it "entropy" for two reasons: there is a similar formula in
Boltzmann's statistical mechanics and you can win more arguments using the
name "entropy" since no one knows what it really is. \cite[pp. 2-3]%
{tribus:30years} How does the Shannon formula $\sum_{i=1}^{m}p_{i}\ln\left(
1/p_{i}\right)  $ (using natural logs) arise in Boltzmann's statistical mechanics?

The context is $n$ particles that can be in $m$ different states (e.g., energy
levels) with a \textit{configuration} (or macrostate) being defined by having
$n_{i}$ particles in the $i^{th}$ state so $\sum_{i=1}^{m}n_{i}=n$. If all the
$m^{n}$ possible assignments ("microstates") of the $n$ particles to the $m$
states are equiprobable, then Boltzmann's idea was that the system would
evolve to the macrostate that had the highest probability. Since the number of
microstates for any configuration is the multinomial coefficient $\binom
{n}{n_{1},...,n_{m}}=\frac{n!}{n_{1}!...n_{m}!}$, the larger the multinomial
coefficient, the larger the probability of that configuration. Hence to find
the equilibrium configuration, the problem is to maximize the multinomial
coefficient subject to the relevant constraints. In addition to $\sum
_{i=1}^{m}n_{i}=n$, each of the $m$ states would have an associated energy
level $\varepsilon_{i}$ and the total energy $\sum_{i=1}^{m}n_{i}%
\varepsilon_{i}$ should equal a constant value $E$. Where did the Shannon
formula come from in Boltzmann's nineteenth-century statistical mechanics?

Since the natural log is a monotonic transformation, it is equivalent to
maximize $\ln\left(  \frac{n!}{n_{1}!...n_{m}!}\right)  $. Moreover, the log
gives an additive quantity to be associated with the extensive quantity of
entropy in thermodynamics. However, maximizing $\ln\left(  \frac{n!}%
{n_{1}!...n_{m}!}\right)  $ subject to the constraints is not very
analytically tractable due to the presence of the factorials $n!$ and $n_{i}%
!$. But there is the Stirling infinite series expression for $\ln\left(
n!\right)  $ and for large $n$, just the first few terms in the series will
give a "for all practical purposes" good approximation. In particular, the
first two terms give the approximation: $n\ln\left(  n\right)  -n\approx
\ln\left(  n!\right)  $. Using that numerical approximation, normalizing by
dividing by $n$, and ignoring the physical Boltzmann's constant, yields a
familiar expression as the approximation:%
\begin{align}
S  &  =\frac{1}{n}\ln\left(  \frac{n!}{n_{1}!...n_{m}!}\right)  =\frac{1}%
{n}\left[  \ln(n!)-\sum_{i=1}^{m}\ln(n_{i}!)\right] \nonumber\\
&  \approx\frac{1}{n}\left[  n\left[  \ln\left(  n\right)  -1\right]
-\sum_{i=1}^{m}n_{i}\left[  \ln\left(  n_{i}\right)  -1\right]  \right]
=\frac{1}{n}\left[  n\ln(n)-\sum_{i}n_{i}\ln(n_{i})\right] \nonumber\\
&  =\frac{1}{n}\left[  \sum n_{i}\ln\left(  n\right)  -\sum n_{i}\ln\left(
n_{i}\right)  \right]  =-\frac{1}{n}\sum_{i}n_{i}\ln\left(  \frac{n_{i}}%
{n}\right) \nonumber\\
&  =-\sum_{i=1}^{m}p_{i}\ln\left(  p_{i}\right)  =\sum_{i=1}^{m}p_{i}%
\ln\left(  1/p_{i}\right)  =H_{e}\left(  p\right)  \text{ where }p_{i}%
=n_{i}/n. \tag{3.26}%
\end{align}

\noindent That is how the two-term Stirling approximation brings the Shannon
formula into the statistical mechanics of Boltzmann (and Gibbs). It should be
noted that the two probability distributions are quite different. For the
exact maximal configuration $\left(  n_{1},...,n_{m}\right)  $, there are
$\frac{n!}{n_{1}!...n_{m}!}$ terms in the equiprobable distribution over the
microstates. For the two-term Stirling approximate probability distribution,
there are only $m$ terms $\left(  p_{1},...,p_{m}\right)  $. It is the total
quantity $\frac{1}{n}\ln\left(  \frac{n!}{n_{1}!...n_{m}!}\right)  $ that is
approximated by the Shannon formula $H_{e}\left(  p\right)  $ for large $n$.

And by taking more terms in the Stirling approximation, one information
theorist notes that one would have an even better approximation \cite[p.
2]{mackay:info}, and a prominent physical chemist notes that $\ln\left(
n!\right)  \approx\sqrt{\ln\left(  2\pi\right)  }+\left(  n+\frac{1}%
{2}\right)  \ln\left(  n\right)  -n$ is a much better approximation \cite[p.
533]{atkins:phy-chem}. But neither uses those formulas since the purpose at
hand is analytical tractability, not better approximations, and the Shannon
formula leads to a very nice development in statistical mechanics--in
particular to the beautiful partition function $Z$ that connects statistical
mechanics to thermodynamics. Unfortunately, the role of what became later
known as the Shannon formula as a very convenient numerical approximation to
Boltzmann entropy is often `forgotten' in the literature where one even sees
expressions like "Shannon-Boltzmann entropy" \cite[p. 11]{cover:eit} or
"Boltzmann-Gibbs-Shannon entropy" (e.g., \cite{ramshaw;bgs}). \noindent
Perhaps nowhere else in mathematical physics has a \textit{numerical}
approximation been attributed such \textit{conceptual} significance.

Another way to emphasize the conceptual difference is to consider a small $n$
example where we can compute both entropies since the original Boltzmann
problem is a tractable integer programming problem \textit{not using any
approximation}. Consider an example of $n=10$ particles with three possible
energy levels of $\varepsilon=\left(  \varepsilon_{1},\varepsilon
_{2},\varepsilon_{3}\right)  =\left(  1,2,3\right)  $ and a total energy of
$E=22$. For $n_{i}$ as the number of particles at energy level $i$, the energy
constraint is $\sum_{i=1}^{3}\varepsilon_{i}n_{i}=E$ and of course $\sum
_{i=1}^{3}n_{i}=n$.\footnote{The example was inspired by Eric Johnson's
excellent treatment of Boltzmann's entropy \cite{johnson:anxiety}.} There are
only four non-negative integer solutions satisfying the two constraints:

\begin{center}%
\begin{tabular}
[c]{|c|c|c|c|}\hline
$n_{1}$ & $n_{2}$ & $n_{3}$ & $\frac{n!}{n_{1}!n_{2}!n_{3}!}$\\\hline\hline
$1$ & $6$ & $3$ & $840$\\\hline
$2$ & $4$ & $4$ & $3150$\\\hline
$3$ & $2$ & $5$ & $2520$\\\hline
$4$ & $0$ & $6$ & $210$\\\hline
\end{tabular}

Table 3: Feasible integer solutions
\end{center}

\noindent The exact Boltzmann solution giving the maximum multinomial
coefficient is $\left(  n_{1},n_{2},n_{3}\right)  =\left(  2,4,4\right)  $ and
the (normalized) Boltzmann entropy is:%
\begin{equation}
S=\frac{1}{10}\ln\left(  \frac{10!}{2!4!4!}\right)  =\frac{1}{10}\ln\left(
3150\right)  =\frac{1}{10}8.055=0.8055 \tag{3.27}%
\end{equation}

\noindent while maximizing the usual Shannon approximation gives the
non-integer result $\left(  n_{1},n_{2},n_{3}\right)  =\left(
2.3837,3.2326,4.3837\right)  $ (to four decimal places) with the Shannon
entropy of $H_{e}\left(  p\right)  =1.0684$ (where $p_{i}=n_{i}/n$). The
probability distribution in the Boltzmann case has $3150$ equal terms with the
value $\frac{1}{3150}$ while the probability distribution in the "Shannon
case" has $3$ terms, $\frac{1}{10}\left(  2.3837,3.2326,4.3837\right)  $. The
maximization of the multinomial coefficient (or its normalized logarithm) and
the Shannon expression are obviously different for low $n$. But for the
enormous number of particles in a system of statistical mechanics, that
numerical difference fades into insignificance--unless one forgets about it
altogether and attaches \textit{conceptual} significance to the Shannon
formula in Boltzmannian statistical mechanics.

\subsection{MaxEntropy with which entropy for discrete distributions?}

Edwin T. Jaynes \cite{jaynes:pt} started a whole "MaxEntropy" subdiscipline in
information theory by arguing that the classical indifference principle used
to give an equiprobable probability distribution (in the lack of other
knowledge) should be generalized to other more constrained contexts by
choosing the probabilities that maximize the Shannon entropy subject to those
constraints. His motivation was based, in significant part, on attaching
conceptual significance to the maximizing of Shannon entropy in Boltzmannian
statistical mechanics:

\begin{quotation}
\noindent the `method of the most probable distribution' dating back to
Boltzmann ... which turns out in the end to be mathematically equivalent to
maximum [Shannon] entropy. \cite[p. 441]{jaynes:pt}
\end{quotation}

The question naturally arises: "What about maximizing logical entropy subject
to the same constraints?" If there are no constraints, then maximizing both
entropies yields the classical result of the equiprobable distribution, i.e.,
the indifference principle. But when there are constraints, then the two
maximums yield different probability distributions.

Consider a function $X:U\rightarrow%
\mathbb{R}
$ with values $X\left(  u_{i}\right)  =x_{i}$ for $i=1,...,n$ with unknown
probabilities $p=\left(  p_{1},...,p_{n}\right)  $. A standard discrete
MaxEntropy problem is to find the "best" probabilities so that the average
value $\sum_{i=1}^{n}p_{i}x_{i}=m$ for some given value of $m$ (which must be
between the maximum and minimum values of the $x_{i}$).

Where "best" is defined by maximizing the Shannon entropy, the procedure is to
maximize the Lagrangian:%
\begin{equation}
\mathcal{L}=-\sum p_{i}\ln\left(  p_{i}\right)  +\lambda\left(  1-\sum
p_{i}\right)  +\tau\left(  m-\sum p_{i}x_{i}\right)  \tag{3.28}%
\end{equation}

\noindent so the first-order conditions are:%
\begin{equation}
\partial\mathcal{L}/\partial p_{i}=-\ln\left(  p_{i}\right)  -p_{i}\frac
{1}{p_{i}}-\lambda-x_{i}\tau=0 \tag{3.29}%
\end{equation}

\noindent where it should be noted at the outset that the use of the log
function $\ln\left(  p_{i}\right)  $ (and the term $1/p_{i}$ in the
first-order conditions) assumes $p_{i}\neq0$ for all $i$. Then exponentiating
gives: $p_{i}=e^{-\left(  1+\lambda+\tau x_{i}\right)  }$. Substituting into
the constraints to determine the Lagrange multipliers:%
\begin{equation}
1=\sum p_{i}=\sum e^{-\left(  1+\lambda+\tau x_{i}\right)  }=e^{-(1+\lambda
)}\sum e^{-\tau x_{i}}soe^{1+\lambda}=\sum e^{-\tau x_{i}} \tag{3.30}%
\end{equation}

\noindent which yields $p_{MaxH}$ in terms of the Lagrange multiplier $\tau$
as:%
\begin{equation}
p_{i}=e^{-\tau x_{i}}/\sum_{j=1}^{n}e^{-\tau x_{j}}. \tag{3.31}%
\end{equation}

\noindent And $m=\sum x_{i}e^{-\left(  1+\lambda+\tau x_{i}\right)  }=\sum
x_{i}e^{-\tau x_{i}}/\sum e^{-\tau x_{i}}$. Rather than trying to solve
directly for $\tau$, it is best to let $w=e^{-\tau}$ and then numerically
solve for a real root $w$ (aside from $w=0$) of the equation:%
\begin{equation}
\sum_{i}x_{i}w^{x_{i}}-m\left(  \sum_{i}w^{x_{i}}\right)  =0. \tag{3.32}%
\end{equation}

\noindent Given such a real $w$, $\tau=-\ln\left(  w\right)  $ and then the
$p_{MaxH}=(p_{1},...,p_{n})$ for maximizing Shannon entropy $H\left(
p\right)  $ are determined by the above formula: $p_{i}=e^{-\tau x_{i}}%
/\sum_{j=1}^{n}e^{-\tau x_{j}}$. Moreover, it is clear from the formula that
all the $p_{i}$ are positive (and sum to $1$).

Where "best" is defined by maximizing logical entropy, the procedure is to
solve the quadratic programming problem of maximizing $h(p)=1-\sum_{i}%
p_{i}^{2}$ subject to the same constraints $\sum_{i}p_{i}x_{i}=m$ and
$\sum_{i}p_{i}=1$ plus the additional non-negativity constraints $0\leq p_{i}$
for $i=1,...,n$. For a certain range of values of $m$, the non-negativity
constraints will be automatically satisfied so one can approach that part of
the problem using the Lagrangian approach.%
\begin{equation}
\mathcal{L}\left(  p_{1},...,p_{n}\right)  =1-\sum_{i}p_{i}^{2}-\lambda\left(
1-\sum p_{i}\right)  +\tau\left(  m-\sum p_{i}x_{i}\right)  \tag{3.33}%
\end{equation}

\noindent so the first-order conditions are:%
\begin{equation}
\partial\mathcal{L}/\partial p_{i}=-2p_{i}+\lambda-\tau x_{i}=0 \tag{3.34}%
\end{equation}
so%
\begin{equation}
p_{i}=\frac{1}{2}\left(  \lambda-\tau x_{i}\right)  . \tag{3.35}%
\end{equation}

\noindent Using the first constraint:%
\begin{equation}
1=\sum p_{i}=\frac{1}{2}\left(  n\lambda-\tau\sum x_{i}\right)  =\frac{n}%
{2}\lambda-\frac{1}{2}\tau\sum x_{i}, \tag{3.36}%
\end{equation}

\noindent and using the second constraint:%
\begin{equation}
m=\sum p_{i}x_{i}=\sum x_{i}\frac{1}{2}\left(  \lambda-\tau x_{i}\right)
=\lambda\frac{1}{2}\sum x_{i}-\tau\frac{1}{2}\sum x_{i}^{2} \tag{3.37}%
\end{equation}

\noindent so we have two linear equations that can be used to solve for the
Lagrange multipliers $\lambda$ and $\tau$. Before going forward, it is useful
to consider the mean and variance of the $x_{i}$'s if they were equiprobable.
Then $\mu=\frac{\sum x_{i}}{n}$ and $Var\left(  X\right)  =E\left(
X^{2}\right)  -\mu^{2}=\frac{\sum x_{i}^{2}}{n}-\mu^{2}=\sigma^{2}$ so
$nVar\left(  X\right)  =\sum x_{i}^{2}-n\mu^{2}$. Then the two equations are:%
\begin{equation}
1=\frac{n}{2}\lambda-\frac{n\mu}{2}\tau\text{ and }m=\frac{n\mu}{2}%
\lambda-\frac{n}{2}\left[  Var\left(  X\right)  +\mu^{2}\right]  \tau.
\tag{3.38}%
\end{equation}

\noindent After a bit of algebra, one arrives at the informative formula for
the $p_{i}$ in $p_{Maxh}=\left(  p_{1},...,p_{n}\right)  $ that results from
maximizing logical entropy subject to the same constraints:%
\begin{equation}
p_{i}=\frac{1}{n}+\frac{\left(  \mu-m\right)  \left(  \mu-x_{i}\right)
}{nVar\left(  X\right)  }=\frac{1}{n}+\frac{1}{n}\left(  \frac{m-\mu}{\sigma
}\right)  \left(  \frac{x_{i}-\mu}{\sigma}\right)  . \tag{3.39}%
\end{equation}

\noindent Since all the operations in the formula are rational (e.g., no
square roots, not to mention transcendental functions), the probabilities are
all rational if all the $x_{i}$ are rational. One test of intuitiveness is: if
$x_{i}$ is equal to the equiprobable mean $\mu$, then shouldn't that $p_{i}$
equal the equiprobable value $\frac{1}{n}$ regardless of the other values?
That is true as we see from the formula for $p_{i}$. If any $x_{i}=\mu$ then
that $p_{i}=\frac{1}{n}$ and if $m=\mu$, then all the $p_{i}=\frac{1}{n}$, the
equiprobable solution. The condition for all the $p_{i}\geq0$ is that
$\frac{\left(  \mu-m\right)  \left(  \mu-x_{i}\right)  }{nVar\left(  X\right)
}\geq-\frac{1}{n}$ or $\left(  \mu-m\right)  \left(  \mu-x_{i}\right)
\geq-Var\left(  X\right)  $ for all $i$. If that condition is not satisfied
for some $p_{i}$, then the non-negativity constraints must be enforced by
using quadratic programming techniques instead of the Lagrangian technique
used above.\footnote{Microsoft Excel with the Solver application is
sufficient. For a thorough treatment, see \cite{kaplan:max-min} or
\cite{best:quadprog}.}

One of the best-known examples is Jaynes's Brandeis dice problem (\cite[p.
47]{jaynes:mit-symposium} or \cite[p. 427]{papoulis:p&s}). If a die was fair,
then the average of the equiprobable outcomes is $\mu=3.5$. But suppose that
it is a given constraint that the average outcome is $4.5$, then what is the
"best" estimate of the probabilities for the six sides?

To maximize Shannon entropy, the $x_{i}$'s are $1,2,3,4,5,6$ so the equation
to be numerically solved is:%
\begin{equation}
\sum_{i=1}^{6}iw^{i}-m\left(  \sum_{i=1}^{6}w^{i}\right)  =0 \tag{3.40}%
\end{equation}

\noindent for $m=4.5$. In addition to $w=0$, the relevant real root to four
decimal places is $w=1.449\,3$. Then $\tau=-\ln\left(  1.449\,3\right)
=-0.371\,08$ and the Jaynes solution for the probabilities to four decimal
places is:%
\begin{equation}
p_{MaxH}=\left(  0.0543,0.0788,0.1142,0.1654,0.2398,0.3475\right)  .
\tag{3.41}%
\end{equation}

To maximize logical entropy, $\mu=\frac{7}{2}$, $m=\frac{9}{2}=4.5$, and
$Var\left(  X\right)  =\frac{35}{12}$, so the formula $p_{i}=\frac{1}{n}%
+\frac{\left(  \mu-m\right)  \left(  \mu-x_{i}\right)  }{nVar\left(  X\right)
}$ can be used to solve for the \textit{rational} maximum logical entropy
solution:%
\begin{align}
p_{Maxh}  &  =\frac{1}{210}\left(  5,17,29,41,53,65\right) \nonumber\\
&  =\left(  0.0238,0.0810,0.1381,0.1952,0.2524,0.3095\right)  . \tag{3.42}%
\end{align}

\noindent In this case, $\left(  \mu-m\right)  \left(  \mu-x_{i}\right)
=\left(  \frac{7}{2}-\frac{9}{2}\right)  \left(  \frac{7}{2}-x_{i}\right)
=-\left(  \frac{7}{2}-x_{i}\right)  \geq-Var\left(  X\right)  =-\frac{35}{12}%
$. The RHS\ is the smallest for $x_{1}=1$ where $-\left(  \frac{7}{2}-\frac
{2}{2}\right)  =-\frac{30}{12}\geq-\frac{35}{12}$ so all the probabilities are
positive. Equality holds when $\mu-m=-\frac{35}{12}\frac{2}{5}=-\frac{7}{6}$
or $m=\frac{7}{2}+\frac{7}{6}=\frac{28}{6}=4\frac{2}{3}$. Hence for any
$m>4\frac{2}{3}$, $p_{1}$ and possibly other $p_{i}$ will be $0$ so quadratic
programming must be used. A little calculation shows that $\frac{7}{3}$ is the
lower bound so that for $m\,<\frac{7}{3}$, there will be some zero
probabilities. For instance, for $m=5$, the probabilities for logical entropy
are: $p_{Maxh}=\frac{1}{10}\left(  0,0,1,2,3,4\right)  $, while the Jaynes
solution is $p_{MaxH}=\left(
0.0205,0.0385,0.0723,0.1357,0.2548,0.4781\right)  $ to four decimal places.

It is interesting that the only alternative to maximizing Shannon entropy that
Jaynes considers \cite[pp. 345-6]{jaynes:pt} is minimizing $\sum_{i}p_{i}^{2}$
which is the same as maximizing logical entropy $h\left(  p\right)
=1-\sum_{i}p_{i}^{2}$. But then he criticizes it because some of the $p_{i}$
may be negative if one uses only the Lagrangian method.

\begin{quotation}
\noindent The formal solution for minimum $\sum_{i}p_{i}^{2}$ lacks the
property of non-negativity. We might try to patch this up in an \textit{ad
hoc} way by replacing the negative values by zero and adjusting the other
probabilities to keep the constraint satisfied. \cite[p. 346]{jaynes:pt}
\end{quotation}

\noindent It is unclear if Jaynes was aware of the field of quadratic
programming which was well-developed in the 1960s \cite[p. 490]{dantzig:lp}
and which hardly proceeds in "an \textit{ad hoc} way by replacing the negative
values by zero and adjusting the other probabilities to keep the constraint satisfied."

Clearly the two solutions are different in general \footnote{For $n=2$, the
two solutions are identical but diverge in general for $n\geq3$.} and each one
maximizes the corresponding type of entropy. How can one determine which
probability distribution is "best"? One criterion that immediately suggests
itself is the distribution $\left(  p_{1},...,p_{n}\right)  $ of numbers that
is the most uniform in the sense of having the least variance $Var\left(
p\right)  $ where each of the numbers $p_{i}$ is considered equally probable.
The minimum is $Var\left(  p\right)  =0$ for the uniform probability
distribution which maximizes both entropies in the absence of constraints. In
the two cases where $m=4.5$ and $m=5$, the logical entropy maximizing
distribution $p_{Maxh}$ has the lower variance $Var\left(  p\right)  $. But is
that true in general?

At first, it seems rather intractable to prove in general which of the two
distributions has least variance in the discrete case since the Jaynes
solution involves finding the roots of a high-degree polynomial. But there is
an easy and general proof that the logical entropy solution has a variance
less than (or equal to) the Jaynes solution when both are maximized subject to
the same constraints--and similarly for being closest to the uniform
distribution in terms of the usual notion of Euclidean distance in $%
\mathbb{R}
^{n}$.

\begin{proposition}
[3.1]$Var(p_{Maxh})\leq Var\left(  p_{MaxH}\right)  $.
\end{proposition}

\textbf{Proof}: For any constraint set on the probability distributions
$p=\left(  p_{1},...,p_{n}\right)  $, minimize the variance itself over all
the feasible distributions (rather than maximize either of the two
entropies--or any other entropy for that matter), and then show that the
minimum variance distribution $p_{MinVar}$is the same as the maximum logical
entropy distribution $p_{Maxh}$. The equality $p_{MinVar}=p_{Maxh}$ is shown
by computing the relationship between $Var\left(  p\right)  $ and $h\left(
p\right)  $. Looking at $\left(  p_{1},...,p_{n}\right)  $ as just a set of
equiprobable numbers with $\sum_{i}p_{i}=1$, it has the variance:%
\begin{align}
Var\left(  p\right)   &  =\sum_{i=1}^{n}\frac{1}{n}\left(  p_{i}-\frac{1}%
{n}\right)  ^{2}=E\left(  p^{2}\right)  -E\left(  p\right)  ^{2}\nonumber\\
&  =\frac{1}{n}\sum_{i}p_{i}^{2}-\left(  \frac{1}{n}\sum_{i}p_{i}\right)
^{2}=\frac{1}{n}\sum_{i}p_{i}^{2}-\left(  \frac{1}{n}\right)  ^{2}=\frac{1}%
{n}\left[  \left(  1-\frac{1}{n}\right)  -h\left(  p\right)  \right]
\tag{3.43}%
\end{align}

\noindent since $\sum_{i=1}^{n}p_{i}^{2}=1-h\left(  p\right)  $. Since
$h\left(  p\right)  $ appears with a negative sign in the expression for
$Var\left(  p\right)  $, minimizing $Var\left(  p\right)  $ is the same as
maximizing $h\left(  p\right)  $ over the set of feasible probability
distributions, so $p_{MinVar}=p_{Maxh}$. $\square$

\begin{corollary}
[3.1]$p_{Maxh}$ minimizes the (Euclidean) distance to the uniform distribution
$\left(  \frac{1}{n},...,\frac{1}{n}\right)  $.
\end{corollary}

\textbf{Proof}: Minimizing Euclidean distance is the same as minimizing the
distance squared and $\sum_{i=1}^{n}\left(  p_{i}-\frac{1}{n}\right)
^{2}=\left(  1-\frac{1}{n}\right)  -h\left(  p\right)  $. $\square$

There is another specialized notion of `distance', namely the
\textit{Kullback-Leibler divergence }\cite{kull-lib:div} $D\left(
p||q\right)  =\sum_{i=1}^{n}p_{i}\log\left(  \frac{p_{i}}{q_{i}}\right)  $
($p$ and $q$ are probability distributions on the same index set with all
$q_{i}>0$) which is neither symmetrical nor satisfies the triangle inequality.
But $D\left(  p||\left(  \frac{1}{n},...,\frac{1}{n}\right)  \right)
=\log\left(  n\right)  -H\left(  p\right)  $ so that maximizing $H\left(
p\right)  $ subject to the constraints is equivalent to minimizing the
Kullback-Leibler divergence of $p$ from the uniform distribution.

The corresponding asymmetrical divergence formula for logical entropy, also
for probability distributions $p$ and $q$ where $q_{i}>0$ for all $i$, is the
\textit{directed logical divergence}:
\begin{equation}
d^{\ast}\left(  p||q\right)  :=\sum_{i=1}^{n}\frac{1}{q_{i}}\left(
q_{i}-p_{i}\right)  ^{2}=\sum_{i}p_{i}\left(  \frac{p_{i}}{q_{i}}-1\right)
=\sum_{i}\frac{p_{i}^{2}}{q_{i}}-1\geq0 \tag{3.44}%
\end{equation}

\begin{center}
with equality iff $p=q$.
\end{center}

\noindent Another way to prove non-negativity is to note that $\left(
q_{i}-p_{i}\right)  \left(  1-\frac{p_{i}}{q_{i}}\right)  \geq0$ since both
terms are negative or both are non-negative, and $\sum_{i}\left(  q_{i}%
-p_{i}\right)  \left(  1-\frac{p_{i}}{q_{i}}\right)  =\sum_{i}\frac{p_{i}^{2}%
}{q_{i}}-1$. Since the KL divergence uses probability ratios inside the log
term, we do the same for dit-bit transform so that: $1-\frac{p_{i}}{q_{i}%
}\rightsquigarrow\log\left(  \frac{1}{p_{i}/q_{i}}\right)  $ and thus:%
\begin{equation}
-d^{\ast}\left(  p||q\right)  =\sum_{i}p_{i}\left(  1-\frac{p_{i}}{q_{i}%
}\right)  \rightsquigarrow\sum_{i}p_{i}\log\left(  \frac{1}{p_{i}/q_{i}%
}\right)  =\sum_{i}p_{i}\log\left(  \frac{q_{i}}{p_{i}}\right)  =-D\left(
p||q\right)  \tag{3.45}%
\end{equation}

\noindent so $d^{\ast}\left(  p||q\right)  \rightsquigarrow D\left(
p||q\right)  $, i.e., the KL divergence is the dit-bit transform of the
directed logical divergence. What is the probability distribution closest to
the uniform distribution using the directed logical divergence?
\begin{align}
d^{\ast}\left(  p||\left(  \frac{1}{n},...,\frac{1}{n}\right)  \right)   &
=\sum_{i}p_{i}\left(  np_{i}-1\right)  =n\sum_{i}p_{i}^{2}-1\nonumber\\
&  =n\left[  \left(  1-\frac{1}{n}\right)  -h\left(  p\right)  \right]
=n\sum_{i=1}^{n}\left(  p_{i}-\frac{1}{n}\right)  ^{2} \tag{3.46}%
\end{align}

\noindent so it is the logical entropy solution that is the closest to the
uniform distribution by the logical notion of directed divergence.

\subsection{Metrical logical entropy = (twice) variance}

The above results suggest a broader connection between the usual notion of the
variance of a random variable and the logical entropy of "differences" when
the differences have metrical significance. The logical entropy $h\left(
X\right)  $ of a random variable $X:U\rightarrow%
\mathbb{R}
$ with $n$ distinct values $\left(  x_{1},...,x_{n}\right)  $ with the
probabilities $p=\left(  p_{1},...,p_{n}\right)  $ is computed as $h\left(
X\right)  =\sum_{i\neq j}p_{i}p_{j}$ which only takes notice of when values
are the same or different. Logical entropy in that sense is a special case of
C. R. Rao's notion of quadratic entropy $\sum_{i,j}d_{ij}p_{i}p_{j}$, where
$d_{ij}$ is a non-negative "distance function" such that $d_{ii}=0$ and
$d_{ij}=d_{ji}$ (\cite{rao:div}, \cite{rao:qe}), for the logical distance
function $d_{ij}=1-\delta_{ij}$, the complement of the Kronecker delta. A
natural \textit{metrical} distance function is the Euclidean distance squared
$d_{ij}=\left(  x_{i}-x_{j}\right)  ^{2}$.

\begin{proposition}
[3.2]$\sum_{j\neq i}p_{i}p_{j}\left(  x_{i}-x_{j}\right)  ^{2}=2Var\left(
X\right)  $.\footnote{This formula goes back at least to \cite[p.
42]{kendall:as} and probably further.}
\end{proposition}

\noindent Proof: Firstly, since for $i=j$, $\left(  x_{i}-x_{j}\right)
^{2}=0$, we can sum over all $i,j$.%
\begin{align}
\sum_{j\neq i}p_{i}p_{j}\left(  x_{i}-x_{j}\right)  ^{2}  &  =\sum_{i,j}%
p_{i}p_{j}\left(  x_{i}-x_{j}\right)  ^{2}\nonumber\\
&  =\sum_{i,j}p_{i}p_{j}\left(  x_{i}^{2}-2x_{i}x_{j}+x_{j}^{2}\right)
=E\left(  X^{2}\right)  -2E\left(  X\right)  ^{2}+E\left(  X^{2}\right)
=2Var\left(  X\right)  .\square\tag{3.47}%
\end{align}

\noindent It was previously noted that when counting distinctions $\left(
u_{i},u_{j}\right)  \in\operatorname{dit}\left(  \pi\right)  $, both $\left(
u_{i},u_{j}\right)  $ and $\left(  u_{j},u_{i}\right)  $ are included. If only
the distinctions $\left(  u_{i},u_{j}\right)  $ for $i<j$ are counted, then
one get half the number as is evident in the logical entropy box diagrams such
as Figure 2.

\begin{corollary}
[3.2]$\sum_{i<j}p_{i}p_{j}\left(  x_{i}-x_{j}\right)  ^{2}=Var\left(
X\right)  $. $\square$
\end{corollary}

Thus the variance of a metrical random variable $X$ is the average distance
squared between the values in an unordered pair of independent trials.

The result extends to covariances as well. Consider two real-valued random
variables $X$ with distinct values $x_{i}$ for $i=1,...,n$ and \ $Y$ with
distinct values $y_{j}$ for $j=1,...,m$ with the joint probability
distribution $p\left(  x_{i},y_{j}\right)  :X\times Y\rightarrow%
\mathbb{R}
$. Two ordered draws from $X\times Y$ gives two ordered pairs: $\left(
x_{i},y_{j}\right)  $ and $\left(  x_{i^{\prime}},y_{j^{\prime}}\right)  $.
For this bivariate distribution, the generalization of $\sum_{j\neq i}%
p_{i}p_{j}\left(  x_{i}-x_{j}\right)  ^{2}$ is:%
\begin{equation}
\sum_{\left(  i,j\right)  \neq\left(  i^{\prime},j^{\prime}\right)  }p\left(
x_{i},y_{j}\right)  p\left(  x_{i^{\prime}},y_{j^{\prime}}\right)  \left(
x_{i}-x_{i^{\prime}}\right)  \left(  y_{j}-y_{j^{\prime}}\right)  \tag{3.48}%
\end{equation}

\begin{center}
Metrical logical entropy for bivariate distributions of metrical random variables
\end{center}

\noindent which is no longer a special case of quadratic entropy since
$\left(  x_{i}-x_{i^{\prime}}\right)  \left(  y_{i}-y_{i^{\prime}}\right)  $
can be negative. The (unordered) two-draw notion of metrical variation for a
bivariate distribution reproduces the usual notion of covariance $Cov\left(
X,Y\right)  =E\left(  XY\right)  -E\left(  X\right)  E\left(  Y\right)  $.

\begin{proposition}
[3.3]$\sum_{\left(  i,j\right)  \neq\left(  i^{\prime},j^{\prime}\right)
}p\left(  x_{i},y_{j}\right)  p\left(  x_{i^{\prime}},y_{j^{\prime}}\right)
\left(  x_{i}-x_{i^{\prime}}\right)  \left(  y_{j}-y_{j^{\prime}}\right)
=2Cov\left(  X,Y\right)  $.
\end{proposition}

\noindent Proof: Since $\left(  x_{i}-x_{i^{\prime}}\right)  \left(
y_{j}-y_{j^{\prime}}\right)  =0$ if $i=i^{\prime}$ or $j=j^{\prime}$, we can
sum over all $i,j$. Abbreviating $p\left(  x_{i},y_{j}\right)  =p_{ij}$, we
have:%
\begin{align}
&  \sum_{i,j,i^{\prime},j^{\prime}}p_{ij}p_{i^{\prime}j^{\prime}}\left(
x_{i}-x_{i^{\prime}}\right)  \left(  y_{j}-y_{j^{\prime}}\right) \nonumber\\
&  =\sum_{i,j,i^{\prime},j^{\prime}}p_{ij}p_{i^{\prime}j^{\prime}}\left[
x_{i}y_{j}-x_{i}y_{j^{\prime}}-x_{i^{\prime}}y_{j}+x_{i^{\prime}}y_{j^{\prime
}}\right] \nonumber\\
&  =\sum_{i,j,i^{\prime},j^{\prime}}p_{ij}p_{i^{\prime}j^{\prime}}x_{i}%
y_{j}-\sum_{i,j,i^{\prime},j^{\prime}}p_{ij}p_{i^{\prime}j^{\prime}}%
x_{i}y_{j^{\prime}}-\sum_{i,j,i^{\prime},j^{\prime}}p_{ij}p_{i^{\prime
}j^{\prime}}x_{i^{\prime}}y_{j}+\sum_{i,j,i^{\prime},j^{\prime}}%
p_{ij}p_{i^{\prime}j^{\prime}}x_{i^{\prime}}y_{j^{\prime}}. \tag{3.49}%
\end{align}

\noindent Then using:%
\begin{equation}
\sum_{i,j,i^{\prime},j^{\prime}}p_{ij}p_{i^{\prime}j^{\prime}}x_{i}y_{j}%
=\sum_{i,j}p_{ij}x_{i}y_{j}\sum_{i^{\prime},j^{\prime}}p_{i^{\prime}j^{\prime
}}=\sum_{ij}p_{ij}x_{i}y_{j}=E\left(  XY\right)  , \tag{3.50}%
\end{equation}
and%
\begin{align}
\sum_{i,j,i^{\prime},j^{\prime}}p_{ij}p_{i^{\prime}j^{\prime}}x_{i}%
y_{j^{\prime}}  &  =\sum_{i,j^{\prime}}x_{i}y_{^{j^{\prime}}}\sum_{i^{\prime}%
}\sum_{j}p_{ij}p_{i^{\prime}j^{\prime}}=\sum_{i,j^{\prime}}x_{i}%
y_{^{j^{\prime}}}\sum_{i^{\prime}}p_{i}p_{i^{\prime}j^{\prime}}\nonumber\\
&  =\sum_{i,j/}p_{i}x_{i}y_{j^{\prime}}p_{j^{\prime}}=\left(  \sum_{i}%
p_{i}x_{i}\right)  \left(  \sum_{j^{\prime}}p_{j^{\prime}}y_{j^{\prime}%
}\right)  =E\left(  X\right)  E\left(  Y\right)  \tag{3.51}%
\end{align}

\begin{center}

\end{center}

\noindent and similarly for the other cases, so we have:%
\begin{align}
&  \sum_{\left(  i,j\right)  \neq\left(  i^{\prime},j^{\prime}\right)
}p\left(  x_{i},y_{j}\right)  p\left(  x_{i^{\prime}},y_{j^{\prime}}\right)
\left(  x_{i}-x_{i^{\prime}}\right)  \left(  y_{j}-y_{j^{\prime}}\right)
\nonumber\\
&  =E\left(  XY\right)  -E\left(  X\right)  E\left(  Y\right)  -E\left(
Y\right)  E\left(  X\right)  +E\left(  XY\right)  =2Cov(X,Y).\square\tag{3.52}%
\end{align}

\begin{center}

\end{center}

\noindent The linear ordering on indices $i$ and $j$ can be extended to the
linear lexicographic (or dictionary) ordering on ordered pairs of indices
where $\left(  i,j\right)  <\left(  i^{\prime},j^{\prime}\right)  $ if
$i<i^{\prime}$ or if $i=i^{\prime}$, then $j<j^{\prime}$. Then for each pair
of distinct ordered pairs $\left(  i,j\right)  \neq\left(  i^{\prime
},j^{\prime}\right)  $, either $\left(  i,j\right)  <\left(  i^{\prime
},j^{\prime}\right)  $ or $\left(  i^{\prime},j^{\prime}\right)  <\left(
i,j\right)  $ but not both, so $\left(  i,j\right)  <\left(  i^{\prime
},j^{\prime}\right)  $ picks out half the cases of $\left(  i,j\right)
\neq\left(  i^{\prime},j^{\prime}\right)  $.

\begin{corollary}
[3.3]$\sum_{\left(  i,j\right)  <\left(  i^{\prime},j^{\prime}\right)
}p\left(  x_{i},y_{i}\right)  p\left(  x_{i^{\prime}},y_{j^{\prime}}\right)
\left(  x_{i}-x_{i^{\prime}}\right)  \left(  y_{j}-y_{j^{\prime}}\right)
=Cov\left(  X,Y\right)  $. $\square$
\end{corollary}

\noindent In the switch from logical entropy to metrical logical entropy, the
interpretation switches from being a two-draw probability (and thus always
non-negative) to being a two-draw average metrical quantity which, like the
covariance, might be positive or negative.

Thus logical entropy connects naturally with the notions of variance and
covariance in statistics. Although beyond the scope of this paper, the
metrical logical entropy for a discrete random variable $g\left(  X\right)  $,
$\sum_{j\neq i}p_{i}p_{j}\left(  g\left(  x_{i}\right)  -g\left(
x_{j}\right)  \right)  ^{2}=2Var\left(  g\left(  X\right)  \right)  $, shows
how to generalize to the logical entropy of a continuous random variable
$g\left(  X\right)  $ where $X$ has the probability density $f\left(
x\right)  $:%
\begin{equation}
h\left(  g\left(  X\right)  \right)  =\int\int f\left(  x\right)  f\left(
x^{\prime}\right)  \left(  g\left(  x\right)  -g\left(  x^{\prime}\right)
\right)  ^{2}dx^{\prime}dx=2Var\left(  g\left(  X\right)  \right)  .
\tag{3.53}%
\end{equation}

\noindent The interpretation of $h\left(  g\left(  X\right)  \right)  $ is the
average (Euclidean) distance squared between the values of $g\left(  X\right)
$ on two independent trials--which is twice the variance.

\section{Quantum logical entropy}

\subsection{Logical entropy via density matrices}

The transition from `classical' (i.e., non-quantum) logical entropy to quantum
logical entropy is facilitated by reformulating logical entropy using density
matrices over the real numbers. A stepping stone in that reformulation is the
notion of an incidence matrix of a binary relation. For a finite $U=\left\{
u_{1},...,u_{n}\right\}  $, a binary relation $R$ on $U$ is a subset
$R\subseteq U\times U$. The $n\times n$ incidence matrix $In\left(  R\right)
$ is defined by:
\begin{equation}
In\left(  R\right)  _{ij}=\left\{
\begin{array}
[c]{c}%
1\text{ if }\left(  u_{i},u_{j}\right)  \in R\\
0\text{ if }\left(  u_{i},u_{j}\right)  \notin R\text{.}%
\end{array}
\right.  . \tag{4.1}%
\end{equation}

\noindent Then the incidence matrix associated with a partition $\pi=\left\{
B_{1},...,B_{m}\right\}  $ is $In\left(  \operatorname{indit}\left(
\pi\right)  \right)  $, the incidence matrix of the partition's inditset,
i.e., the associated equivalence relation. And then for equiprobable points in
$U$, the density matrix $\rho\left(  \pi\right)  $ associated with $\pi$ is
the incidence matrix $In\left(  \operatorname{indit}\left(  \pi\right)
\right)  $ rescaled to be of trace $1$ (trace = sum of diagonal elements):%
\begin{equation}
\rho\left(  \pi\right)  =\frac{1}{n}In\left(  \operatorname{indit}\left(
\pi\right)  \right)  . \tag{4.2}%
\end{equation}

\noindent Each off-diagonal element has two associated diagonal elements in
its row and column. If an off-diagonal element in $In\left(
\operatorname{indit}\left(  \pi\right)  \right)  $ or $\rho\left(  \pi\right)
$ is non-zero, then the corresponding diagonal elements are for elements
$u_{i},u_{j}\in B_{k}$ for some block $B_{k}\in\pi$.

For $U$ with point probabilities $p=\left(  p_{1},...,p_{n}\right)  $, the
density matrix $\rho\left(  \pi\right)  $ can be constructed block by block.
For a block $B_{i}\in\pi$, let $\left\vert B_{i}\right\rangle $ be the column
vector with the $j^{th}$ entry being $\sqrt{\frac{p_{j}}{\Pr\left(
B_{i}\right)  }}$ if $u_{j}\in B_{i}$ and otherwise $0$. Then the $\rho\left(
B_{i}\right)  $ is the $n\times n$ matrix formed by the product of column
vector $\left\vert B_{i}\right\rangle $ times its row vector transpose
$\left\vert B_{i}\right\rangle ^{t}$, and the \textit{density matrix}
$\rho\left(  \pi\right)  $ is the probability-weighted sum:%
\begin{equation}
\rho\left(  \pi\right)  =\sum_{B_{i}\in\pi}\Pr\left(  B_{i}\right)
\rho\left(  B_{i}\right)  . \tag{4.3}%
\end{equation}

\noindent Then each $jk$ entry in $\rho\left(  \pi\right)  $ is:%
\begin{equation}
\rho\left(  \pi\right)  _{jk}=\left\{
\begin{array}
[c]{c}%
\sqrt{p_{j}p_{k}}\text{ if }\left(  u_{j},u_{k}\right)  \in
\operatorname{indit}\left(  \pi\right) \\
0\text{ otherwise.}%
\end{array}
\right.  . \tag{4.4}%
\end{equation}

\noindent These values are the square roots of the unshaded squares in the
logical entropy box diagrams, e.g., figures 2-5.

For instance, if $\pi=\left\{  \left\{  u_{1},u_{3}\right\}  ,\left\{
u_{2},u_{4}\right\}  \right\}  $, then:%
\begin{equation}
\rho\left(  \pi\right)  =%
\begin{bmatrix}
p_{1} & 0 & \sqrt{p_{1}p_{3}} & 0\\
0 & p_{2} & 0 & \sqrt{p_{2}p_{4}}\\
\sqrt{p_{3}p_{1}} & 0 & p_{3} & 0\\
0 & \sqrt{p_{4}p_{2}} & 0 & p_{4}%
\end{bmatrix}
\tag{4.5}%
\end{equation}

\noindent where the non-zero off-diagonal elements indicate which elements are
in the same block of the partition. With a suitable interchange of rows and
columns, the matrix would become block-diagonal--where the entries squared
correspond to the values of the unshaded squares in the box diagram for
logical entropy. The density matrix is symmetric, has trace $1$, and all
non-negative elements.

The most important calculation for our purposes is the trace of the square
$\rho\left(  \pi\right)  ^{2}$ of a density matrix. Consider a diagonal
element $\left(  \rho\left(  \pi\right)  ^{2}\right)  _{jj}$ where $u_{j}\in
B_{i}$ which is the product of the $j^{th}$ row times the $j^{th}$ column of
$\rho\left(  \pi\right)  $:%
\begin{equation}
\left(  \rho\left(  \pi\right)  ^{2}\right)  _{jj}=\sum_{u_{k}\in B_{i}}%
\sqrt{p_{j}p_{k}}\sqrt{p_{k}p_{j}}=p_{j}\sum_{u_{k}\in B_{i}}p_{k}=p_{j}%
\Pr\left(  B_{i}\right)  . \tag{4.6}%
\end{equation}

\noindent Then summing over all those diagonal elements for $u_{j}\in B_{i}$
gives $\sum_{u_{j}\in B_{i}}p_{j}\Pr\left(  B_{i}\right)  =\Pr\left(
B_{i}\right)  ^{2}$. These block probabilities squared were the values
assigned to the unshaded blocks in the box diagrams for logical entropy.
Finally summing over all the diagonal elements yields the basic result about
the trace of $\rho\left(  \pi\right)  ^{2}$:%
\begin{equation}
\operatorname{tr}\left[  \rho\left(  \pi\right)  ^{2}\right]  =\sum_{B_{i}%
\in\pi}\Pr\left(  B_{i}\right)  ^{2}. \tag{4.7}%
\end{equation}

\noindent This result immediately yields the translation of the logical
entropy $h\left(  \pi\right)  $ into the density matrix formalism:%
\begin{equation}
h\left(  \pi\right)  =1-\operatorname{tr}\left[  \rho\left(  \pi\right)
^{2}\right]  \tag{4.8}%
\end{equation}

\noindent i.e., the sum of the shaded squares in the box diagrams for logical entropy.

We will define the tensor product of matrices by considering the example of a
$2\times2$ matrix $A$ times a $3\times3$ matrix $B$:%
\begin{align}
A\otimes B  &  =%
\begin{bmatrix}
a_{11} & a_{12}\\
a_{21} & a_{22}%
\end{bmatrix}
\otimes%
\begin{bmatrix}
b_{11} & b_{12} & b_{13}\\
b_{21} & b_{22} & b_{23}\\
b_{31} & b_{32} & b_{33}%
\end{bmatrix}
=%
\begin{bmatrix}
a_{11}B & a_{12}B\\
a_{21}B & a_{22}B
\end{bmatrix}
\nonumber\\
&  =%
\begin{array}
[c]{c}%
\left(  1,1\right) \\
\left(  1,2\right) \\
\left(  1,3\right) \\
\left(  2,1\right) \\
\left(  2,2\right) \\
\left(  2,3\right)
\end{array}%
\begin{bmatrix}
a_{11}b_{11} & a_{11}b_{12} & a_{11}b_{13} & a_{12}b_{11} & a_{12}b_{12} &
a_{12}b_{13}\\
a_{11}b_{21} & a_{11}b_{22} & a_{11}b_{23} & a_{12}b_{21} & a_{12}b_{22} &
a_{12}b_{23}\\
a_{11}b_{31} & a_{11}b_{32} & a_{11}b_{33} & a_{12}b_{31} & a_{12}b_{32} &
a_{12}b_{33}\\
a_{21}b_{11} & a_{21}b_{12} & a_{21}b_{13} & a_{22}b_{11} & a_{22}b_{12} &
a_{22}b_{13}\\
a_{21}b_{21} & a_{21}b_{22} & a_{21}b_{23} & a_{22}b_{21} & a_{22}b_{22} &
a_{22}b_{23}\\
a_{21}b_{31} & a_{21}b_{32} & a_{21}b_{33} & a_{22}b_{31} & a_{22}b_{32} &
a_{22}b_{33}%
\end{bmatrix}
. \tag{4.9}%
\end{align}

\noindent In particular, it might be noted that all the diagonal elements have
the form $a_{ii}b_{jj}$ but their (row,column) designators are $\left(
i,j\right)  \left(  i,j\right)  $. Thus $a_{11}b_{33}$ is the diagonal element
in the $\left(  1,3\right)  $ row and the $\left(  1,3\right)  $ column, i.e.,
a diagonal element of the tensor product $A\otimes B$.

One can take the tensor product of an $n\times n$ density matrix $\rho\left(
\pi\right)  $ (where $\pi=f^{-1}$ for some $f:U\rightarrow%
\mathbb{R}
$) with itself to obtain a $n^{2}\times n^{2}$ matrix whose diagonal elements
are $\left(  \rho\left(  \pi\right)  \otimes\rho\left(  \pi\right)  \right)
_{\left(  i,j\right)  \left(  i,j\right)  }=p_{i}p_{j}$. Let
$P_{\operatorname{dit}\left(  \pi\right)  }$ be the $n^{2}\times n^{2}$
diagonal (projection) matrix with diagonal elements $\left(
P_{\operatorname{dit}\left(  \pi\right)  }\right)  _{\left(  i,j\right)
\left(  i,j\right)  }=\chi_{\operatorname{dit}\left(  \pi\right)  }\left(
u_{i},u_{j}\right)  $. Then the matrix product $P_{\operatorname{dit}\left(
\pi\right)  }\rho\left(  \pi\right)  \otimes\rho\left(  \pi\right)  $ will
have the non-zero diagonal elements $p_{i}p_{j}$ for $\left(  u_{i}%
,u_{j}\right)  \in\operatorname{dit}\left(  \pi\right)  $, and thus:%
\begin{equation}
h\left(  f^{-1}\right)  =h\left(  \pi\right)  =\sum_{\left(  u_{i}%
,u_{j}\right)  \in\operatorname{dit}\left(  \pi\right)  }p_{i}p_{j}%
=\operatorname{tr}\left[  P_{\operatorname{dit}\left(  \pi\right)  }%
\rho\left(  \pi\right)  \otimes\rho\left(  \pi\right)  \right]  . \tag{4.10}%
\end{equation}

\noindent That formula will carry over to the quantum case.

In general, a density matrix $\rho$ is said to represent a \textit{pure state}
if $\operatorname{tr}\left[  \rho^{2}\right]  =1$, and otherwise a
\textit{mixed state}. For partitions, the only pure state density matrix is
$\rho\left(  \mathbf{0}_{U}\right)  $, the density matrix of the indiscrete
partition $\mathbf{0}_{U}=\left\{  U\right\}  $ on $U$ which has zero logical entropy.

Given another partition $\sigma=\left\{  C_{1},...,C_{m^{\prime}}\right\}  $
on $U$, the join partition $\pi\vee\sigma$ is the partition whose blocks are
all the non-empty intersections $B_{i}\cap C_{j}$ for $B_{i}\in\pi$ and
$C_{j}\in\sigma$. Then $\operatorname{dit}\left(  \pi\vee\sigma\right)
=\operatorname{dit}\left(  \pi\right)  \cup\operatorname{dit}\left(
\sigma\right)  $ so that $\operatorname{indit}\left(  \pi\vee\sigma\right)
=\operatorname{indit}\left(  \pi\right)  \cap\operatorname{indit}\left(
\sigma\right)  $. The logical entropy $h\left(  \pi\vee\sigma\right)  $, also
the joint logical entropy $h\left(  \pi,\sigma\right)  $, is: $h\left(
\pi\vee\sigma\right)  =1-\sum_{B_{i}\in\pi,C_{j}\in\sigma}\Pr\left(  B_{i}\cap
C_{j}\right)  ^{2}$. This has an elegant formulation in the density matrix
formalism which implies the earlier result since $\pi\vee\pi=\pi$.

\begin{lemma}
[4.1]$h\left(  \pi\vee\sigma\right)  =1-\operatorname{tr}\left[  \rho\left(
\pi\right)  \rho\left(  \sigma\right)  \right]  $.
\end{lemma}

\noindent\textbf{Proof}: The $k^{th}$ diagonal entry in $\rho\left(
\pi\right)  \rho\left(  \sigma\right)  $ is the scalar product $\sum_{j}%
\rho\left(  \pi\right)  _{kj}\rho\left(  \sigma\right)  _{jk}$ with
$\rho\left(  \pi\right)  _{kj}=\sqrt{p_{k}p_{j}}$ if \text{, }$\left(
u_{j},u_{k}\right)  \in\operatorname{indit}\left(  \pi\right)  $ and otherwise
$0$, and similarly for $\rho\left(  \sigma\right)  _{jk}$. Hence the only
non-zero terms in that sum are for $\left(  u_{k},u_{j}\right)  \in
\operatorname{indit}\left(  \pi\right)  \cap\operatorname{indit}\left(
\sigma\right)  =\operatorname{indit}\left(  \pi\vee\sigma\right)  $. Hence%
\begin{equation}
\operatorname{tr}\left[  \rho\left(  \pi\right)  \rho\left(  \sigma\right)
\right]  =\sum_{\left(  u_{j},u_{k}\right)  \in\operatorname{indit}\left(
\pi\vee\sigma\right)  }p_{j}p_{k}=1-\sum_{\left(  u_{j},u_{k}\right)
\in\operatorname{dit}\left(  \pi\vee\sigma\right)  }p_{j}p_{k}=1-h\left(
\pi\vee\sigma\right)  \tag{4.11}%
\end{equation}
so%
\begin{equation}
h\left(  \pi\vee\sigma\right)  =1-\operatorname{tr}\left[  \rho\left(
\pi\right)  \rho\left(  \sigma\right)  \right]  .\square\tag{4.12}%
\end{equation}

In coding theory, the difference-based notion of distance between two $0,1$
$n$-vectors is the \textit{Hamming distance} \cite[p. 66]{mceliece:info} which
is just the number of places where the corresponding entries in the two
vectors are different. If we think of the $0,1$ $n$-vectors as characteristic
functions of subsets $S$ and $T$ of an $n$-element set, then the Hamming
distance is the cardinality of the symmetric difference: $\left\vert
S-T\right\vert +\left\vert T-S\right\vert =\left\vert S\cup T\right\vert
-\left\vert S\cap T\right\vert $. This motivates the definition of the
\textit{logical distance} (or \textit{Hamming distance}) between two
partitions as: $h\left(  \pi|\sigma\right)  +h\left(  \sigma|\pi\right)
=h\left(  \pi\vee\sigma\right)  -m\left(  \pi,\sigma\right)  $, the product
probability measure on the dits that in one partition but not the other. But
there is the \textit{Hilbert-Schmidt distance measure}, $\operatorname{tr}%
\left[  \left(  \rho-\tau\right)  ^{2}\right]  $ between density matrices
$\rho$ and $\tau$ which does not mention logical entropy at all (see
\cite{tamir-cohen:holevo}). Taking the two density matrices as $\rho\left(
\pi\right)  $ and $\rho\left(  \sigma\right)  $, we have the following result
that the logical (Hamming) distance between partitions is the Hilbert-Schmidt
distance between the partitions.

\begin{proposition}
[4.1]$\operatorname{tr}\left[  \left(  \rho\left(  \pi\right)  -\rho\left(
\sigma\right)  \right)  ^{2}\right]  =h\left(  \pi|\sigma\right)  +h\left(
\sigma|\pi\right)  $.
\end{proposition}

\noindent\textbf{Proof}: $\operatorname{tr}\left[  \left(  \rho\left(
\pi\right)  -\rho\left(  \sigma\right)  \right)  ^{2}\right]
=\operatorname{tr}\left[  \rho\left(  \pi\right)  ^{2}\right]
-\operatorname{tr}\left[  \rho\left(  \pi\right)  \rho\left(  \sigma\right)
\right]  -\operatorname{tr}\left[  \rho\left(  \sigma\right)  \rho\left(
\pi\right)  \right]  +\operatorname{tr}\left[  \rho\left(  \sigma\right)
^{2}\right]  $ so:%
\begin{align}
\operatorname{tr}\left[  \left(  \rho\left(  \pi\right)  -\rho\left(
\sigma\right)  \right)  ^{2}\right]   &  =2\left[  1-\operatorname{tr}\left[
\rho\left(  \pi\right)  \rho\left(  \sigma\right)  \right]  \right]  -\left(
1-\operatorname{tr}\left[  \rho\left(  \pi\right)  ^{2}\right]  -\left(
1-\operatorname{tr}\left[  \rho\left(  \sigma\right)  ^{2}\right]  \right)
\right) \nonumber\\
&  =2h\left(  \pi\vee\sigma\right)  -h\left(  \pi\right)  -h\left(
\sigma\right)  =h\left(  \sigma|\pi\right)  +h\left(  \pi|\sigma\right)
.\square\tag{4.13}%
\end{align}

\noindent One point of developing these results in this classical case, is
that the same theorems hold, \textit{mutatis mutandis}, for quantum logical
entropy \cite{ell:quantum-entropy}.

Another set of classical results about logical entropy that extend to the
quantum case are concerned with the quantum notion of (projective) measurement
which is described by the L\"{u}ders mixture operation \cite[p. 279]%
{auletta:qm}. For the partition $\sigma=\left\{  C_{1},...,C_{m^{\prime}%
}\right\}  $ on $U$, let $P_{C}$ be the diagonal $n\times n$ projection matrix
whose diagonal entries are just the characteristic function $\chi_{C}(u_{i})$
for $C\in\sigma$. Then the \textit{L\"{u}ders mixture operation of performing
a `}$\sigma$\textit{-measurement'} on $\rho\left(  \pi\right)  $ is defined
as: $\sum_{C\in\sigma}P_{C}\rho\left(  \pi\right)  P_{C}$.

\begin{theorem}
[4.1 L\"{u}ders mixture operation = partition join operation]$\sum_{C\in
\sigma}P_{C}\rho\left(  \pi\right)  P_{C}=\rho\left(  \pi\vee\sigma\right)  $.
\end{theorem}

\noindent Proof: A non-zero entry in $\rho\left(  \pi\right)  $ has the form
$\rho\left(  \pi\right)  _{jk}=\sqrt{p_{j}p_{k}}$ iff there is some block
$B\in\pi$ such that $\left(  u_{j},u_{k}\right)  \in B\times B$, i.e., if
$u_{j},u_{k}\in B$. The matrix operation $P_{C}\rho\left(  \pi\right)  $ will
preserve the entry $\sqrt{p_{j}p_{k}}$ if $u_{j}\in C$, otherwise the entry is
zeroed. And if the entry was preserved, then the further matrix operation
$\left(  P_{C}\rho\left(  \pi\right)  \right)  P_{C}$ will preserve the entry
$\sqrt{p_{j}p_{k}}$ if $u_{k}\in C$, otherwise it is zeroed. Hence the entries
$\sqrt{p_{j}p_{k}}$ in $\rho\left(  \pi\right)  $ that are preserved in
$P_{C}\rho\left(  \pi\right)  P_{C}$ are the entries where both $u_{j}%
,u_{k}\in B$ for some $B\in\pi$ and $u_{j},u_{k}\in C$. These are the entries
in $\rho\left(  \pi\vee\sigma\right)  $ corresponding to the non-empty blocks
$B\cap C$ of $\pi\vee\sigma$ for some $B\in\pi$, so summing over $C\in\sigma$
gives the result: $\sum_{C\in\sigma}P_{C}\rho\left(  \pi\right)  P_{C}%
=\rho\left(  \pi\vee\sigma\right)  $. $\square$

Thus projective quantum measurement is modeled classically by the
distinction-creating partition join. Hence the logical information
\textit{created} by the $\sigma$-measurement of $\rho\left(  \pi\right)  $ is
$h\left(  \sigma\vee\pi\right)  -h\left(  \pi\right)  =h\left(  \sigma
|\pi\right)  $. Moreover, this increase in logical entropy can be computed
from the changes in the entries in the density matrices. A non-zero
off-diagonal entry in a density matrix $\rho\left(  \pi\right)  $ indicates
that the $u_{j}$ and $u_{k}$ for the corresponding diagonal elements must
`cohere' together in the same block of $\pi$. If such a non-zero off-diagonal
element of $\rho\left(  \pi\right)  $ was zeroed in the transition to the
density matrix $\rho\left(  \pi\vee\sigma\right)  $ of the $\sigma
$-measurement result, then it means that $u_{j}$ and $u_{k}$ were `decohered'
by $\sigma$, i.e., were in different blocks of $\sigma$.

\begin{corollary}
[4.1]The sum of all the squares $p_{j}p_{k}$ of all the non-zero off-diagonal
entries $\sqrt{p_{j}p_{j}}$ of $\rho\left(  \pi\right)  $ that were zeroed in
the L\"{u}ders mixture operation that transforms $\rho\left(  \pi\right)  $
into $\sum_{C\in\sigma}P_{C}\rho\left(  \pi\right)  P_{C}=\rho\left(  \pi
\vee\sigma\right)  $ is $h\left(  \pi\vee\sigma\right)  -h\left(  \pi\right)
=h\left(  \sigma|\pi\right)  $.
\end{corollary}

\noindent Proof: All the entries $\sqrt{p_{j}p_{j}}$ that got zeroed were for
ordered pair $\left(  u_{j},u_{k}\right)  $ that were indits of $\pi$ but not
indits of $\pi\vee\sigma$, i.e., $\left(  u_{j},u_{k}\right)  \in
\operatorname{indit}\left(  \pi\right)  \cap\operatorname{indit}\left(
\pi\vee\sigma\right)  ^{c}=\operatorname{dit}\left(  \pi\right)  ^{c}%
\cap\operatorname{dit}\left(  \pi\vee\sigma\right)  =\operatorname{dit}\left(
\pi\vee\sigma\right)  -\operatorname{dit}\left(  \pi\right)  $. The sum of
products $p_{j}p_{k}$ for those pairs $\left(  u_{j},u_{k}\right)  $ is just
the product probability measure on that set $\operatorname{dit}\left(  \pi
\vee\sigma\right)  -\operatorname{dit}\left(  \pi\right)  $ which is $h\left(
\pi\vee\sigma|\pi\right)  $. And since $\operatorname{dit}\left(  \pi\right)
\subseteq\operatorname{dit}\left(  \pi\vee\sigma\right)  $, the measure on
$\operatorname{dit}\left(  \pi\vee\sigma\right)  -\operatorname{dit}\left(
\pi\right)  $ is $h\left(  \pi\vee\sigma|\pi\right)  =h\left(  \pi\vee
\sigma\right)  -h\left(  \pi\right)  =h\left(  \sigma|\pi\right)  $. $\square$

\noindent It might be noted that nothing about logical entropy was used in the
definition of the L\"{u}ders mixture operation that describes the "$\sigma
$-measurement of $\rho\left(  \pi\right)  $." Yet the logical information
created by the $\sigma$-measurement of $\rho\left(  \pi\right)  $ is $h\left(
\sigma|\pi\right)  $, the logical information that is in $\sigma$ over and
above the information in $\pi$. And that logical entropy $h\left(  \sigma
|\pi\right)  $ can be computed directly from the terms in the density matrix
$\rho\left(  \pi\right)  $ that were zeroed in the L\"{u}ders operation.

Now we are ready to make the transition to quantum logical information theory
where all the corresponding results will hold.

\subsection{Linearizing `classical' to quantum logical entropy}

One of the developers of quantum information theory, Charles Bennett, said
that information was fundamentally about distinguishability.

\begin{quotation}
\noindent\lbrack Information] is the notion of distinguishability abstracted
away from what we are distinguishing, or from the carrier of information.
\ldots And we ought to develop a theory of information which generalizes the
theory of distinguishability to include these quantum properties\ldots
\ \cite[pp. 155-157]{bennett:qinfo}\ 
\end{quotation}

\noindent Given a normalized vector $\left\vert \psi\right\rangle $ in an
$n$-dimensional Hilbert space $V$, a pure state density matrix is formed as
$\rho\left(  \psi\right)  =\left\vert \psi\right\rangle \left\langle
\psi\right\vert =\left\vert \psi\right\rangle \left(  \left\vert
\psi\right\rangle \right)  ^{\dagger}$ (where $\left(  {}\right)  ^{\dagger}$
is the conjugate-transpose) and mixed state density matrix is a probability
mixture $\rho=\sum_{i}p_{i}\rho\left(  \psi_{i}\right)  $ of pure state
density matrices. Any such density matrix always has a spectral decomposition
into the form $\rho=\sum p_{i}\rho\left(  \psi_{i}\right)  $ where the
different vectors $\psi_{i}$ and $\psi_{i^{\prime}}$ are orthogonal. The
general definition of the \textit{quantum logical entropy of a density matrix}
is: $h\left(  \rho\right)  =1-\operatorname{tr}\left[  \rho^{2}\right]  $,
where if $\rho$ is a pure state if and only if $\operatorname{tr}\left[
\rho^{2}\right]  =1$ so $h\left(  \rho\right)  =0$, and $\operatorname{tr}%
[\rho^{2}]<1$ for mixed states so $1>h\left(  p\right)  >0$ for mixed states.
The formula $h\left(  \rho\right)  =1-\operatorname*{tr}\left[  \rho
^{2}\right]  $ is hardly new. Indeed, $\operatorname*{tr}\left[  \rho
^{2}\right]  $ is usually called the \textit{purity} of the density matrix so
the complement $1-\operatorname*{tr}\left[  \rho^{2}\right]  $ has been called
the \textquotedblleft mixedness\textquotedblright\ \cite[p. 5]{jaeger:q-info}
or \textquotedblleft impurity\textquotedblright\ of the state $\rho$. The
seminal paper of Manfredi and Feix \cite{manfredi:wigner} approaches the same
formula $1-\operatorname*{tr}\left[  \rho^{2}\right]  $ (which they denote as
$S_{2}$) from the advanced viewpoint of Wigner functions, and they present
strong arguments for this notion of quantum entropy.

While that definition is an easy generalization of the classical one
formulated using density matrices, our goal is to develop quantum logical
entropy in a manner that brings out the analogy with classical logical entropy
and relates it closely to quantum measurement.

There is a method (or what Gian-Carlo Rota would call a "yoga") to
\textit{linearize} set concepts to vector-space concepts:

\begin{center}
\textbf{The Yoga of Linearization}:

Apply the set concept to a basis-set of a vector space

(i.e., treat the basis set as a set universe $U$) and

whatever is generated is the corresponding vector-space concept.
\end{center}

\noindent For instance, there is the classical Boolean logic of subsets, and a
subset of a basis set generates a subspace so the Boolean logic of subsets
linearizes to vector spaces as the logic of subspaces, and specializing to
Hilbert spaces yields the usual quantum logic of subspaces
\cite{birk-vonn:logic-of-qm}.

In view of the category-theoretic duality of subsets of a set and partitions
on a set (e.g., the image subset of the codomain and the inverse-image
partition on the domain of a set function), there is a dual `classical' logic
of partitions on a set (\cite{ell:lop}, \cite{ell:intropartitions}). To
linearize the set concept of a partition to vector spaces, one considers a set
partition on a basis set of a vector space and then sees what it generates.
Each block generates a subspace and the set of subspaces corresponding to the
blocks form a direct-sum decomposition (DSD) of the vector space. A
\textit{direct-sum decomposition} of a vector space $V$ is a set $\left\{
V_{i}\right\}  _{i\in I}$ of subspaces such that $V_{i}\cap\sum_{i^{\prime
}\neq i}V_{i}=\left\{  0\right\}  $ (where $\sum_{i^{\prime}\neq i}V_{i}$ is
the subspace generated by those $V_{i^{\prime}}$, and $\left\{  0\right\}  $
is the zero subspace), and which span the space $V$ and is written
$V=\oplus_{i\in I}V_{i}$. Then every non-zero vector $v\in V$ is a unique sum
of vectors from the subspaces $\left\{  V_{i}\right\}  $. That is the
vector-space version of characterizing a partition $\pi$ on a set $U$ as a
collection of subsets $B_{i}$ (blocks) of $U$ such that every non-empty subset
$S\subseteq U$ is uniquely expressed as a union of subsets of the blocks,
i.e., $S\cap B_{i}$ for $B_{i}\in\pi$. Hence the logic of partitions
linearizes to the dual logic of DSDs of a vector space which specialized to
Hilbert spaces yields the quantum logic of DSDs \cite{ell:dsd-logic} dual to
the usual Birkhoff-von-Neumann quantum logic of subspaces.

Another basic set concept is the notion of a numerical attribute
$f:U\rightarrow\mathbb{K}$ that evaluates the points of $U$ in a field
$\mathbb{K}$. Taking $U$ to be a basis set of a vector space $V$ over the
field $\mathbb{K}$, the corresponding vector-space notion that can be seen as
generated is the notion of a \textit{diagonalizable linear operator}
$F:V\rightarrow V$ defined by $Fu=f\left(  u\right)  u$ linearly extended to
$V$. The values of $f$ linearize to the eigenvalues of $F$, the constant sets
of $f$ linearize to the eigenvectors of $F$, and the set of constant sets for
a specific value linearizes to the eigenspace of eigenvectors for that
eigenvalue. For instance, if we let "$rS$" stand for assigning the value $r$
to each element of the subset $S\subseteq U$, then the set version of the
eigenvector equation $Fv=\lambda v$ is $f\left(  S\right)  =rS$.

The Cartesian product of two basis sets of two vector spaces (same base field)
generates the tensor product of the two vector spaces--so the linearization of
the direct or Cartesian product of sets is not the direct product (as might be
suggested by category theory) but the tensor product of vector spaces. And the
cardinality of sets linearizes to the dimension of vector spaces and so forth
as illustrated in Table 4. Those examples show how the set-based classical
logical information theory will linearize to vector spaces and particularly to
Hilbert spaces for the logical version of quantum information
theory.\footnote{Since set-concepts can be formulated in vector spaces over $%
\mathbb{Z}
_{2}$, that means there is a pedagogical or `toy' model of quantum mechanics
over $%
\mathbb{Z}
_{2}$, i.e., over sets \cite{ell:qm-over-sets}.}

\begin{center}%
\begin{tabular}
[c]{|c|c|}\hline
Set concept & Vector-space concept\\\hline\hline
Universe set $U$ & Basis set of a space $V$\\\hline
Cardinality of a set $U$ & Dimension of a space $V$\\\hline
Subset of a set $U$ & Subspace of a space $V$\\\hline
Partition of a set $U$ & Direct-sum decomposition of a space $V$\\\hline
Numerical attribute $f:U\rightarrow\mathbb{K}$ & Diagonalizable linear op.
$F:V\rightarrow V$\\\hline
Value $r$ in image $f\left(  U\right)  $ of $f$ & Eigenvalue $\lambda_{i}$ of
$F$\\\hline
Constant set $S$ of $f$ & Eigenvector $v$ of $F$\\\hline
Set of constant $r$-sets $\wp\left(  f^{-1}\left(  r\right)  \right)  $ &
Eigenspace $V_{i}$ of $\lambda_{i}$\\\hline
Direct product of sets & Tensor product of spaces\\\hline
Elements $\left(  u_{k},u_{k^{\prime}}\right)  $ of $U\times U$ & Basis
vectors $u_{k}\otimes u_{k^{\prime}}$ of $V\otimes V$\\\hline
\end{tabular}

Table 4: Linearization of set concepts to corresponding vector-space concepts.
\end{center}

Let $F:V\rightarrow V$ be a self-adjoint (or Hermitian) operator (observable)
on a $n$-dimensional Hilbert space $V$ with the real eigenvalues $\phi
_{1},\dots,\phi_{I}$, and let $U=\left\{  u_{1},\dots,u_{n}\right\}  $ be an
orthonormal (ON) basis of eigenvectors of $F$. The quantum version of a "dit"
is a "qudit." A \textit{qudit }is \textit{defined by the DSD of eigenspaces of
an observable}, just as classically, a distinction or dit is defined by the
partition $\left\{  f^{-1}\left(  r\right)  \right\}  _{r\in f\left(
U\right)  }$ determined a numerical attribute $f:U\rightarrow%
\mathbb{R}
$. Then, there is a set partition $\pi=\left\{  B_{i}\right\}  _{i=1,\dots,I}$
on the ON basis $U$ so that $B_{i}$ is a basis for the eigenspace of the
eigenvalue $\phi_{i}$ and $\left\vert B_{i}\right\vert $ is the
\textquotedblleft multiplicity\textquotedblright\ (dimension of the
eigenspace) of the eigenvalue $\phi_{i}$ for $i=1,\dots,I$. The eigenspaces
$V_{i}$ generated by the blocks $B_{i}$ for the eigenvalues $\phi_{i}$ form a
direct-sum decomposition of $V$. Note that the real-valued numerical attribute
or eigenvalue function $f:U\rightarrow\mathbb{R}$ that takes each eigenvector
in $u_{j}\in B_{i}\subseteq U$ to its eigenvalue $\phi_{i}$ so that
$f^{-1}\left(  \phi_{i}\right)  =B_{i}$ contains all the information in the
self-adjoint operator $F:V\rightarrow V$ since $F$ can be reconstructed by
defining it on the basis $U$ as $Fu_{j}=f\left(  u_{j}\right)  u_{j}$. The
important information-theoretic aspect of the eigenvalues is not their
numerical value but when they are the same or different, and that information
is there in the eigenspaces $\left\{  V_{i}\right\}  _{i\in I}$ of the
direct-sum decomposition.\footnote{That is why the quantum logic of DSDs
\cite{ell:dsd-logic} is essentially the quantum logic of observables--in much
the same sense that the logic of partitions on $U$ is essentially the logic of
numerical attributes $f:U\rightarrow%
\mathbb{R}
$ on $U$.}

Classically, a \textit{dit of the partition} $\left\{  f^{-1}\left(  \phi
_{i}\right)  \right\}  _{i\in I}$ on $U$, defined by $f:U\rightarrow%
\mathbb{R}
$, is a pair $\left(  u_{k},u_{k^{\prime}}\right)  \in U\times U$ of points in
distinct blocks of the partition, i.e., $f\left(  u_{k}\right)  \neq f\left(
u_{k^{\prime}}\right)  $. Hence, a \textit{qudit of } $F$ is a pair $\left(
u_{k},u_{k^{\prime}}\right)  $ (interpreted as $u_{k}\otimes u_{k^{\prime}}\in
V\otimes V$) of vectors in the eigenbasis distinguished by $F$, i.e.,
$f\left(  u_{k}\right)  \neq f\left(  u_{k^{\prime}}\right)  $ for the
eigenvalue function $f:U\rightarrow%
\mathbb{R}
$. Let $G:V\rightarrow V$ be another self-adjoint operator on $V$, which
commutes with $F$ so that we may then assume that $U$ is an orthonormal basis
of simultaneous eigenvectors of $F$ and $G$ \cite[p. 177]{hoff-kunze:la1}. The
assumption that $F$ and $G$ commute plays the role of considering partitions
$\pi=f^{-1}$ for $f:U\rightarrow%
\mathbb{R}
$ and $\sigma=g^{-1}$ for $g:U\rightarrow%
\mathbb{R}
$ being defined on the same universe $U$. Let $\left\{  \gamma_{j}\right\}
_{j\in J}$ be the set of eigenvalues of $G$, and let $g:U\rightarrow
\mathbb{R}$ be the eigenvalue function so a pair $\left(  u_{k},u_{k^{\prime}%
}\right)  $ is a \textit{qudit of }$G$ if $g\left(  u_{k}\right)  \neq
g\left(  u_{k^{\prime}}\right)  $, i.e., if the two eigenvectors have distinct
eigenvalues of $G$.

As Kolmogorov suggested;

\begin{quotation}
\noindent Information theory must precede probability theory, and not be based
on it. By the very essence of this discipline, the foundations of information
theory have a finite combinatorial character. \cite[p. 39]{kolmogor:combfound}
\end{quotation}

In classical logical information theory, information is defined prior to
probabilities by certain subsets (e.g., ditsets and differences between and
intersections of ditsets) or, in~the quantum case, quantum information is
defined by certain subspaces prior to the introduction of any probabilities
(unlike the case with Shannon or von Neumann entropies). Since the transition
from classical to quantum logical information theory is straightforward, it
will be first presented in a table (which does not involve any probabilities),
where the qudits $\left(  u_{k},u_{k^{\prime}}\right)  $ are interpreted as
$u_{k}\otimes u_{k^{\prime}}$. The \textit{qudit space}, the vector-space
analogue of the ditset, associated with $F$ (the vector-space analogue of
$f:U\rightarrow%
\mathbb{R}
$) is the subspace $\left[  qudit\left(  F\right)  \right]  \subseteq V\otimes
V$ generated by the qudits $u_{k}\otimes u_{k^{\prime}}$ of $F$.

\begin{center}%
\begin{tabular}
[c]{|c|c|}\hline
Classical Logical Information & Quantum Logical Information\\\hline\hline
$f,g:U\rightarrow%
\mathbb{R}
$ & Commuting self-adjoint ops. $F,G$\\\hline
$U=\left\{  u_{1},...,u_{n}\right\}  $ & ON basis simultaneous eigenvectors
$F,G$\\\hline
Values $\left\{  \phi_{i}\right\}  _{i\in I}$ of $f$ & Eigenvalues $\left\{
\phi_{i}\right\}  _{i\in I}$ of $F$\\\hline
Values $\left\{  \gamma_{j}\right\}  _{j\in J}$ of $g$ & Eigenvalues $\left\{
\gamma_{j}\right\}  _{j\in J}$ of $G$\\\hline
Partition $\left\{  f^{-1}\left(  \phi_{i}\right)  \right\}  _{i\in I}$ &
Eigenspace DSD of $F$\\\hline
Partition $\left\{  g^{-1}\left(  \gamma_{j}\right)  \right\}  _{j\in J}$ &
Eigenspace DSD of $G$\\\hline
dits of $\pi:\left(  u_{k},u_{k^{\prime}}\right)  \in U^{2}$, $f\left(
u_{k}\right)  \neq f\left(  u_{k^{\prime}}\right)  $ & Qudits of $F$:
$u_{k}\otimes u_{k^{\prime}}\in V\otimes V$, $f\left(  u_{k}\right)  \neq
f\left(  u_{k^{\prime}}\right)  $\\\hline
dits of $\sigma:\left(  u_{k},u_{k^{\prime}}\right)  \in U^{2}$, $g\left(
u_{k}\right)  \neq g\left(  u_{k^{\prime}}\right)  $ & Qudits of $G$:
$u_{k}\otimes u_{k^{\prime}}\in V\otimes V$, $g\left(  u_{k}\right)  \neq
g\left(  u_{k^{\prime}}\right)  $\\\hline
$\operatorname{dit}\left(  \pi\right)  \subseteq U\times U$ & $\left[
Qudit\left(  F\right)  \right]  $ = subspace gen. by qudits of $F$\\\hline
$\operatorname{dit}\left(  \sigma\right)  \subseteq U\times U$ & $\left[
Qudit\left(  G\right)  \right]  $ = subspace gen. by qudits of $G$\\\hline
$\operatorname{dit}\left(  \pi\right)  \cup\operatorname{dit}\left(
\sigma\right)  \subseteq U\times U$ & $\left[  Qudit\left(  F\right)  \cup
Qudit\left(  G\right)  \right]  \subseteq V\otimes V$\\\hline
$\operatorname{dit}\left(  \pi\right)  -\operatorname{dit}\left(
\sigma\right)  \subseteq U\times U$ & $\left[  Qudit\left(  F\right)
-Qudit\left(  G\right)  \right]  \subseteq V\otimes V$\\\hline
$\operatorname{dit}\left(  \pi\right)  \cap\operatorname{dit}\left(
\sigma\right)  \subseteq U\times U$ & $\left[  Qudit\left(  F\right)  \cap
Qudit\left(  G\right)  \right]  \subseteq V\otimes V$\\\hline
\end{tabular}

Table 5: Ditsets and qudit subspaces without probabilities.
\end{center}

If $F=\lambda I$ is a scalar multiple of the identity $I$ (the vector-space
analogue of a constant function $f:U\rightarrow%
\mathbb{R}
$), then it has no qudits, so its qudit subspace $\left[  qudit\left(  \lambda
I\right)  \right]  $ is the zero subspace (the analogue of the empty ditset of
the indiscrete partition). The Common Dits Theorem says that any two non-empty
ditsets have a non-zero intersection. In the quantum case, this means any two
non-zero qudit spaces $\left[  qudit\left(  F\right)  \right]  $ and $\left[
qudit\left(  G\right)  \right]  $ for commuting $F$ and $G$ have a non-zero
intersection, i.e., have a non-zero mutual information space. That is, for
commuting $F$ and $G$, there are always two simultaneous eigenvectors $u_{k}$
and $u_{k^{\prime}}$ that have different eigenvalues both by $F$ and by $G$.

The observables do not provide the point probabilities in a measurement; the
probabilities come from the pure (normalized) state $\psi$ being measured. Let
$\left\vert \psi\right\rangle =\sum_{j=1}^{n}\left\langle u_{j}|\psi
\right\rangle \left\vert u_{j}\right\rangle =\sum_{j=1}^{n}\alpha
_{j}\left\vert u_{j}\right\rangle $ be the resolution of $\left\vert
\psi\right\rangle $ in terms of the orthonormal basis $U=\left\{  u_{1}%
,\dots,u_{n}\right\}  $ of simultaneous eigenvectors for $F$ and $G$. Then,
$p_{j}=\alpha_{j}\alpha_{j}^{\ast}$ ($\alpha_{j}^{\ast}$ is the complex
conjugate of $\alpha_{j}$) for $j=1,\dots,n$ are the point probabilities on
$U$, and the pure state density matrix $\rho\left(  \psi\right)  =\left\vert
\psi\right\rangle \left\langle \psi\right\vert $ (where $\left\langle
\psi\right\vert =\left\vert \psi\right\rangle ^{\dagger}$ is the
conjugate-transpose) has the entries: $\rho_{jk}\left(  \psi\right)
=\alpha_{j}\alpha_{k}^{\ast}$, so the diagonal entries $\rho_{jj}\left(
\psi\right)  =\alpha_{j}\alpha_{j}^{\ast}=p_{j}$ are the point probabilities.
Then we have the Table 5 giving the remaining parallel development with the
probabilities provided by the pure state $\psi$ where we write $\rho\left(
\psi\right)  ^{\dagger}\rho\left(  \psi\right)  $ as $\rho\left(  \psi\right)
^{2}$.

The definition of quantum logical entropy%
\begin{equation}
h\left(  F:\psi\right)  =\operatorname{tr}\left[  P_{\left[  qudit\left(
F\right)  \right]  }\rho\left(  \psi\right)  \otimes\rho\left(  \psi\right)
\right]  \tag{4.14}%
\end{equation}

\noindent is just the quantum version of the formulation of the classical
logical entropy%
\begin{equation}
h\left(  f^{-1}\right)  =h\left(  \pi\right)  =\sum_{\left(  u_{i}%
,u_{j}\right)  \in\operatorname{dit}\left(  \pi\right)  }p_{i}p_{j}%
=\operatorname{tr}\left[  P_{\operatorname{dit}\left(  \pi\right)  }%
\rho\left(  \pi\right)  \otimes\rho\left(  \pi\right)  \right]  \tag{4.15}%
\end{equation}

\noindent for $f:U\rightarrow%
\mathbb{R}
$ with the point probabilities $\left(  p_{1},...,,p_{n}\right)  $ on $U$ and
thus $p\times p$ on $U\times U$. The tensor product $\rho\left(  \psi\right)
\otimes\rho\left(  \psi\right)  $ is an $n^{2}\times n^{2}$ matrix with the
diagonal entries $\left(  \rho\left(  \psi\right)  \otimes\rho\left(
\psi\right)  \right)  _{\left(  j,k\right)  ,\left(  j,k\right)  }=\rho\left(
\psi\right)  _{jj}\rho\left(  \psi\right)  _{kk}=p_{j}p_{k}$ where
$p_{j}=\alpha_{j}\alpha_{j}^{\ast}$ for $\left\vert \psi\right\rangle
=\sum_{j=1}^{n}\left\langle u_{j}|\psi\right\rangle \left\vert u_{j}%
\right\rangle =\sum_{j=1}^{n}\alpha_{j}\left\vert u_{j}\right\rangle $ where
$U=\left\{  u_{1},...,u_{n}\right\}  $ is an ON basis of eigenvectors of the
observable $F$. The $n^{2}\times n^{2}$ diagonal projection matrix $P_{\left[
qudit\left(  F\right)  \right]  }$ has a diagonal element $\left(  P_{\left[
qudit\left(  F\right)  \right]  }\right)  _{\left(  j,k\right)  \left(
j,k\right)  }=1$ if $u_{j}\otimes u_{k}\in\left[  qudit\left(  F\right)
\right]  $. i.e., if the eigenvectors $u_{j}$ and $u_{k}$ have different
eigenvalues, and $0$ otherwise. Hence the product $P_{\left[  qudit\left(
F\right)  \right]  }\rho\left(  \psi\right)  \otimes\rho\left(  \psi\right)  $
will pick out the products $p_{j}p_{k}$ for $u_{j}\otimes u_{k}\in\left[
qudit\left(  F\right)  \right]  $ and the trace will sum them. Hence we have
the result that:%
\begin{align}
h\left(  F:\psi\right)   &  =\operatorname{tr}\left[  P_{\left[  qudit\left(
F\right)  \right]  }\rho\left(  \psi\right)  \otimes\rho\left(  \psi\right)
\right] \nonumber\\
&  =\sum_{j,k}\left\{  p_{j}p_{k}:u_{j}\otimes u_{k}\in\left[  qudit\left(
F\right)  \right]  \right\}  =\sum_{j,k}\left\{  p_{j}p_{k}:f\left(
u_{j}\right)  \neq f\left(  u_{k}\right)  \right\}  \tag{4.16}%
\end{align}

\noindent where $f:U\rightarrow%
\mathbb{R}
$ is the eigenvalue function taking each eigenvector to its eigenvalue.

With those preliminaries, the definitions in Table 6 might be better motivated
and the statements clearer.

\begin{center}%
\begin{tabular}
[c]{|c|c|}\hline
`Classical' Logical Entropy & Quantum Logical Entropy\\\hline\hline
Pure state density matrix, e.g., $\rho\left(  \mathbf{0}_{U}\right)  $ & Pure
state density matrix $\rho\left(  \psi\right)  $\\\hline
$U=\left\{  u_{1},...,u_{n}\right\}  $ & ON basis simultaneous eigenvectors
$F,G$\\\hline
$p\times p$ on $U\times U$ & $\rho\left(  \psi\right)  \otimes\rho\left(
\psi\right)  $ on $V\otimes V$\\\hline
$h\left(  \mathbf{0}_{U}\right)  =1-\operatorname{tr}\left[  \rho\left(
\mathbf{0}_{U}\right)  ^{2}\right]  =0$ & $h\left(  \rho\left(  \psi\right)
\right)  =1-\operatorname{tr}\left[  \rho\left(  \psi\right)  ^{2}\right]
=0$\\\hline
$h\left(  \pi\right)  =p\times p\left(  \operatorname{dit}\left(  \pi\right)
\right)  $ & $h\left(  F:\psi\right)  =\operatorname{tr}\left[  P_{\left[
qudit\left(  F\right)  \right]  }\rho\left(  \psi\right)  \otimes\rho\left(
\psi\right)  \right]  $\\\hline
$h\left(  \pi,\sigma\right)  =p\times p\left(  \operatorname{dit}\left(
\pi\right)  \cup\operatorname{dit}\left(  \sigma\right)  \right)  $ &
$h\left(  F,G:\psi\right)  =\operatorname{tr}\left[  P_{\left[  qudit\left(
F\right)  \cup qudit\left(  G\right)  \right]  }\rho\left(  \psi\right)
\otimes\rho\left(  \psi\right)  \right]  $\\\hline
$h\left(  \pi|\sigma\right)  =p\times p\left(  \operatorname{dit}\left(
\pi\right)  -\operatorname{dit}\left(  \sigma\right)  \right)  $ & $h\left(
F|G:\psi\right)  =\operatorname{tr}\left[  P_{\left[  qudit\left(  F\right)
-qudit\left(  G\right)  \right]  }\rho\left(  \psi\right)  \otimes\rho\left(
\psi\right)  \right]  $\\\hline
$m\left(  \pi,\sigma\right)  =p\times p\left(  \operatorname{dit}\left(
\pi\right)  \cap\operatorname{dit}\left(  \sigma\right)  \right)  $ &
$m\left(  F,G:\psi\right)  =\operatorname{tr}\left[  P_{\left[  qudit\left(
F\right)  \cap qudit\left(  G\right)  \right]  }\rho\left(  \psi\right)
\otimes\rho\left(  \psi\right)  \right]  $\\\hline
$h\left(  \pi\right)  =h\left(  \pi|\sigma\right)  +m\left(  \pi
,\sigma\right)  $ & $h\left(  F:\psi\right)  =h\left(  F|G:\psi\right)
+m\left(  F,G:\psi\right)  $\\\hline
$h\left(  \pi\right)  =$ $2$-draw prob. diff. $f$-values & $h\left(
F:\psi\right)  =2$-meas. prob. diff. $F$-eigenvalues\\\hline
$\rho\left(  \pi\right)  =\sum_{i}P_{B_{i}}\rho\left(  \mathbf{0}_{U}\right)
P_{B_{i}}$ & $\hat{\rho}\left(  \psi\right)  =\sum_{i}P_{V_{i}}\rho\left(
\psi\right)  P_{V_{i}}$\\\hline
$h\left(  \pi\right)  =1-\operatorname{tr}\left[  \rho\left(  \pi\right)
^{2}\right]  $ & $h\left(  F:\psi\right)  =1-\operatorname{tr}\left[
\hat{\rho}\left(  \psi\right)  ^{2}\right]  $\\\hline
$h\left(  \pi\right)  =$ sum sq. zeroed $\rho\left(  \mathbf{0}_{U}\right)
\rightsquigarrow\rho\left(  \pi\right)  $ & $h\left(  F:\psi\right)  =$ sum
ab. sq. zeroed $\rho\left(  \psi\right)  \rightsquigarrow\hat{\rho}\left(
\psi\right)  $\\\hline
\end{tabular}

Table 6: Probabilities applied to ditsets and qudit spaces.
\end{center}

\subsection{Some basic results about quantum logical entropy}

A self-adjoint operator $F$ on $V$, i.e., an observable, alone defines the
eigenvalue partition $f^{-1}=\left\{  f^{-1}\left(  \phi_{i}\right)  \right\}
_{i\in I}$ on a basis $U$ of ON eigenvectors for $F$. But the points have no
probabilities associated with them. The probabilities are supplied by a
normalized vector $\left\vert \psi\right\rangle \in V$ as $p_{i}=\alpha
_{i}\alpha_{i}^{\ast}$ for $\alpha_{i}=\left\langle u_{i}|\psi\right\rangle $.
Then we have a completely classical situation, a set partition $f^{-1}$ on a
set $U$ with point probabilities provided by $\left\vert \psi\right\rangle $
which will be denoted $\pi\left(  F:\psi\right)  $. Hence that partition will
have a (classical) logical entropy $h\left(  \pi\left(  F:\psi\right)
\right)  $. Since the blocks in that $\pi\left(  F:\psi\right)  $ partition on
$U$ are the sets of basis vectors each for a certain eigenvalue, the
probabilities for those blocks are $\sum_{j}\left\{  p_{j}:f\left(
u_{j}\right)  =\phi_{i}\right\}  =\Pr\left(  f^{-1}\left(  \phi_{i}\right)
\right)  =\Pr\left(  B_{i}\right)  $ for $i=1,...,I.$ Hence we have:%
\begin{align}
h\left(  \pi\left(  F:\psi\right)  \right)   &  =1-\sum_{i\in I}\Pr\left(
f^{-1}\left(  \phi_{i}\right)  \right)  ^{2}=1-\sum_{i\in I}\left(  \sum
_{j}\left\{  p_{j}:f\left(  u_{j}\right)  =\phi_{i}\right\}  \right)
^{2}\nonumber\\
&  =1-\sum_{i}\left(  \sum_{f\left(  u_{j}\right)  =\phi_{i}}p_{j}^{2}%
+\sum_{j\neq k}\left\{  p_{j}p_{k}:f\left(  u_{j}\right)  =\phi_{i}=f\left(
u_{k}\right)  \right\}  \right) \nonumber\\
&  =\sum_{j,k}\left\{  p_{j}p_{k}:f\left(  u_{j}\right)  \neq f\left(
u_{k}\right)  \right\}  =h\left(  F:\psi\right)  =\operatorname{tr}\left[
P_{\left[  qudit\left(  F\right)  \right]  }\rho\left(  \psi\right)
\otimes\rho\left(  \psi\right)  \right]  \tag{4.17}%
\end{align}

And there is another way to arrive at this logical entropy, namely perform the
$F$-measurement on the pure state density matrix $\rho\left(  \psi\right)  $.
The results of the $F$-measurement is given by the L\"{u}ders mixture
operation \cite[p. 279]{auletta:qm} on the density matrix $\rho\left(
\psi\right)  $. The block $B_{i}\in\pi\left(  F:\psi\right)  $ generates the
eigenspace $V_{i}$ corresponding to the eigenvalue $\phi_{i}$ so $P_{V_{i}}$
is the projection matrix to that eigenspace for $i=1,...,I$. Then the
L\"{u}ders mixture operation, representing the $F$-measurement of $\psi$,
gives the mixed state density matrix:%
\begin{equation}
\hat{\rho}\left(  \psi\right)  =\sum_{i\in I}P_{V_{i}}\rho\left(  \psi\right)
P_{V_{i}}\text{.} \tag{4.18}%
\end{equation}

To show that $h\left(  \hat{\rho}\left(  \psi\right)  \right)
=1-\operatorname*{tr}\left[  \hat{\rho}\left(  \psi\right)  ^{2}\right]
=h\left(  \pi\left(  F:\psi\right)  \right)  $ for $\hat{\rho}\left(
\psi\right)  =\sum_{i=1}^{I}P_{V_{i}}\rho\left(  \psi\right)  P_{V_{i}}$, we
need to compute $\operatorname*{tr}\left[  \hat{\rho}\left(  \psi\right)
^{2}\right]  $. An off-diagonal element in $\rho_{jk}\left(  \psi\right)
=\alpha_{j}\alpha_{k}^{\ast}$ of $\rho\left(  \psi\right)  $ survives (i.e.,
is not zeroed and has the same value) the L\"{u}ders operation if and only if
$f\left(  u_{j}\right)  =f\left(  u_{k}\right)  $. Hence, the $j$-th diagonal
element of $\hat{\rho}\left(  \psi\right)  ^{2}$ is:%
\begin{equation}
\sum_{k=1}^{n}\left\{  \alpha_{j}^{\ast}\alpha_{k}\alpha_{j}\alpha_{k}^{\ast
}:\phi\left(  u_{j}\right)  =\phi\left(  u_{k}\right)  \right\}  =\sum
_{k=1}^{n}\left\{  p_{j}p_{k}:f\left(  u_{j}\right)  =f\left(  u_{k}\right)
\right\}  =p_{j}\Pr\left(  B_{i}\right)  \tag{4.19}%
\end{equation}

\noindent where $u_{j}\in B_{i}$. Then, grouping the $j$-th diagonal elements
for $u_{j}\in B_{i}$ gives $\sum_{u_{j}\in B_{i}}p_{j}\Pr\left(  B_{i}\right)
=\Pr\left(  B_{i}\right)  ^{2}$. Hence, the whole trace is:
$\operatorname*{tr}\left[  \hat{\rho}\left(  \psi\right)  ^{2}\right]
=\sum_{i=1}^{I}\Pr\left(  B_{i}\right)  ^{2}$, and thus:%
\begin{equation}
h\left(  \hat{\rho}\left(  \psi\right)  \right)  =1-\operatorname*{tr}\left[
\hat{\rho}\left(  \psi\right)  ^{2}\right]  =1-\sum_{i=1}^{I}\Pr\left(
B_{i}\right)  ^{2}=h\left(  F:\psi\right)  \text{.} \tag{4.20}%
\end{equation}

\noindent This completes the proof of the following theorem which shows the
different ways to characterize $h\left(  F:\psi\right)  $.

\begin{theorem}
[4.2]$h\left(  F:\psi\right)  =h\left(  \pi\left(  F:\psi\right)  \right)
=h\left(  \hat{\rho}\left(  \psi\right)  \right)  $.$\square$
\end{theorem}

Like the classical join operation on partitions, \textit{quantum measurement
creates distinctions}, i.e., turns coherences into \textquotedblleft
decoherences\textquotedblright,\footnote{This notion of "decoherence" is used
in an older sense, not the recent sense given by the work of Zurek
\cite{zurek:decoh} and others.} which, classically, is the operation of
distinguishing elements by classifying them according to some attribute like
classifying the faces of a die by their parity. The fundamental theorem about
quantum logical entropy and projective measurement, in the density {matrix}
version, shows how the quantum logical entropy created (starting with
$h\left(  \rho\left(  \psi\right)  \right)  =0$ for the pure state $\psi$) by
the measurement can be computed directly from the coherences of $\rho\left(
\psi\right)  $ that are decohered in $\hat{\rho}\left(  \psi\right)  $.

\begin{theorem}
[4.3 Fundamental theorem about quantum measurement and logical entropy.]The
increase in quantum logical entropy, $h\left(  \hat{\rho}\left(  \psi\right)
\right)  =h\left(  F:\psi\right)  $ due to the $F$-measurement of the pure
state $\psi$ is the sum of the absolute squares of the non-zero off-diagonal
terms (coherences) in $\rho\left(  \psi\right)  $ (represented in an ON basis
of $F$-eigenvectors) that are zeroed (`decohered') in the post-measurement
L\"{u}ders mixture density matrix $\hat{\rho}\left(  \psi\right)  =\sum
_{i=1}^{I}P_{V_{i}}\rho\left(  \psi\right)  P_{V_{i}}$.
\end{theorem}

\textbf{Proof:} $h\left(  \hat{\rho}\left(  \psi\right)  \right)  -h\left(
\rho\left(  \psi\right)  \right)  =\left(  1-\operatorname*{tr}\left[
\hat{\rho}\left(  \psi\right)  ^{2}\right]  \right)  -\left(
1-\operatorname*{tr}\left[  \rho\left(  \psi\right)  ^{2}\right]  \right)
=\sum_{j,k}\left(  \left\vert \rho_{jk}\left(  \psi\right)  \right\vert
^{2}-\left\vert \hat{\rho}\left(  \psi\right)  \right\vert ^{2}\right)  $
since $\operatorname{tr}\left[  \rho^{2}\right]  =\sum_{i,j}\left\vert
\rho_{ij}\right\vert ^{2}$ is the sum of the absolute squares of all the
elements of $\rho$ \cite[p. 77]{fano:density}. If $u_{j}$ and $u_{k}$ are a
qudit of $F$, then and only then are the corresponding off-diagonal terms
zeroed by the L\"{u}ders mixture operation $\sum_{i=1}^{I}P_{V_{i}}\rho\left(
\psi\right)  P_{V_{i}}$ to obtain $\hat{\rho}\left(  \psi\right)  $ from
$\rho\left(  \psi\right)  $. $\square$

\textbf{Example}: For a simple quantum example, consider a system with two
spin-observable $\sigma$ eigenstates$\ \left\vert \uparrow\right\rangle $ and
$\left\vert \downarrow\right\rangle $ (like electron spin up or down along the
$z$-axis) where the given normalized superposition state is $\left\vert
\psi\right\rangle =\alpha_{\uparrow}\left\vert \uparrow\right\rangle
+\alpha_{\downarrow}\left\vert \downarrow\right\rangle =%
\begin{bmatrix}
\alpha_{\uparrow}\\
\alpha_{\downarrow}%
\end{bmatrix}
$ so the density matrix is $\rho\left(  \psi\right)  =%
\begin{bmatrix}
p_{\uparrow} & \alpha_{\uparrow}\alpha_{\downarrow}^{\ast}\\
\alpha_{\downarrow}\alpha_{\uparrow}^{\ast} & p_{\downarrow}%
\end{bmatrix}
$ where $p_{\uparrow}=\alpha_{\uparrow}\alpha_{\uparrow}^{\ast}$ and
$p_{\downarrow}=\alpha_{\downarrow}\alpha_{\downarrow}^{\ast}$. Using the
L\"{u}ders mixture operation, the measurement of that spin-observable $\sigma$
goes from the pure state $\rho\left(  \psi\right)  $ to%
\begin{align}
&  P_{\uparrow}\rho\left(  \psi\right)  P_{\uparrow}+P_{\downarrow}\rho\left(
\psi\right)  P_{\downarrow}\nonumber\\
&  =%
\begin{bmatrix}
1 & 0\\
0 & 0
\end{bmatrix}%
\begin{bmatrix}
p_{\uparrow} & \alpha_{\uparrow}\alpha_{\downarrow}^{\ast}\\
\alpha_{\downarrow}\alpha_{\uparrow}^{\ast} & p_{\downarrow}%
\end{bmatrix}%
\begin{bmatrix}
1 & 0\\
0 & 0
\end{bmatrix}
+%
\begin{bmatrix}
0 & 0\\
0 & 1
\end{bmatrix}%
\begin{bmatrix}
p_{\uparrow} & \alpha_{\uparrow}\alpha_{\downarrow}^{\ast}\\
\alpha_{\downarrow}\alpha_{\uparrow}^{\ast} & p_{\downarrow}%
\end{bmatrix}%
\begin{bmatrix}
0 & 0\\
0 & 1
\end{bmatrix}
\nonumber\\
&  =%
\begin{bmatrix}
p_{\uparrow} & 0\\
0 & p_{\downarrow}%
\end{bmatrix}
=\hat{\rho}\left(  \psi\right)  . \tag{4.21}%
\end{align}

\noindent The logical entropy of any pure state such as $\rho\left(
\psi\right)  $ is $0$. The logical entropy of $\hat{\rho}\left(  \psi\right)
$ is $h\left(  \hat{\rho}\left(  \psi\right)  \right)  =1-\operatorname{tr}%
\left[  \hat{\rho}\left(  \psi\right)  ^{2}\right]  =1-p_{\uparrow}%
^{2}-p_{\downarrow}^{2}$. The entries that were zeroed in the L\"{u}ders
mixture operation were the two off-diagonal elements $\alpha_{\uparrow}%
\alpha_{\downarrow}^{\ast}$ and $\alpha_{\downarrow}\alpha_{\uparrow}^{\ast}$
so the sum of their absolute squares is $2\alpha_{\uparrow}\alpha_{\downarrow
}^{\ast}\alpha_{\downarrow}\alpha_{\uparrow}^{\ast}=2p_{\uparrow}%
p_{\downarrow}$ which equals $1-p_{\uparrow}^{2}-p_{\downarrow}^{2}$ since
$1=\left(  p_{\uparrow}+p_{\downarrow}\right)  ^{2}=p_{\uparrow}%
^{2}+p_{\downarrow}^{2}+2p_{\uparrow}p_{\downarrow}$.

\section{Conclusions}

The underlying thesis is that information is defined in terms of distinctions,
differences, distinguishability, and diversity--or, with similar uses of the
di-prefix (which means "two"), discriminations, divisions, or
differentiations. Yet those are all vague concepts so this notion of
information-as-distinctions is made precise using the basic mathematical
concept that represents differences and non-differences (or equivalences),
namely partitions (including the inverse-image partitions of random
variables). The elements in the same block of a partition are similar or
equivalent (block = equivalence class), and the ordered pairs of elements in
\textit{different} blocks are the distinctions or dits. Hence logical entropy
measures information-as-distinctions by the probability measure on
distinctions, so the logical entropy of a partition is the probability that a
distinction of the partition is obtained in two independent draws from the
underling universe set of elements. This notion of information-as-distinctions
then encompasses the Shannon notion of entropy as the average minimum number
of binary partitions (bits) that have to be joined to make the same
distinctions of the partition. Moreover, there is the dit-bit transform that
derives all of Shannon's definitions of entropy, joint entropy, conditional
entropy, and mutual information from the corresponding definitions of logical
entropy that are based on logical entropy being defined as a (probability)
measure in the sense of measure theory. A few applications were discussed;
distinguishing the Boltzmann and Shannon entropies, developing the MaxEntropy
method with logical entropy, and showing how the metrical notion of logical
entropy gives the notion of variance in statistical theory.

There is a method, linearization, to lift set-based concepts to the
corresponding vector-space concepts, and that provides the method to develop
the corresponding quantum notions from the `classical' or non-quantum notions
of logical entropy. There are two equivalent formulations of quantum
mechanics; one using wave functions and the other using density matrices
\cite[p. 102]{nielsen-chuang:bible}. But only one of those formulations maps
naturally to the mathematics of partitions, namely the density matrix formulation.

At the beginning of our presentation, density matrices were foreshadowed by
the box diagrams representing logical entropy. The box diagrams led to the
incidence matrices for $\operatorname{indit}\left(  \pi\right)  $, or the
complementary ones for $\operatorname{dit}\left(  \pi\right)  $, and then
point probabilities are introduced into the matrices so that when normalized
by their trace, the matrices are density matrices. In that manner, a
reformulation of the classical logical entropy framework is first presented
using density matrices over the real numbers to foreshadow the later quantum
results over the complex numbers. Every density matrix over the complex
numbers has a spectral decomposition into a probability mixture of orthogonal
pure states which correspond classically to the disjoint blocks and block
probabilities of a partition.

The fundamental theorem for logical entropy and measurement shows there is a
simple, direct and quantitative connection between density matrices and
logical entropy. The theorem directly connects the changes in the density
matrix due to a projective measurement (sum of absolute squares of zeroed
off-diagonal terms) with the increase in logical entropy due to the
$F$-measurement $h\left(  F:\psi\right)  =h\left(  \hat{\rho}\left(
\psi\right)  \right)  $ (where $h\left(  \rho\left(  \psi\right)  \right)  =0$
for the pure state $\psi$). Moreover, the quantum logical entropy has a simple
\textquotedblleft two-draw probability\textquotedblright\ interpretation,
i.e.,~$h\left(  F:\psi\right)  =h\left(  \hat{\rho}\left(  \psi\right)
\right)  $ is the probability that two independent $F$-measurements of $\psi$
will yield distinct $F$-eigenvalues, i.e., will yield a qudit of $F$. In
contrast, the von Neumann entropy has no such simple interpretation, and there
seems to be no such intuitive connection between pre- and post-measurement
density matrices and von Neumann entropy, although von Neumann entropy also
increases in a projective measurement \cite[Theorem 11.9, p. 515]%
{nielsen-chuang:bible}.

This direct quantitative connection between state discrimination and quantum
logical entropy reinforces the judgment of Boaz Tamir and Eliahu Cohen
(\cite{tamir-cohen:logentqm}; \cite{tamir-cohen:holevo}) that quantum logical
entropy is a natural and informative entropy concept for quantum mechanics.

\begin{quotation}
\noindent We find this framework of partitions and distinction most suitable
(at least conceptually) for describing the problems of quantum state
discrimination, quantum cryptography and in general, for discussing quantum
channel capacity. In these problems, we are basically interested in a distance
measure between such sets of states, and this is exactly the kind of knowledge
provided by logical entropy [Reference to \cite{ell:cd}]. \cite[p.
1]{tamir-cohen:logentqm}
\end{quotation}

In summary, the basic idea of information as distinctions, differences,
distinguishability, and diversity is naturally quantified at the `classical'
level in terms of logical entropy and then naturally linearized to the quantum
notion of logical entropy.\footnote{For further developments beyond the scope
of this paper see \cite{ell:quantum-entropy}, \cite{ell:NF4IT},
\cite{tamir-et-al:qle}, and the other papers in this issue.}

No funding, no conflicts, and no acknowledgements.

\end{document}